\documentclass[twocolumn,aps,pra,showpacs,superscriptaddress]{revtex4}
\usepackage[latin9]{inputenc}
\usepackage{amsmath}
\usepackage{amssymb}
\usepackage{graphicx}
\usepackage{float}

\usepackage{color}
\usepackage{braket} 

\usepackage[colorlinks, linkcolor=blue, citecolor=blue, urlcolor=blue, breaklinks=blue]{hyperref}

\makeatletter

\@ifundefined{textcolor}{}
{%
 \definecolor{BLACK}{gray}{0}
 \definecolor{WHITE}{gray}{1}
 \definecolor{RED}{rgb}{1,0,0}
 \definecolor{GREEN}{rgb}{0,1,0}
 \definecolor{BLUE}{rgb}{0,0,1}
 \definecolor{CYAN}{cmyk}{1,0,0,0}
 \definecolor{MAGENTA}{cmyk}{0,1,0,0}
 \definecolor{YELLOW}{cmyk}{0,0,1,0}
}

\definecolor{mygreen}{rgb}{0,0.5,0}
\definecolor{myblue}{rgb}{0,0,0.75}
\definecolor{mymagenta}{cmyk}{0,1,0,0.12}

\usepackage{umoline}

\newcommand{\be}{\begin{equation}}
\newcommand{\eeq}{\end{equation}}
\newcommand{\bea}{\begin{eqnarray}}
\newcommand{\eea}{\end{eqnarray}}

\def\app#1#2{%
  \mathrel{%
    \setbox0=\hbox{$#1\sim$}%
    \setbox2=\hbox{%
      \rlap{\hbox{$#1\propto$}}%
      \lower1.1\ht0\box0%
    }%
    \raise0.25\ht2\box2%
  }%
}

\usepackage{color}\usepackage{dsfont}

\usepackage{rotating}\usepackage{braket}\usepackage{amsthm}\usepackage{bbm}

\usepackage[abs]{overpic}

\makeatother

\begin{document}

\title{Trapped-ion quantum simulation of excitation transport: disordered, noisy, and long-range connected quantum networks}

\date{\today}

\author{N.~Trautmann}
\email{nils.trautmann@physik.tu-darmstadt.de}
\affiliation{Institut f\"ur Angewandte Physik, Technische Universit\"at Darmstadt,D-64289,
  Germany}

\author{P.~Hauke}
\email{philipp.hauke@kip.uni-heidelberg.de} 
\affiliation{Kirchhoff-Institute for Physics, Heidelberg University, 69120 Heidelberg, Germany}
\affiliation{Institute for Theoretical Physics, Heidelberg University, 69120 Heidelberg, Germany}
\affiliation{Institute for Quantum Optics and Quantum Information of the Austrian Academy of Sciences, 6020 Innsbruck, Austria}
\affiliation{Institute for Theoretical Physics, University of Innsbruck, 6020 Innsbruck, Austria}

\begin{abstract}

The transport of excitations governs fundamental properties of matter. 
Particularly rich physics emerges in the interplay between disorder and environmental noise, even in small systems such as photosynthetic biomolecules. 
Counterintuitively, noise can enhance coherent quantum transport, which has been proposed as a mechanism behind the high transport efficiencies observed in photosynthetic complexes. This effect has been called "environment-assisted quantum transport" (ENAQT). 
Here, we propose a quantum simulation of the excitation transport in an open quantum network, taking advantage of the high controllability of current trapped-ion experiments. 
Our scheme allows for the controlled study of various different aspects of the excitation transfer, ranging from the influence of static disorder and interaction range, over the effect of Markovian and non-Markovian dephasing, to the impact of a continuous insertion of excitations. 
Our proposal discusses experimental error sources and realistic parameters, showing that it can be implemented in state-of-the-art ion-chain experiments.  
\end{abstract}
\maketitle

\section{Introduction\label{sec:Introduction}}
The way how excitations propagate through a network defines the fundamental properties of matter from large solids to small molecules. 
Extremely rich physics can be at play even in apparently simple systems, especially if coupled to an outside environment, such as happens in photosynthesic complexes \cite{Fassioli2013,HuelgaPlenio2013,Lambert2013}. 
In such biomolecules, photon energy is absorbed in pigments of a photosynthetic antenna, creating an exciton quasiparticle. 
The exciton is then transferred to a reaction center where the energy is harvested in a biochemical process. 
The surprisingly high efficiency of this energy transfer triggered several decades of active research (see, e.g., \cite{hu1997pigment,ritz2001kinetics,novoderezhkin2004energy,cho2005exciton}). 
After experiments demonstrated the presence of long-lived coherences in the dynamics of the Fenna--Matthews--Olson (FMO) complex \cite{engel2007evidence,Lee2007a,Panitchayangkoon2010}, various theoretical investigations suggested that quantum dynamical processes are of major importance for the excitation transport in such biological systems \cite{mohseni2008environment,plenio2008dephasing,caruso2009highly,Thorwart2009,chen2011excitation,chin2013role,delRey2013,mohseni2014energy}. 
These studies found that the Anderson localization of excitations, induced by static disorder within the network, can be lifted by dephasing, induced by coupling to the environment. The result is an unexpectedly large transfer efficiency, termed environment-assisted quantum transport (ENAQT). 
Whether this effect actually appears in biomolecules is, however, disputed, because the illuminating sunlight is incoherent \cite{Fassioli2013}. Moreover, the effect depends on the precise way the transfer from the network to the reaction center is modeled \cite{Leon2014}. 
Recently, model experiments have started investigating ENAQT in small networks of photonic wave-guides \cite{Viciani2015,Biggerstaff2016}, classical electrical oscillators \cite{Leon2015}, and superconducting qubits \cite{Mostame2012,Potocnik2017}. 
Proposals exist also to analyse ENAQT in embedded Rydberg aggregates \cite{Schempp2015,Schoenleber2015,Genkin2016,Schoenleber2016}, where first studies on the quantum transport under dissipation have already been performed \cite{Guenter2013}.

\begin{figure}
	\begin{centering}
		\begin{overpic}[width=7cm,tics=5]
			{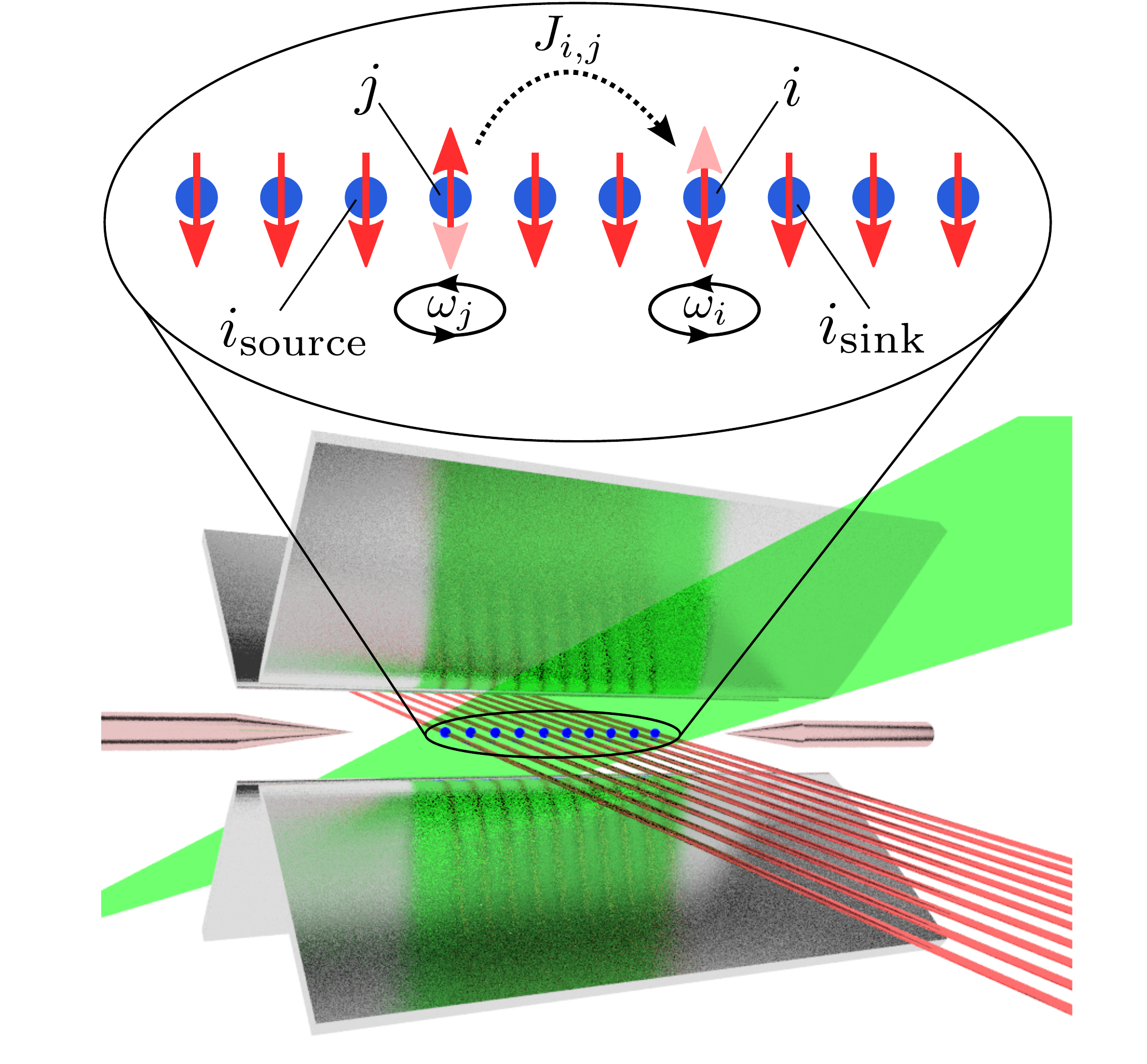}
			\put(0,170){ \large a)}
			\put(0,115){ \large b)}
		\end{overpic}
		\par\end{centering}
	\caption{Proposed quantum-simulation setup to study excitation transfer in the interplay between engineered disorder, dephasing, and long-range hopping. 
		(a) Spin model of the open quantum network. 
		Spins at sites $i$ have disordered on-site energies $\hbar\omega_{i}$, and are coupled to other spins with hopping strength $J_{ij}$. 
		The considered scenario is the propagation of an initial excitation injected at site $i_{\text{source}}$ to the target site $i_{\text{sink}}$. 
		(b) Illustration of the proposed ion-trap implementation.
		The spins are mapped onto internal, electric degrees of freedom of ions (blue dots) arranged in a linear chain. 
		The quantum simulation requires laser beams with single-site addressability (red)
		to locally adjust AC-Stark shifts, which induce disorder and dephasing, as well as a broad laser beam  
		illuminating the entire ion chain (green), which drives  the interactions $J_{ij}$. \label{fig:Setup}}
\end{figure}

Here, we discuss how the precisely controllable platform of trapped ions can be used to realize engineered noise, and thus investigate ENAQT with various highly tunable parameters. These include disorder strength, environmental dephasing rate, noise spectrum, and hopping range of excitations. Importantly, the trapped-ion platform allows also for control of injection and extraction rates, enabling a systematic study of the effect of interactions on the quantum transport. 
Rather than aiming at a quantitative modeling of ENAQT in biomolecules, our aim in this work is to illustrate the broad range of phenomena that can be explored in trapped-ion quantum simulation. 
Our numerical studies should be of independent interest for understanding the excitation transfer in quantum networks. Our theoretical proposal is accompanied by a detailed discussion of the experimental error sources, showing that the proposed quantum simulation can be implemented in state-of-the-art experiments. 
The engineered noise brings a new dimension to transport experiments in trapped ions or cold gases, which have investigated clean \cite{Cheneau2012,jurcevic2014quasiparticle,Richerme2014,Krinner2015} as well as disordered systems \cite{Roati2008,Billy2008,Kondov2011,Semeghini2014,Schreiber2015,smith2015many}, but mostly viewing environmental noise as a nuisance. Here, we are interested in the beneficial effects of engineered noise for enhanced quantum transport. 

The quantum network we are interested in is represented by a spin model [Fig.~\ref{fig:Setup}(a)], which can be mapped onto internal, electric degrees of freedom of the ions [Fig.~\ref{fig:Setup}(b)]. 
Tunable hoppings of excitations are generated by spin-dependent forces, induced by lasers and transmitted by phonons \cite{sorensen1999quantum,Porras2004a,Friedenauer2008,kim2009entanglement,Britton2012,jurcevic2014quasiparticle}. 
We focus on linear chains with interactions that have approximately a power-law distance dependence, as are natural in various trapped-ion experiments \cite{kim2009entanglement,Britton2012,jurcevic2014quasiparticle}, but the proposal can also be applied to other realizable geometries \cite{Bermudez2012b,Britton2012,Bohnet2016,Richerme2016}. 
In particular, the irregular networks found in biological molecules could be designed in special trap designs such as segmented Paul traps \cite{Zippilli2014} or surface traps \cite{Schmied2009b}, as well as by engineering the interactions via time-periodic driving \cite{Nevado2017} or additional laser frequencies \cite{Korenblit2012}.  
Addressable AC-Stark shifts have been shown to enable the generation of programmable disorder \cite{smith2015many} and have been proposed for the simulation of dephasing in adiabatic quantum optimization \cite{Hauke2015}. Amplitude and phase modulations have also been used to simulate a qubit in a dephasing environment \cite{Soare2014}. 
These ingredients render trapped ions a highly versatile platform for investigating the excitation transfer in open quantum networks.

Below, we will use detailed numerical investigations to discuss how the quantum transport depends on various parameters that become accessible in this platform, especially hopping range, disorder strength, Markovian and non-Markovian dephasing, and non-linear effects appearing at large injection rates of excitations. We now briefly summarize the main effects for each of these parameters. 

A series of recent works has demonstrated that the propagation of excitations in regular networks with power-law hoppings depends crucially on the hopping range \cite{hauke2013spread,Eisert2013,Foss-Feig2015,Cevolani2015}:  
If the hopping strength diminishes fast with distance, the propagation is bound to effective sound cones{, similar to what happens for exponentially localized interactions \cite{Lieb1972}. 
In the opposite limit of infinite-range hopping, destructive interference between hopping paths suppresses the transfer efficiency \cite{caruso2009highly}. 
For large but finite hopping range, this effect is weakened, leading to extremely slow excitation modes balanced by other modes with divergent propagation speed \cite{hauke2013spread,Cevolani2015}. 
In our numerical studies we find that this co-existence increases transfer efficiency at short times as compared to short-range interactions and decreases it at longer times. 
Disorder has a non-trivial effect on this behavior:  
While large amounts of disorder leads to Anderson localization \cite{anderson1958absence} or many-body localization \cite{smith2015many}, weak disorder counteracts the destructive interference induced by the long-range hopping \cite{caruso2009highly}, thus actually improving the transfer efficiency.  
Trapped-ion experiments have already observed the fast excitation modes appearing with power-law interactions \cite{Richerme2014,jurcevic2014quasiparticle}, as well as the obstruction of thermalization due to long-range interactions \cite{neyenhuis2016observation} as well as many-body localization. Here, we propose to perform similar experiments in view of the efficiency of excitation transport and in the interplay with engineered noise.  

Although noise destroys the linear superpositions that are a main feature of quantum mechanics, it is known that dephasing can lift Anderson localization and thus lead to ENAQT \cite{mohseni2008environment,caruso2009highly,plenio2008dephasing}. When the dephasing becomes too strong, however, a quantum Zeno dynamics sets in that freezes the excitation \cite{misra1977zeno}. Thus, maximum transfer efficiencies are attained in an intermediate regime of dephasing. 
Moreover, the character of the noise can be adjusted from Markovian to non-Markovian. We compare Markovian noise to a simple non-Markovian process, where we find that---while the maximum achievable transfer efficiency is similar for both types of noise---the non-Markovian dephasing can yield high transfer efficiencies in a broader parameter range.
Recent theoretical investigations have shown that non-Markovian baths 
play a crucial role in ENAQT. They enhance coherences \cite{Thorwart2009,chin2013role,Jesenko2013} and baths that are structured to fit the energy spectrum of the network can strongly improve the energy transfer \cite{chen2011excitation,delRey2013,mohseni2014energy}. 
This advantage disappears, however, at long times if the transfer to the reaction center is the only loss mechanism \cite{Jesenko2013}, showing the delicate interplay between different dissipative and coherent effects. 
In biomolecules, structured non-Markovian noise appear through coupling to phonon modes. Here, they can be designed by hand by adjusting the power spectra of the engineered dephasing. 

Finally, the presented scheme can also be used to investigate non-linear effects in the spin network, which appear in the presence of multiple excitations. We investigate the dynamics of a driven-dissipative system in which the excitations are continuously injected into the system by incoherently coupling a source site to an infinite-temperature heat bath. The injection of excitations from a heat bath is of particular significance for simulating the dynamics of biomolecules, as it resembles the incoherent absorption of photons by photosynthetic systems \cite{Fassioli2013}. We find that there exists a finite value of the driving that yields optimal transfer rates. 

The body of this article is structured as follows. 
First, in Sec.~\ref{sec:Model}, we introduce the model of the quantum network that we propose to simulate. 
We show how this model can be implemented in the ion chain in Sec.~\ref{sec:Quantum_Simulation}. 
In Sec.~\ref{sec:Numerical_Results}, we present numerical studies for various scenarios of excitation transfer that may be simulated in the ion chain. 
In Sec.~\ref{sec:Experimental_Error_Sources}, we address the robustness of the simulation towards possible sources of errors in a realistic experimental setup. 
Finally, in Sec.~\ref{sec:Conclusion}, we present our conclusions.

\section{Model of the quantum network\label{sec:Model}}

In this section, we introduce the model that we use for studying the excitation transfer, depicted in Fig.~\ref{fig:Setup}(a). Its realization in a chain of trapped ions is discussed in the next section. 
In photosynthetic complexes, the transport is governed by exciton quasiparticles. 
Their bosonic commutation relations are naturally matched by the internal pseudo-spins manipulated in trapped-ion setups. 
The dynamics of such a spin network, consisting of $N$ connected sites, is modeled by the Hamiltonian 
\be
\hat{H}=\hat{H}_{J}+\hat{H}_{\omega_{i}}
\eeq
with 
\begin{eqnarray}
\hat{H}_{J} & = & \hbar\sum_{i<j}J_{ij}\hat{\sigma}_{i}^{+}\hat{\sigma}_{j}^{-}+\text{H.c.}\,,\\
\hat{H}_{\omega_{i}} & = & \hbar\sum_{i}\omega_{i}\hat{\sigma}_{i}^{+}\hat{\sigma}_{i}^{-}\,. 
\end{eqnarray}
Here, $\hat{\sigma}_{i}^{+}$($\hat{\sigma}_{i}^{-}$) are spin raising
(lowering) operators for site $i$, $\hbar\omega_{i}$ is the on-site
excitation energy, and $J_{ij}$ denote the coupling strengths between spin $i$ and $j$. 
We denote the eigenvectors of the operator $\hat{\sigma}_{i}^{+}\hat{\sigma}_{i}^{-}$, which counts the presence/absence of an excitation at site $i$, with $\mid\uparrow\rangle_{i}$ for the eigenvalue $1$ and $\mid\downarrow\rangle_{i}$ for the eigenvalue $0$. 

In the following, we focus mainly on a situation where an exciton quasiparticle has just been generated. We model this situation through an initial state with only the spin at site $i_{\text{source}}$
in state $\mid\uparrow\rangle$ and all the other spins in state $\mid\downarrow\rangle$, i.e., 
\be
\label{eq:initialState}
\ket{\psi\left(t_{0}\right)}=\mid\downarrow\rangle_{1}\mid\downarrow\rangle_{2}\dots\mid\uparrow\rangle_{i_{\text{source}}}\mid\downarrow\rangle_{i_{\text{source}}+1}\dots\mid\downarrow\rangle_{N}\,.
\eeq
We are interested in the transfer of the initial excitation to the 
remote site $i_{\text{sink}}$, where the excitation is absorbed and
removed from the quantum network. The absorption of the excitations
is modeled by a Markovian dissipation process, described by the Lindblad
super operator 
\begin{eqnarray}
\mathcal{L}_{\text{diss}}(\hat{\rho}) & = & \frac{\Gamma}{2}\left[-\left\{ \hat{\sigma}_{i_{\text{sink}}}^{+}\hat{\sigma}_{i_{\text{sink}}}^{-},\hat{\rho}\right\} +2\hat{\sigma}_{i_{\text{sink}}}^{-}\rho\hat{\sigma}_{i_{\text{sink}}}^{+}\right]\,,
\end{eqnarray}
with $\Gamma$ being the rate at which an excitation at site $i_{\text{sink}}$
is removed from the quantum network. 

Moreover, we consider the transport behavior under dephasing processes that act independently on each site, as are caused by environmental noise.
If the correlation time of the noise goes to zero, the dephasing process is Markovian, which can be modeled by 
the Lindblad super operator 
\begin{eqnarray}
\label{eq:Ldeph}
\mathcal{L}_{\text{deph}}(\hat{\rho}) & = & \sum_{i}\frac{\gamma_{i}}{2}\left[-\left\{ \hat{\sigma}_{i}^{+}\hat{\sigma}_{i}^{-},\hat{\rho}\right\} +2\hat{\sigma}_{i}^{+}\hat{\sigma}_{i}^{-}\hat{\rho}\hat{\sigma}_{i}^{+}\hat{\sigma}_{i}^{-}\right]
\end{eqnarray}
with $\gamma_{i}$ denoting the dephasing rate at site $i$.
In the following, we will also study non-Markovian dephasing caused by 
noise generated by the Goldstein--Kac telegraph process \cite{goldstein1951diffusion,masoliver1989continuous}.
This process allows us to adjust the bath correlation time, thereby enabling us to study the crossover from Markovian to non-Markovian noise.

In total, the time evolution of the quantum network is described by the Master equation 
\be
\label{eq:Master equation}
\frac{d}{dt}\hat{\rho}=-\frac{i}{\hbar}\left[\hat{H},\hat{\rho}\right]+\mathcal{L}_{\text{diss}}(\hat{\rho})+\mathcal{L}_{\text{deph}}(\hat{\rho})\,.
\eeq

The dynamics of the model discussed above is confined to the single-excitation and zero-excitation sectors, as $\hat{H}$ and $\mathcal{L}_{\text{deph}}$ preserve the excitation number $\bra{\psi\left(t_{0}\right)}\hat{N}_{\rm exc}\ket{\psi\left(t_{0}\right)}=1$ with $\hat{N}_{\rm exc}=\sum_i \hat{\sigma}_{i}^{+}\hat{\sigma}_{i}^{-}$,  and $\mathcal{L}_{\text{diss}}$ can only reduce it. 
Here, interactions between the excitations are of no importance. However, the framework presented in this article can also be used to investigate the non-linear dynamics of a spin network when 
several excitations are present. In Sec.~\ref{sec:drivenSystem}, we study the regime where a large number of excitations is injected by a continuous drive. This driven dynamics can be modeled by the Markovian process 
\begin{eqnarray}
\label{eq:Lsource}
 &  & \mathcal{L}_{\text{source}}(\hat{\rho})=\\
 &  & \phantom{-}\frac{\Gamma_{\text{source}}}{2}\left[-\left\{ \hat{\sigma}_{i_{\text{source}}}^{+}\hat{\sigma}_{i_{\text{source}}}^{-},\hat{\rho}\right\} +2\hat{\sigma}_{i_{\text{source}}}^{-}\rho\hat{\sigma}_{i_{\text{source}}}^{+}\right]\nonumber\\
 &  & -\frac{\Gamma_{\text{source}}}{2}\left[-\left\{ \hat{\sigma}_{i_{\text{source}}}^{-}\hat{\sigma}_{i_{\text{source}}}^{+},\hat{\rho}\right\} +2\hat{\sigma}_{i_{\text{source}}}^{+}\rho\hat{\sigma}_{i_{\text{source}}}^{-}\right]\,,\nonumber
\end{eqnarray}
which describes the incoherent creation and annihilation of particles at site $i_{\text{source}}$ caused by the coupling to an infinite-temperature heat bath, modeling the absorption of photons by photosynthetic systems \cite{mohseni2008environment,plenio2008dephasing,caruso2009highly}.

\section{Mapping of the Model to an ion chain\label{sec:Quantum_Simulation}}

In this section, we discuss an implementation of the above model in
an ion-trap quantum simulation. A schematic representation of such
an ion trap is depicted in Fig.~\ref{fig:Setup}(b). We focus here 
on the setup described in Ref.~\cite{schindler2013quantum} for $^{40}\text{Ca}^{+}$ ions. Similar
considerations also apply to other experimental implementations. We
consider an ion chain with $N$ ions, confined in a linear Paul trap with axial trapping frequency $\omega_{z}$ and radial trapping frequencies $\omega_{x,y}$.
Each site of the spin model is represented by a qubit encoded in the internal level structure of a single ion, $\ket{\uparrow}_{i}=\ket{3^{2}D_{5/2}\, m_{j}=-1/2}_{i}$ and $\ket{\downarrow}_{i}=\ket{4^{2}S_{1/2}\, m_{j}=-1/2}_{i}$.

\subsection{Implementation of the hopping terms\label{sub:Implementation_hopping}}

The hopping of an excitation between sites as described
by $\hat{H}_{J}$ can be generated by a non-local interaction between the
qubits in the ion chain, such as in Mølmer--Sørenson-type protocols \cite{sorensen1999quantum,kim2009entanglement,jurcevic2014quasiparticle}. 
The Mølmer--Sørenson interaction is driven by a laser
field that illuminates the entire ion register uniformly with two frequencies
$\omega_{\pm}=\omega_{0}\pm\Delta$, being $\omega_{0}$ the
frequency of the atomic transition $\ket{\uparrow}\leftrightarrow\ket{\downarrow}$ and $\Delta$ a detuning of the laser fields. By coupling
the transverse vibrational modes of the ion chain to the electronic
state of the ions, one can implement the interaction Hamiltonian 
\be
\label{eq:Hint}
\hat{H}_{\text{int}}=\hbar\sum_{i<j}J_{ij}\sigma_{i}^{x}\sigma_{j}^{x}\,.
\eeq
The coupling strength between ion $i$ and $j$
is given by 
\begin{equation}
J_{ij}=\Omega_{i}\Omega_{j}\frac{\hbar k^{2}}{2m}\sum_{n}\frac{b_{i,n}b_{j,n}}{\Delta^{2}-\nu_{n}^{2}}\label{eq:J_i,j_gate}\,,
\end{equation}
with $m$ the ion mass, $\nu_{n}$ the eigenfrequencies of the transverse phononic
modes, and $b_{i,n}$ the elements of the normal-mode-matrix
\cite{james1998quantum}. 
Further, $k$ is the laser wavenumber and $\Omega_{i}$ are the Rabi frequencies induced by the laser
driving. 
For the sake of simplicity, we assume 
all $\Omega_{i}$ equal. 

In addition to the interaction described by $\hat{H}_{\text{int}}$,
one can realize large, constant on-site excitation energies as described by \cite{Richerme2014,jurcevic2014quasiparticle,Jurcevic2015}
\be
\hat{H}_{\omega_{\text{const}}}=\hbar\omega_{\text{const}}\sum_{i}\hat{\sigma}_{i}^{+}\hat{\sigma}_{i}^{-}\,.
\eeq
This is achieved by shifting the two frequencies of the laser beam
that implement the Mølmer--Sørenson interaction by a frequency $\omega_{\text{const}}$. 
In the limit of large on-site energies
$\omega_{\text{const}}\gg J_{ij}$, we can neglect off-resonant transitions generated by $\hat{\sigma}_{i}^{+}\hat{\sigma}_{j}^{+}$ and $\hat{\sigma}_{i}^{-}\hat{\sigma}_{j}^{-}$, and we obtain 
\be
\label{eq:HintToHJ}
\hat{H}_{\text{int}}+\hat{H}_{\omega_{\text{const}}}\rightarrow\hat{H}_{J}+\hat{H}_{\omega_{\text{const}}}\,.
\eeq
In this limit, the Hamiltonian approximately decouples into sectors with conserved excitation number. 

An interesting feature of the ion-chain implementation is the ability to tune the range of the interaction
encoded in $J_{ij}$, as given by Eq.~\eqref{eq:J_i,j_gate}, by changing the
detuning $\Delta$ \cite{kim2009entanglement,Britton2012,jurcevic2014quasiparticle}. 
The result is an adjustable distance dependence approximating a power-law,
i.e.,  
\begin{equation}
|J_{ij}|\propto\left|\left|\mathbf{x}_{i}-\mathbf{x}_{j}\right|\right|{}^{-\alpha}\,,\label{eq:J_i,j_powerlaw}
\end{equation}
where the $\mathbf{x}_{i}$ denote the equilibrium positions of the ions. 
For an ion chain with almost equidistant ions the power-law simplifies to 
\begin{equation}
|J_{ij}|\propto\left|i-j\right|{}^{-\alpha}\,.\label{eq:iJ_i_j_deal_equidistant}
\end{equation}
This tunability allows us to study networks with different geometrical properties. In principle, the decay exponent $\alpha$ can be tuned between $0$ and $3$, though realistic laser intensities restrict it to the range $\alpha\in\left[0.75,1.75\right]$ while maintaining reasonable coupling strength on the order of $100\;\text{s}^{-1}$ \cite{kim2009entanglement,Britton2012,jurcevic2014quasiparticle}. By addressing the axial center-of-mass mode, it is additionally possible to study the limit $\alpha=0$ \cite{sorensen1999quantum}. In the following, this limit will be of particular interest, as it corresponds to the well-studied model of a fully connected, equally weighted graph \cite{caruso2009highly}, with 
\be
\left|J_{ij}\right|=\left|J_{i^{\prime}j^{\prime}}\right|\text{ for }i\neq j,\; i^{\prime}\neq j^{\prime}\,. 
\eeq

In order to determine the exponent $\alpha$ for a given detuning, we fit the spin-wave dispersion relation in the single-excitation manifold (the eigenvalues of the coupling matrix $J$) for an exact power-law dependence, as given by Eq.~(\ref{eq:J_i,j_powerlaw}),  to the dispersion relation for the experimentally relevant $J_{ij}$ derived from Eq.~(\ref{eq:J_i,j_gate}) \cite{jurcevic2014quasiparticle}. 
The relation between $\alpha$ and the detuning $\Delta$ is illustrated
in Fig.~\ref{fig:exponent-Delta} for realistic experimental parameters. As this figure shows, the power-law dependence is a good approximation for small systems \cite{jurcevic2014quasiparticle}. For large chains, the distance dependence is better described by a combination of power law and exponential decay~\cite{Nevado2014}. In the limiting cases $\alpha=0$ and $\alpha=3$, however, the ideal power law becomes exact \cite{Nevado2014}. Moreover, deviations from the ideal power law are very small in the range $\alpha\in[2,3]$. This behavior is consistent with the fact that in this range the power-law interactions have only a weak effect on, e.g., dispersion relation and dynamics as compared to a system with short-range interactions \cite{hauke2013spread}.

\begin{figure}
\begin{centering}
\begin{overpic}[width=4.15cm,tics=5]
{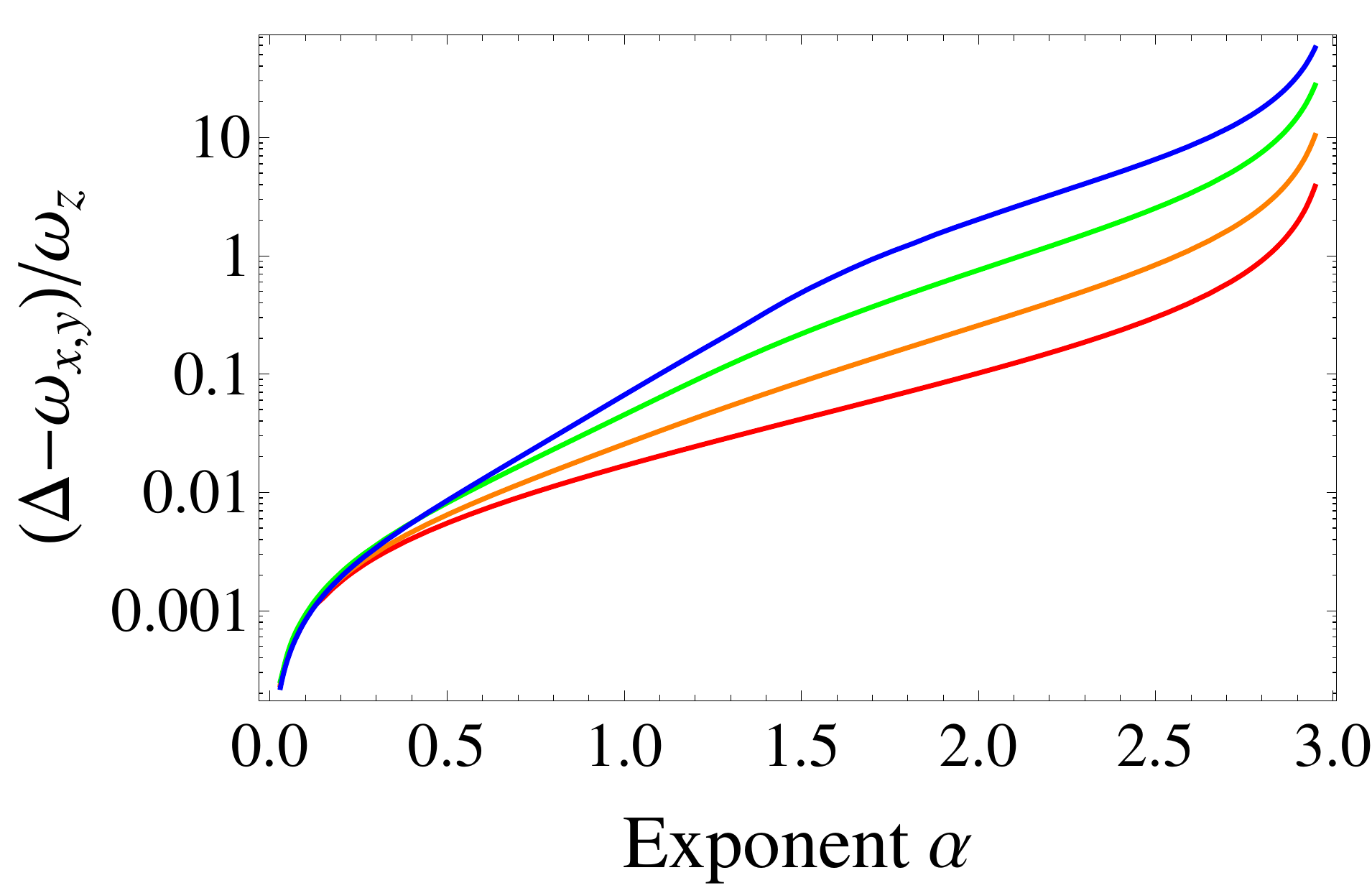}
\put(-5,70){ \large a)}
\end{overpic}\hspace{0.3cm}\begin{overpic}[width=4.15cm,tics=5]
{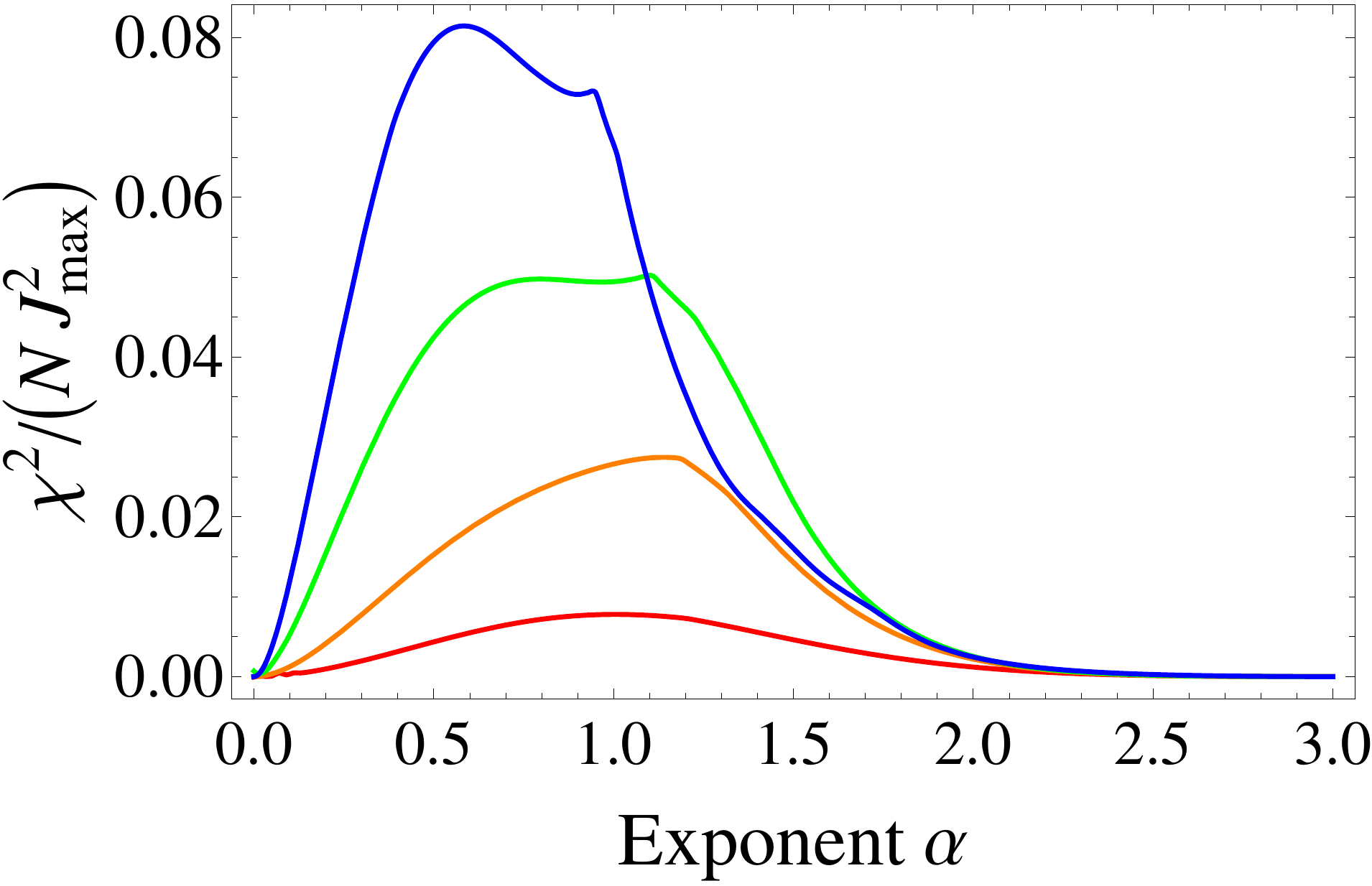}
\put(-5,70){ \large b)}
\end{overpic}\hspace{0.3cm}
\par\end{centering}
\caption{Fitting the experimentally relevant interactions to an ideal power law. (a) Detuning $\Delta$  plotted over the fitted exponent $\alpha$, obtained from fitting the dispersion relation for an exact power law to the dispersion relation calculated for an experimental implementation. The parameters are: $\omega_{x,y}=50\omega_{z}$, for
$N=5, 10, 20, 30$ (red, orange, green, blue). 
(b) The deviations of the fit, measured by $\chi^2$, increase with system size. In the limits $\alpha=0$ and $\alpha=3$, the ideal power law becomes exact, but already in the range $\alpha=2\dots 3$ deviations are small. 
}
\label{fig:exponent-Delta}
\end{figure}

\subsection{Implementation of the on-site energies\label{sub:Implementation_onsite_energies}}

Another important ingredient of our model is the ability to realize on-site energies $\hbar\omega_{i}$ that vary from site to site, used to simulate static disorder. 
By letting them fluctuate over time, they moreover simulate Markovian as well as non-Markovian dephasing. 
Such on-site energies can be generated in an ion chain through the quadratic AC-Stark effect induced by an additional, off-resonant laser field. 
The realization of site-dependent on-site energies requires tightly focused, steerable laser beams, as have been demonstrated in the setup described in \cite{schindler2013quantum}
and have been used to implement static disorder for studies of many-body localization in an ion chain \cite{smith2015many}. 

\subsection{Initial state preparation\label{sub:InitialState}}

The initial state $\ket{\psi_0}$ as given in Eq.~\eqref{eq:initialState} can be realized faithfully through optical pumping and steerable, addressable laser beams \cite{jurcevic2014quasiparticle}.

\subsection{Implementation of $\mathcal{L}_{\text{diss}}$\label{sub:Implementation_decay}}

The Markovian process described by $\mathcal{L}_{\text{diss}}$, which
models the absorption of the excitation at site $i_{\text{sink}}$,
can be implemented by exploiting spontaneous decay processes in the level
structure of the ions. 
A suitable realization is via the quantum operation 
\begin{eqnarray}
\mathcal{E}_{\text{decay}}(\hat{\rho}) & = & \sum_{i\in\left\{ 1,2\right\} }\hat{K}_{i}^{\dagger}\hat{\rho}\hat{K}_{i}
\end{eqnarray}
with the Kraus operators 
\begin{eqnarray}
\hat{K}_{1} & = & \sqrt{p_{\text{decay}}}\,\hat{\sigma}_{+}^{i_{\text{sink}}}\,,\\
\hat{K}_{2} & = & \ket{\downarrow}_{i_{\text{sink}}}\langle\downarrow\mid_{i_{\text{sink}}}+\sqrt{1-p_{\text{decay}}}\ket{\uparrow}_{i_{\text{sink}}}\langle\uparrow\mid_{i_{\text{sink}}}.
\end{eqnarray} 
By interrupting the time evolution after a time $\Delta T$, applying
$\mathcal{E}_{\text{decay}}$, and repeating this process, we recover in the
limit $\Delta T\rightarrow0$ the decay process described by $\mathcal{L}_{\text{decay}}$ with $p_{\text{decay}}=\Gamma\Delta T$. 
The implementation of this amplitude-damping operation in an ion chain is described in Ref.~\cite{schindler2013quantum}. 
There, the tightly focused, steerable laser beams are used to transfer some population from the qubit state $\ket{\uparrow}_{i_{\text{sink}}}$
to the intermediate state $|S^{\prime}\rangle_{i_{\text{sink}}}=\ket{4^{2}S_{1/2}\, m_{j}=1/2}_{i_{\text{sink}}}$
via a partial Rabi flop. The amplitude damping is completed by a laser
field that drives the transition from $|S^{\prime}\rangle_{i_{\text{sink}}}$
to the manifold $4^{2}P_{1/2}$ and pumps the population to the qubit state $\ket{\downarrow}_{i_{\text{sink}}}$
via optical pumping. Single-side addressability of the second laser beam is here not required, since only ions in the intermediate state $|S^{\prime}\rangle_{i_{\text{sink}}}$
are affected. 

For an experimental investigation
of the system dynamics, it may also be of interest to realize the
decay of $\mid\uparrow\rangle_{i_{\rm sink}}$ to a third auxiliary internal level  by pumping the population to a different state of
the ions. In this
way, the simulated amplitude damping  as described by $\mathcal{L}_{\text{diss}}$ can be distinguished from the radiation-field-induced spontaneous decay process acting on the ions.

\subsection{Simulation of dephasing and non-Markovian dynamics\label{sub:non_Markovian}}

In the trapped-ion quantum simulator, controlled noise can be induced by time-dependent laser fields to realize fluctuating on-site energies $\hbar\omega_{i}(t)$ \cite{Soare2014,Hauke2015}. 
A strength of the proposed trapped-ion setup is that it can realize almost arbitrary power spectra, which will allow a quantitative modeling of the structured spectra characteristic of biomolecules \cite{Thorwart2009,chin2013role,Jesenko2013,chen2011excitation,delRey2013,mohseni2014energy}. We focus in the following on a generic and simple noise process generated by the Goldstein--Kac telegraph model \cite{goldstein1951diffusion,masoliver1989continuous}, as has been applied, e.g., to investigate the impact of phase \cite{eberly1984noise} and intensity fluctuations \cite{wodkiewicz1985random} on atom-laser interactions.   
The Goldstein--Kac process simulates Markovian dephasing as induced
by a heat bath with vanishing bath correlation time and modeled by
$\mathcal{L}_{\text{deph}}$, as well as non-Markovian processes associated to heat baths with finite correlation
time such as phonon fluctuations with exponential memory. 
The Goldstein--Kac telegraph process is a dichotomic process (i.e., 
$\omega_{i}(t)\in\left\{ -\omega_{\text{GK}}/2,\omega_{\text{GK}}/2\right\} $), described by the Markovian master equation
\begin{eqnarray}
\label{eq:GoldsteinKac}
 \partial_{t_{2}}P\left[\omega_{i}(t_{2})=\pm\omega_{\text{GK}}/2\,|\,\omega_{i}(t_{1})=\omega\right]  =\qquad\nonumber\\
  \qquad-\lambda_{\pm}P\left[\omega_{i}(t_{2})=\pm\omega_{\text{GK}}/2\,|\,\omega_{i}(t_{1})=\omega\right]\\
  \qquad+\lambda_{\mp}P\left[\omega_{i}(t_{2})=\mp\omega_{\text{GK}}/2\,|\,\omega_{i}(t_{1})=\omega\right]\nonumber\\
  \text{ for }\omega\in\left\{ -\omega_{\text{GK}}/2,\omega_{\text{GK}}/2\right\} \text{ and }t_{2}>t_{1}.\qquad\nonumber
\end{eqnarray}
Here, $\lambda_{\pm}$ are the transition rates of the dichotomic Markovian process. For the sake of simplicity, we assume that $\lambda_{+}=\lambda_{-}\equiv\lambda$
and that at time $t_{0}$ the process attained its equilibrium state
\begin{equation}
P\left[\omega_{i}(t_{0})=\omega_{\text{GK}}/2\right]=P\left[\omega_{i}(t_{0})=-\omega_{\text{GK}}/2\right]=1/2\;.\label{eq:equbilibrium}
\end{equation}
Furthermore, we assume that energies $\hbar\omega_{i}(t)$
of different sites are sampled from independent telegraph processes. The two-time correlation
function is then given by
\be
\langle\langle\omega_{i}(t)\omega_{j}(t+\delta)\rangle\rangle_{T}=\delta_{i,j}\omega_{\text{GK}}^{2}e^{-2\lambda\mid\delta\mid}/4\;,
\eeq
corresponding to a Lorentzian noise spectrum 
\be
S_{i,j}(\omega)=\delta_{i,j}\frac{\omega_{\text{GK}}^{2}}{2\lambda-i\omega}\,.
\eeq
Even though the Goldstein--Kac telegraph process, as given by Eq.~\eqref{eq:GoldsteinKac}, is itself Markovian, the distribution of $\omega_i(t)$ that it generates can be non-Markovian, characterized by a frequency-dependent noise spectrum of $S_{i,j}(\omega)$. 
We recover Markovian dephasing, as described by $\mathcal{L}_{\text{deph}}$, Eq.~\eqref{eq:Ldeph}, in the limit $\lambda\rightarrow\infty$, with associated dephasing rates $\gamma_{i}=\gamma=\frac{\omega_{\text{GK}}^{2}}{2\lambda}$.
Non-Markovian features become unimportant as soon as the rate $\lambda$ is larger than all the other rates and frequencies in the system, i.e., 
$\lambda\gg J_{ij},\Gamma,\Gamma_{\text{source}},\omega_{\text{GK}}$.

\subsection{Implementation of \textmd{$\mathcal{L}_{\text{source}}$}}

The tightly focused steerable laser beam used to implement $\mathcal{L}_{\text{deph}}$
can also be employed in a similar way to generate the Markovian process
described by $\mathcal{L}_{\text{source}}$, Eq.~\eqref{eq:Lsource}. For this, the laser field has to be tuned in resonance to the transition $\ket{\uparrow} _{i_{\text{source}}}\leftrightarrow\ket{\downarrow}_{i_{\text{source}}}$
and the intensity as well as its phase has to be modulated faster than the time scales induced by the rates $\Gamma$, $\Gamma_{\text{source}}$, 
$\gamma$, and $J_{ij}$ in order to simulate the thermal noise
of an infinite-temperature reservoir. This can be done by
implementing suitable step functions as described in Sec.~\ref{sub:non_Markovian}.

\section{Numerical Results\label{sec:Numerical_Results}}

Thus, all terms describing the model of a quantum network given by the Master equation Eq.~\eqref{eq:Master equation} can be realized with existing trapped-ion technology. We postpone the discussion of potential error sources to Sec.~\ref{sec:Experimental_Error_Sources}, and first investigate numerically the excitation transport through a quantum network described by the Master equation Eq.~\eqref{eq:Master equation}. These results not only enable predictions for the ion-chain quantum simulator, they also represent detailed theoretical studies for excitation transfer and ENAQT in systems with power-law interactions.
Our numerical simulations have been performed by propagating the density operator as described by the Lindblad master equation (\ref{eq:Master equation})} using the QuTiP (Quantum Toolbox in Python) package \cite{johansson2012qutip}. 
In the numerical calculations, we focus mainly on ion chains of length $N=10$, which is on the order of system sizes where current experiments have demonstrated individually addressable AC-Stark shifts \cite{smith2015many,lee2016engineering}, but we present results for up to $N=70$ ions. 
We assume that the radial trapping frequency $\omega_{x,y}$ in the linear Paul trap is $20$ times the axial trapping frequency $\omega_{z}$. 
To simplify the following discussion, we moreover introduce the maximum value of the coupling strengths 
\be
J_{\text{max}}=\max_{i<j}\left|J_{ij}\right|
\eeq
and express all other relevant parameters in units of $J_{\text{max}}$.
The calculations are performed for a decay rate at site $i_{\text{sink}}$ of $\Gamma=J_{\text{max}}$. 
Furthermore, we assume that the dephasing rates for all sites are equal, i.e.,  
\be
\gamma=\gamma_{i}\text{ for all }i\in\left\{ 1,2...,N\right\} \,.
\eeq
With the exception of Sec.~\ref{sec:drivenSystem}, we take $\Gamma_{\text{source}}=0$, such that excitations are brought into the system only during the preparation of the initial state $\ket{\psi\left(t_{0}\right)}$.
In order to reduce the impact of boundary effects, we assume that the excitation is initially injected at ion $i_{\text{source}}=N/5+1$ and is removed by the dissipative process at $i_{\text{sink}}=4N/5$ (when comparing different $N$) and $i_{\text{sink}}=7$ (when considering fixed $N=10$), respectively. 
We study the speed of excitation transfer through the ion chain by evaluating the probability for having absorbed the excitation after a certain time at site $i_{\text{sink}}$.

We address several physical regimes of interest. In particular, we study the transfer efficiency in dependence on hopping range, disorder, and Markovian and non-Markovian dephasing. At the end of this section, we will moreover consider a driven system with $\Gamma_{\text{source}}\neq0$, where interactions between excitations play a fundamental role. 

\begin{figure*}
\vspace{0.5cm}
\begin{overpic}[width=5.5cm,tics=5]
{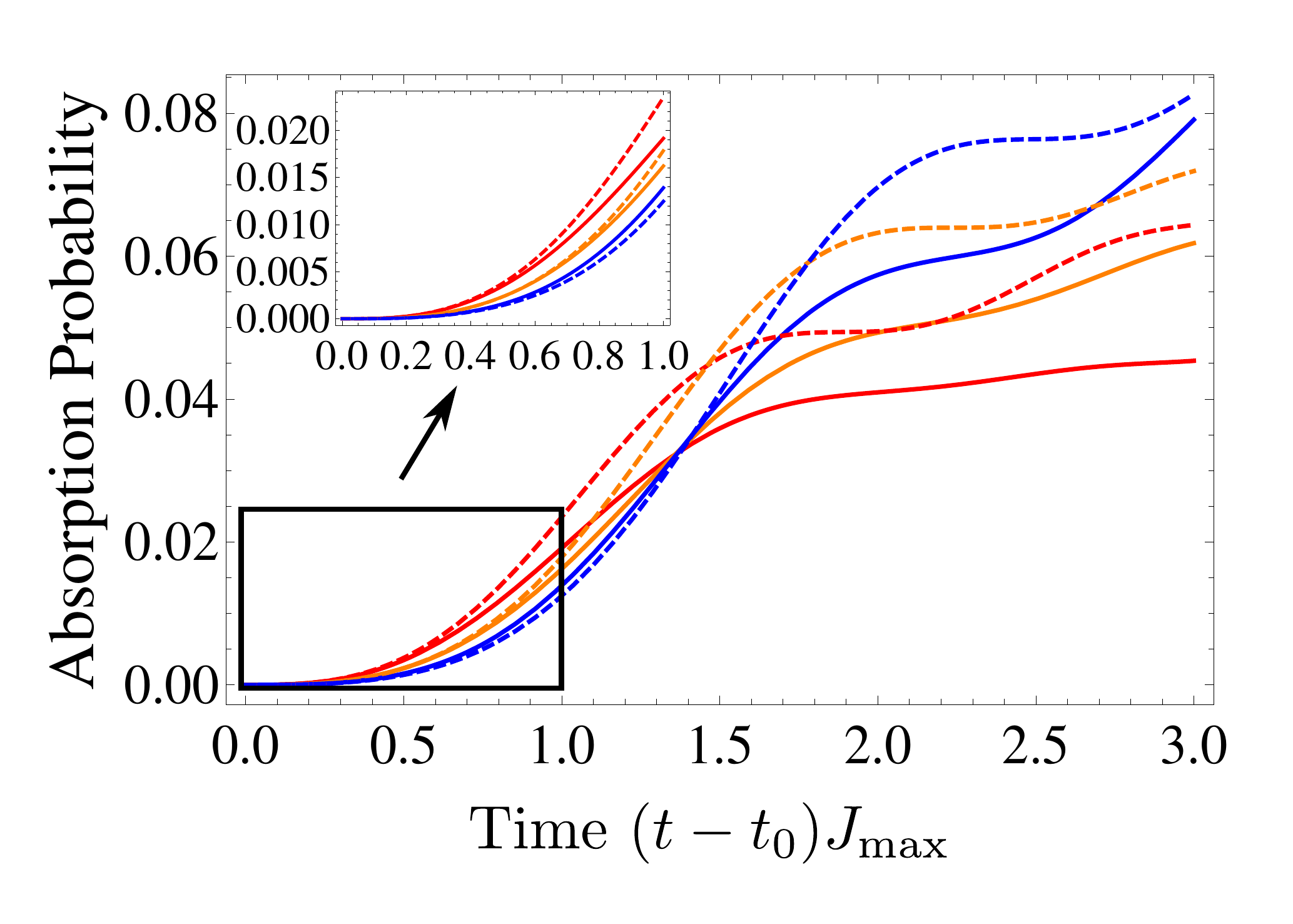}
\put(-10,115){ \large a)}
\put(56,108){ \large $\gamma=0.0 J_{\text{max}}$} 
\end{overpic}\hspace{0.3cm}\begin{overpic}[width=5.5cm,tics=5]
{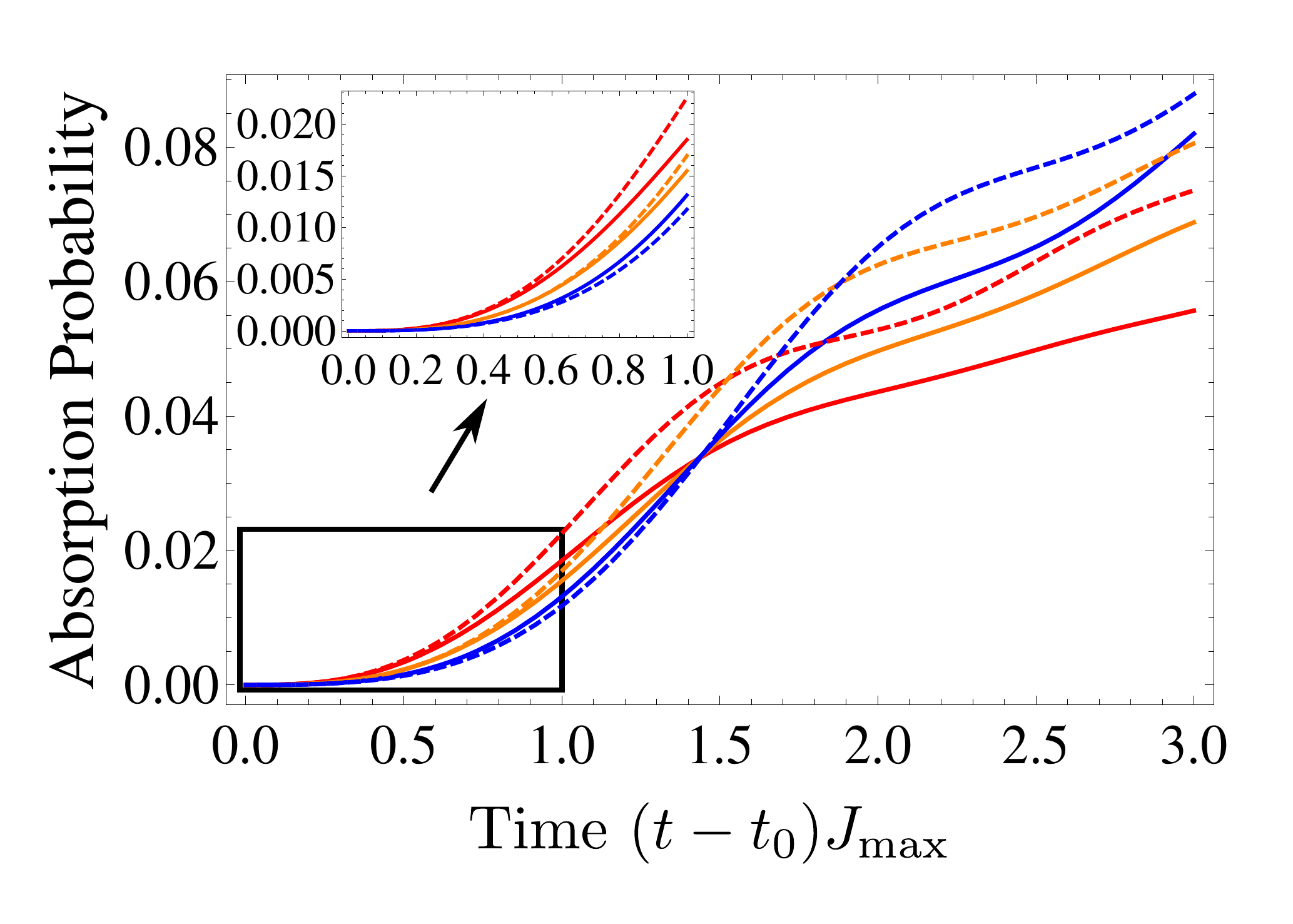}
\put(-10,115){ \large b)}
\put(56,108){ \large $\gamma=0.1 J_{\text{max}}$} 
\end{overpic}\hspace{0.3cm}\begin{overpic}[width=5.5cm,tics=5]
{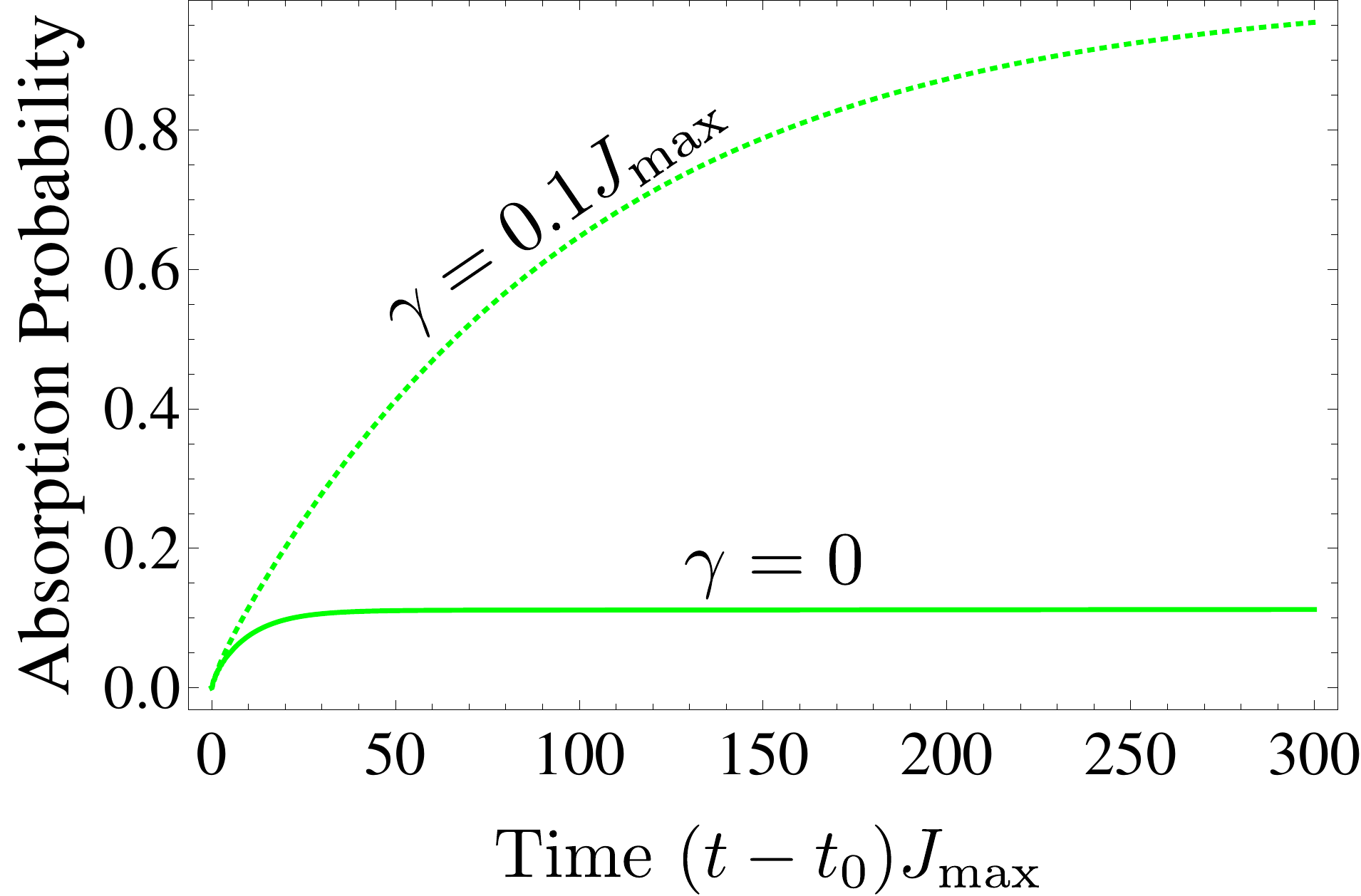}
\put(-10,115){ \large c)} 
\end{overpic}
\caption{
Transfer efficiency, defined as the time-dependent probability for having absorbed the excitation at site $i_{\text{sink}}$, as function of time, comparing different hopping ranges $\alpha$ and dephasing rates. The system is without static disorder and its size is $N=10$ sites.  
(a,b) Comparison between $\alpha=0.8$ (red), $\alpha=1.0$ (orange), 
and $\alpha=1.2$ (blue). Solid lines correspond to realistic couplings in the ion chain (see Eq.~(\ref{eq:J_i,j_gate}))
and dashed lines to an ideal power-law dependence
(see Eq.~(\ref{eq:iJ_i_j_deal_equidistant})). (a) Results for vanishing dephasing. 
Initially, a few fast modes carry part of the excitation earlier to the target site with decreasing $\alpha$ (insets). 
At later times, destructive interference effects decelerate the excitation transfer for small $\alpha$.  
(b) This characteristic behaviour is rather robust against dephasing (data for $\gamma=0.1J_{\text{max}}$). 
(c) Long-time behavior for $\alpha=0$. 
Without dephasing, destructive interference hinders a large part of the excitation from ever reaching the target site (solid line). 
Non-vanishing dephasing cancels the interference effect and facilitates the excitation transfer (dashed line). 
\label{fig:P-time}}
\end{figure*}

\subsection{Influence of the hopping range}

We start our investigation by studying the influence of the hopping range on the speed of the excitation transfer. 
To cleanly extract the influence of the hopping range, we assume for now that no static disorder is present. 

It is known that the dynamics of a one-dimensional spin model quantitatively differs in the three regimes $\alpha<1$, $1<\alpha<2$ and $\alpha>2$
\cite{hauke2013spread,Eisert2013,Foss-Feig2015,Cevolani2015,Richerme2014,jurcevic2014quasiparticle}. In the regime $\alpha>2$,
the propagation of the excitation is confined by a well-defined sound cone, typical of systems with short-range interactions \cite{Lieb1972}. In the regime $1<\alpha<2$ of weakly long-range interactions,
a clear sound cone cannot be defined, because some spin-wave modes develop a weakly divergent speed of propagation. 
A fully non-local behaviour can be observed in the regime $\alpha<1$. 
In this regime, a strong divergence appears in the spin-wave dispersion relation, and a part of the excitation can spread almost instantaneously over the entire network. The total weight of the excitation carried by these divergent modes, however, remains limited. 
Away from the divergence, the dispersion relation flattens out, and the corresponding modes become slower and slower with decreasing $\alpha$, so that extremely fast and extremely slow modes coexist \cite{hauke2013spread,Cevolani2015}.

For our purposes, the transition from $\alpha<1$ to $\alpha>1$ is of particular significance, as this can directly be observed in the short-time behaviour of the quantum network. This is illustrated in Fig.~\ref{fig:P-time}(a), where we compare three different parameters, $\alpha=1.2$ (blue), $\alpha=1$ (orange), and $\alpha=0.8$ (red). 
For comparison, we include ideal power-law interaction dependence as given by Eq.~(\ref{eq:iJ_i_j_deal_equidistant}) (dashed lines), as well as realistic coupling strengths that can be realized in the ion chain, given by Eq.~(\ref{eq:J_i,j_gate}) (solid lines). 
In both cases, the initial excitation transfer accelerates with decreasing $\alpha$. 

As illustrated in Fig.~\ref{fig:P-time}(b), this short-time behaviour is robust against dephasing. 
The chosen value of $\gamma=0.1J_{\text{max}}$ is well beyond natural dephasing rates from experimental imperfections in a state-of-the-art ion-trap experiment \cite{schindler2013quantum} (see Sec.~\ref{sec:Experimental_Error_Sources}).
Even in the presence of such strong dephasing, the characteristic behaviour in the dynamical regimes $\alpha<1$ and $\alpha>1$ can be observed. 

For large times, another effect comes into play. As shown in Ref.~\cite{caruso2009highly}, in case of a fully connected graph with equal coupling strengths, the probability for having absorbed the excitation after $\left(t-t_{0}\right)\rightarrow\infty$ converges against $\frac{1}{N-1}$.
For $t\rightarrow\infty$, the remaining quantum state has zero overlap with $i_{\text{sink}}$, and, hence, a large part of the excitation remains in the network without ever being absorbed at $i_{\text{sink}}$. 
This behaviour, which can be understood as a destructive interference effect, can clearly be observed in the time evolution for $\alpha\rightarrow0$ shown in Fig.~\ref{fig:P-time}(c). 
In the presence of dephasing, however, the destructive interference at $i_{\text{sink}}$ is destroyed and for all $\alpha$ the probability for having absorbed the excitation
converges against unity for $t\rightarrow\infty$. 

We find that the same destructive interference effect also causes a slowing-down of the absorption rate for small but non-zero $\alpha$, as can be observed in Figs.~\ref{fig:P-time}(a) 
and (b). As a result, for not too short times, lower values of $\alpha$ result in a lower absorption efficiency, thus reversing the short-time behaviour. 
This behavior can also be understood in terms of the spin-wave dispersion relation, with its coexistence of fast and slow modes. 

To work out the dependence on the hopping range more clearly, we study the absorption probability at fixed times as a function of the exponent
$\alpha$ and system size. To facilitate comparison between different ion numbers $N$, we choose an ideal power-law dependence of the interactions (see Eq.~(\ref{eq:iJ_i_j_deal_equidistant})). 
As we are interested in transport from one end of the chain to the other, where the distance between source and target sites increases as a function of $N$, we require a suitable scaling of $(t-t_{0})$. 

\begin{figure}
	\vspace{0.7cm}
	\begin{overpic}[width=4.15cm,tics=5]
		{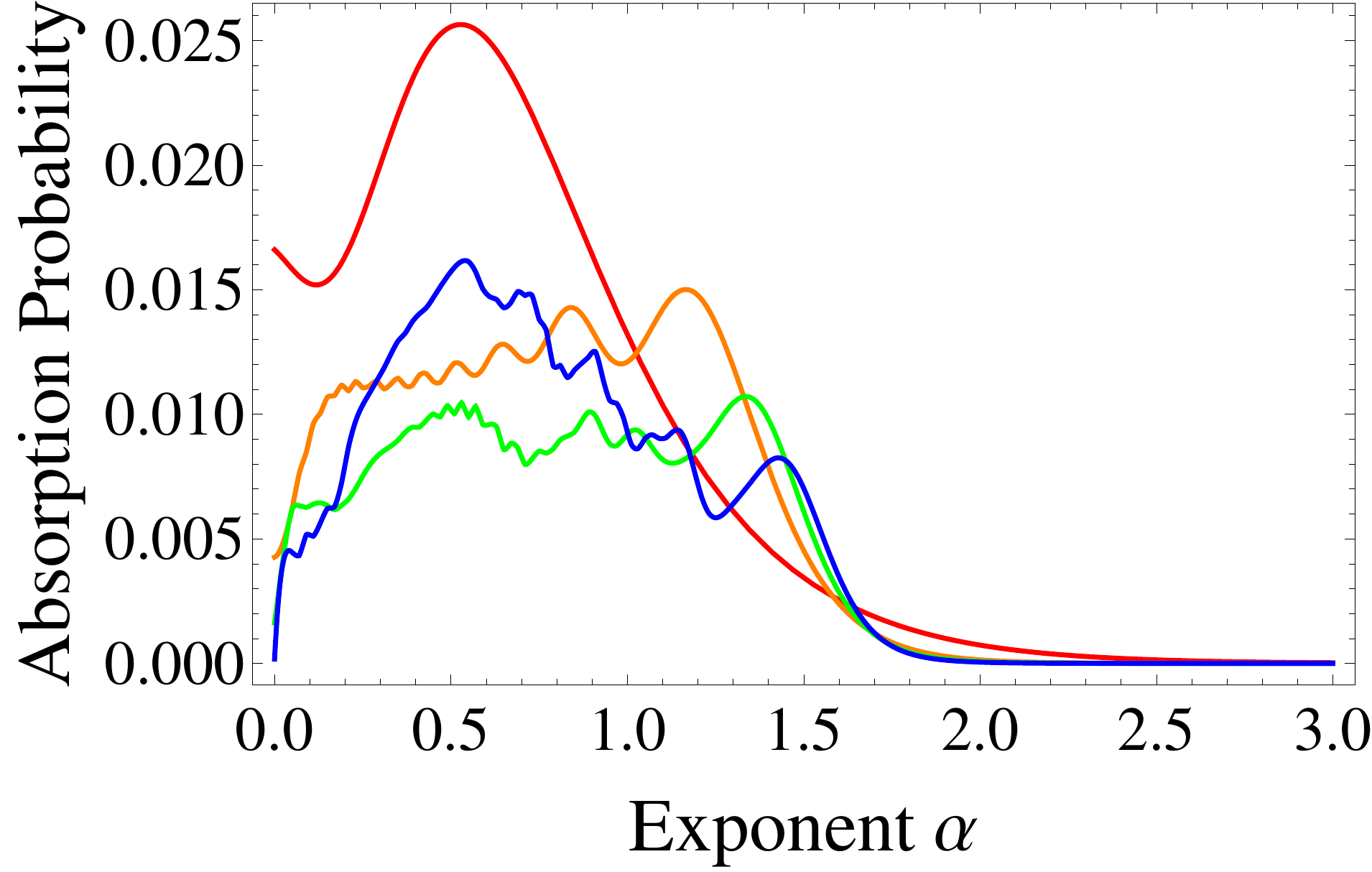}
		\put(-5,80){ \large a)} 
	\end{overpic}
	\hspace{0.1cm}
	\begin{overpic}[width=4.15cm,tics=5]
		{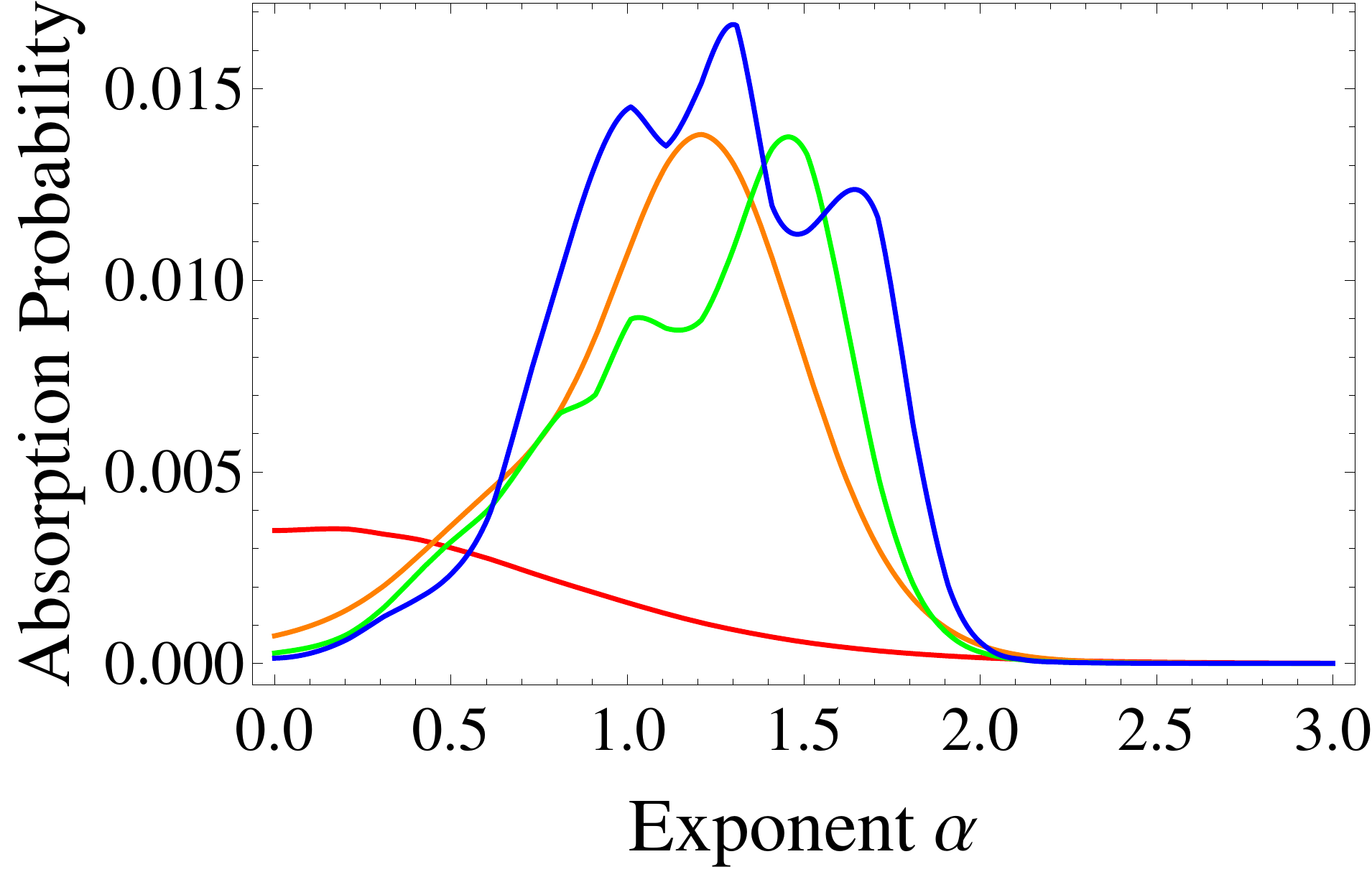}
		\put(-5,80){ \large b)} 
	\end{overpic}\caption{
		System-size dependence of the transfer efficiency as function of hopping range $\alpha$, for vanishing dephasing rate and static disorder. 
		System sizes are $N=10$ (red), $N=30$ (orange), $N=50$ (green), and $N=70$ (blue), and plotted is the probability for having absorbed an excitation (initially created at
		$i_{\text{source}}=N/5+1$) at site $i_{\text{sink}}=4N/5$ after a fixed time $(t-t_{0})$. 
		(a) Considering $(t-t_{0})J_{\text{max}}=N/10$, there appears a rather smooth increase of the absorption probability below $\alpha\lesssim 1.7$ and a sharp drop at $\alpha\to0$. 
		(b) Choosing $(t-t_{0})\left\langle v_{g}\right\rangle =(i_{\text{sink}}-i_{\text{source}})/5$, with $\left\langle v_{g}\right\rangle $ the average group velocity, the absorption probability drastically increases at $\alpha\lesssim 2$ and has a maximum in the range $1\lesssim\alpha\lesssim 2$ (except for the smallest studied system size of $N=10$). 
		Results are for an ideal power-law dependence. 
		\label{fig:P-alpha_finite_size}}
\end{figure}

In Fig.~\ref{fig:P-alpha_finite_size}(a), we take $(t-t_{0})$ simply proportional to $N$. 
For the chosen proportionality constant, the probability for having absorbed the excitation goes to $0$ for large $\alpha$, as in short-range interacting chains the excitation requires a finite propagation time to reach the target site. 
At smaller $\alpha$, fast modes reach the target site even at short times, leading to a non-vanishing absorption probability. In the opposite limit of $\alpha\to 0$, the destructive interference effect suppresses the excitation transfer. As a consequence, large transfer efficiencies are reached in the range $0<\alpha\approx 1.7$. 

Since the time required for an excitation to propagate is given by the group velocity, it can be more convenient to use this quantity to define the size-dependence of $t$. Corresponding results are plotted in Fig.~\ref{fig:P-alpha_finite_size}(b), where we used the average group velocity
\be
\left\langle v_{g}\right\rangle =(N-1)\left(\omega_{N}-\omega_{1}\right)/\pi
\eeq
to fix $(t-t_{0})\left\langle v_{g}\right\rangle =(i_{\text{sink}}-i_{\text{source}})/5$. Here, the $\omega_{1}\leq\omega_{2}\leq...\leq\omega_{N}$
are the eigenfrequencies of the coupling matrix $J$ defining the spin-wave dispersion. 
The time is chosen such that the target site lies outside the sound cone existing at $\alpha>2$, which is determined by the maximum group velocity. 
With this choice, we observe a clear transition between the two regimes $\alpha<2$ and $\alpha>2$. 
For $\alpha<2$, a clear sound cone can not be defined \cite{hauke2013spread}, and part of the excitation can reach the target site already at very short times, even for large systems. 

These results hold true in the short-time regime. For larger $t-t_{0}$ the situation is quite different, as the destructive interference effect and the larger number of slow spin-wave modes dominate the behavior. The probability for absorbing the excitation at $i_{\text{sink}}$ then increases for higher values of $\alpha$.

\subsection{Influence of static disorder\label{subsec:disorder}}

\begin{figure}
\includegraphics[scale=0.3]{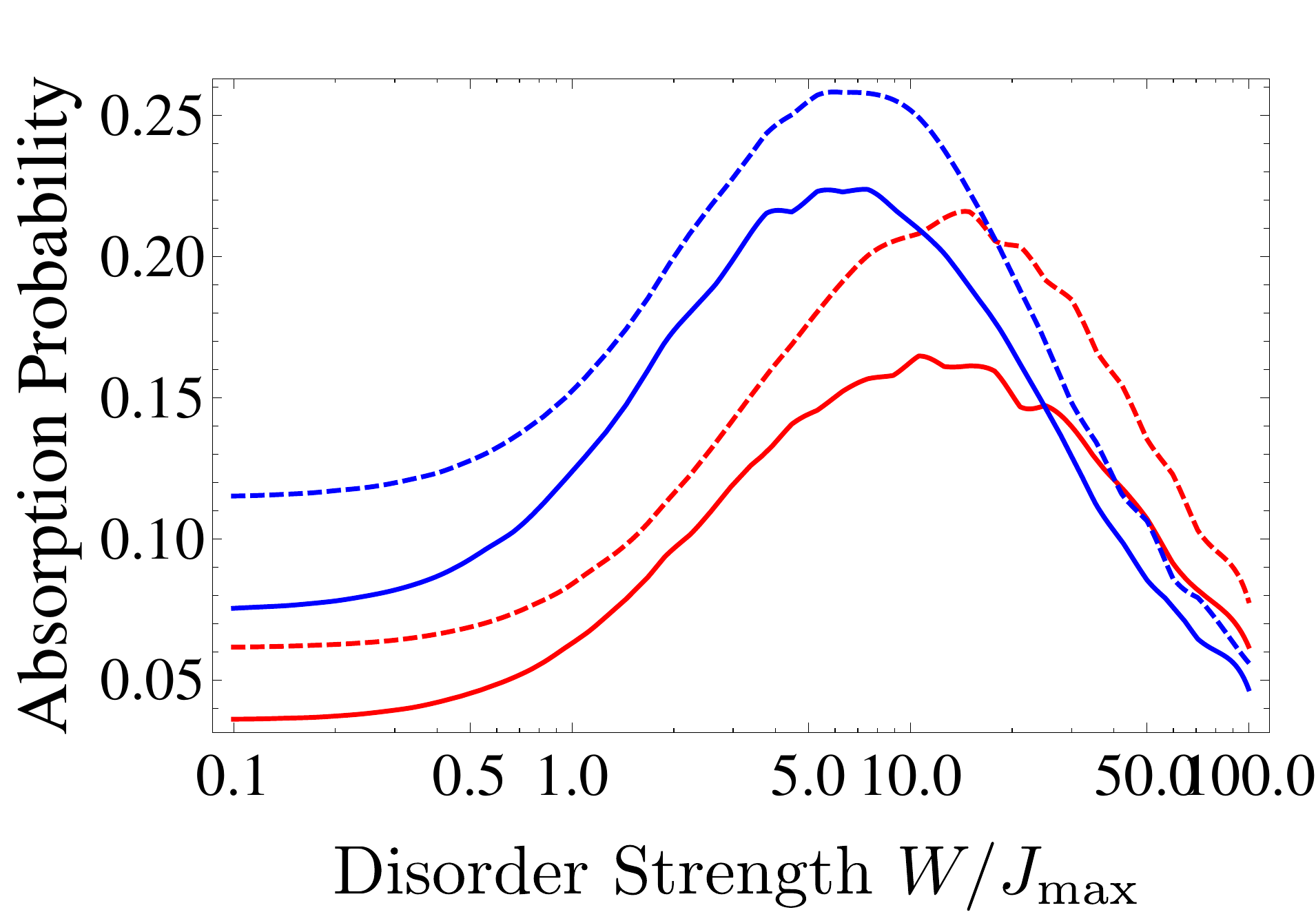}
\caption{
Impact of static disorder on the transfer efficiency, for two dephasing rates (solid lines: $\gamma=0$; dashed: $\gamma=0.1J_{\text{max}}$) and two system sizes (blue: $N=10$ at time $t-t_{0}=10/J_{\text{max}}$; red: $N=20$ at time $t-t_0=20/J_{\text{max}}$; $i_{\text{source}}=N/5+1$ and $i_{\text{sink}}=4N/5$). 
Results are for $\alpha=0$. 
The destructive interference hindering the excitation transfer in clean systems is lifted by small amounts of disorder, while at large disorder Anderson localization obstructs the transfer. Optimal transfer efficiency is achieved in an intermediate range of disorder. 
Markovian dephasing (dashed lines, $\gamma=0.1J_{\text{max}}$) can increase the absorption probability when compared to the noise-free system (solid lines, $\gamma=0$).
The numerical simulations show averages over 1000 randomly generated samples of static disorder. 
\label{fig:P-disorder}}
\end{figure}

Another question that can be addressed by the proposed quantum simulation is the influence of static disorder. 
In the following, we choose the on-site energies randomly and independently
from a uniform distribution over the interval $[-W,W]$, 
with $W$ being the disorder strength. Such a bounded distribution captures the physics of doped semiconductors and alloy models (see \cite{Johri2012} and references therein), but other disorder distributions could also be easily realized in the trapped-ion setup. 
For large $\alpha$, the transport efficiency will be reduced by disorder-induced localization. 
For small $\alpha$, however, we expect a trade-off between the destructive interference described previously,
which is destroyed for finite values of $W$ \cite{caruso2009highly}, and Anderson localization, which sets in at large $W$. The trade-off between
both effects, giving an optimal transfer efficiency at intermediate disorder strengths, is illustrated in Fig.~\ref{fig:P-disorder}.

\subsection{Impact of non-Markovian dephasing}

\begin{figure}
\includegraphics[scale=0.3]{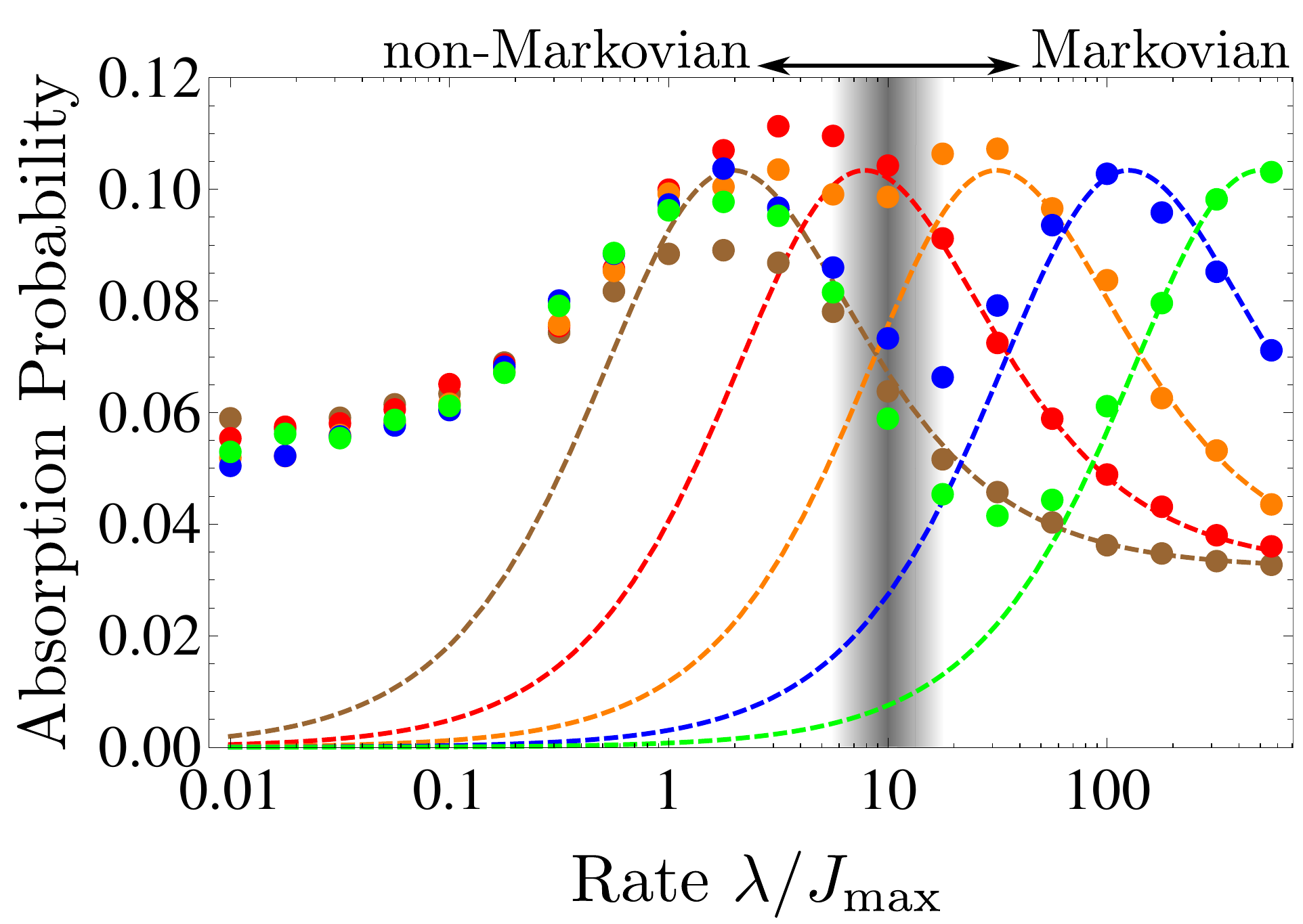}
\caption{
Impact of non-Markovian dephasing on the transfer efficiency, as a function of the rate $\lambda=\lambda_{+}=\lambda_{-}$ characterizing the Goldstein--Kac telegraph process (bullets).
Data are for $\alpha=0$, $N=10$, $t-t_{0}=2.5/J_{\text{max}}$, and $\hbar\omega_{\text{GK}}/J_{\text{max}}=4, 8, 16, 32, 64$ 
(brown, red, orange, blue, and green). 
The absorption probabilities for corresponding Markovian processes with dephasing rate $\gamma=\omega_{\text{GK}}^{2}/\left(2\lambda\right)$ are plotted as dashed lines. 
These illustrate the crossover from the non-Markovian regime ($\lambda\lesssim 10 J_{\text{max}}$) to the Markovian regime ($\lambda\gtrsim 10 J_{\text{max}}$). 
The numerical simulations are performed by averaging over 500 randomly generated samples of noise. 
\label{fig:non-markovian}}
\end{figure}

The impact of non-Markovian environmental noise on excitation transfer in biological systems has recently become an active field of research \cite{Thorwart2009,mohseni2014energy,chen2011excitation,chin2013role,Jesenko2013}. 
Our scheme allows for the investigation of such non-Markovian effects following the ideas outlined in Sec.~\ref{sub:non_Markovian}. 
In the following, we focus on the propagation of an excitation in a fully connected network with $\alpha=0$.  The results of a numerical simulation are illustrated in Fig.~\ref{fig:non-markovian}. 

We employ the Goldstein--Kac telegraph process described in Sec.~\ref{sub:non_Markovian} to model physical scenarios with varying temporal noise correlations. For long persisting temporal correlations, i.e., $\lambda\ll J_{\text{max}},\Gamma,W, \omega_{\text{GK}}$, the noise behaves effectively as static disorder.
Its impact on the system dynamics differs from what has been discussed in Sec.~\ref{subsec:disorder}, since under the dichotomic telegraph process at $\lambda\to 0$ and $\alpha=0$, the network effectively decouples into two infinitely connected subgraphs defined by $\omega_i=\pm\omega_{\text{GK}}/2$. Within each subgraph, transport is unhindered, in contrast to disorder chosen from the uniform distribution over the interval $[-W,W]$.  

With increasing $\lambda$, the noise enhances the absorption probability, similar to the impact of Markovian dephasing discussed above. 
In the opposite limit of $\lambda\gg J_{\text{max}},\Gamma,W,\omega_{\text{GK}}$, where the temporal correlations of the noise are extremely short, one recovers an effectively Markovian dephasing process with associated dephasing rate 
\be
\gamma=\frac{\omega_{\text{GK}}^{2}}{2\lambda}\,.
\label{eq:gammai}
\eeq
When sending $\lambda\rightarrow\infty$ with $\omega_{\text{GK}}=\text{const.}$, the dephasing rate $\gamma\rightarrow0$; the impact of the noise vanishes and the excitation is again localized by the destructive interference effect.
Consequently, the probability for absorbing the excitation attains its maximum for intermediate values of $\lambda$, in a regime dominated by finite temporal noise correlations and non-Markovian behavior. 

In Fig.~\ref{fig:non-markovian}, we also compare these results with a corresponding Markovian process, with the dephasing rate given by Eq.~\eqref{eq:gammai}. 
The maximally achieved values of the absorption probabilities are similar for Markovian and non-Markovian noise. Nevertheless, non-Markovian dephasing can reach higher absorption probabilities over a larger parameter range as compared to its Markovian counterpart. 

\begin{figure}
	\includegraphics[scale=0.3]{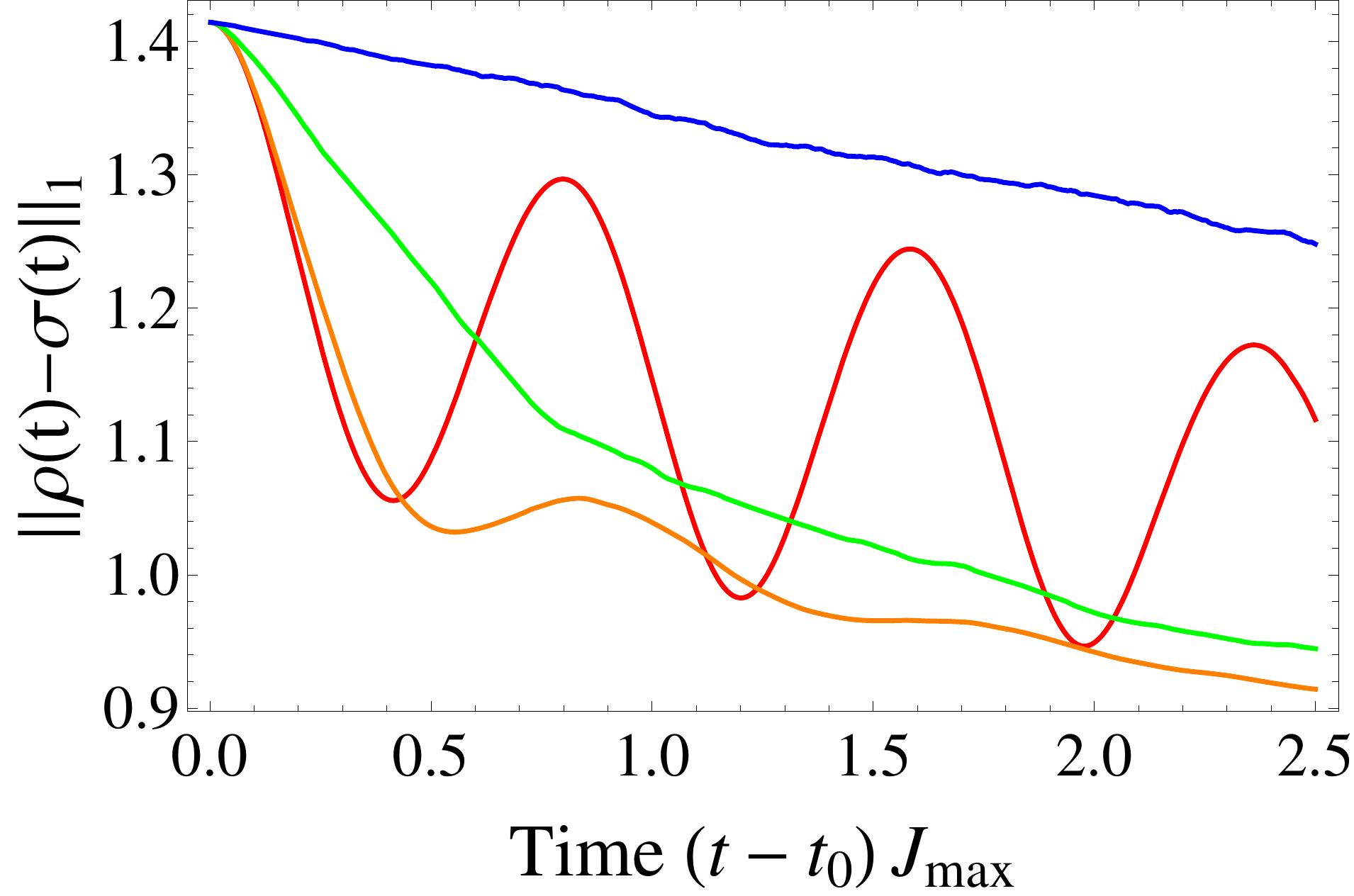}
	\caption{
		Trace distance as measure of non-Markovianity. At low rates $\lambda$, strong recurrences (i.e., positive derivatives) demonstrate the non-Markovian nature of the noise. As the rate is increased, the curves gradually become more monotonic, indicating purely Markovian behavior. 
		Data for $\omega_{GK}/J_{\max}=4$, $\alpha=0$, $N=10$, averages over 150 noise realizations. Red, orange, green, blue curves correspond to $\lambda/J_{\max}=0.1,1,10,100$, respectively. 
		\label{fig:non-markovian_trace_distance}}
\end{figure}

We can quantify the crossover from Markovian to non-Markovian behavior by computing the  distance between two different starting states \cite{Rivas2014,Breuer2016}. At sufficiently long times, all states will converge to $\bigotimes_{i=1}^{10}\ket{\downarrow}_i$. In the Markovian case, this convergence is monotonic, while the memory effects of non-Markovian noise can lead to temporary recurrences of the trace distance. Figure \ref{fig:non-markovian_trace_distance} shows the trace distance for two starting states, $\rho(t=0)=\ket{\psi_1}\bra{\psi_1}$ and $\sigma(t=0)=\ket{\psi_2}\bra{\psi_2}$, with $\ket{\psi_1}=\sigma_2^+\bigotimes_{i=1}^{10} \ket{\downarrow}_i$ and 
$\ket{\psi_2}=\frac{1}{\sqrt{2}}(\mathbb{I}+\sigma_2^+)\bigotimes_{i=1}^{10}\ket{\downarrow}_i$, where there is a single excitation at site 2, respectively half an excitation. As it can be seen in Fig.~\ref{fig:non-markovian_trace_distance}, in the region that is well described by the equivalent Markovian process (Fig.~\ref{fig:non-markovian}), the trace distance decreases monotonically. Upon crossing over into the non-Markovian region, clear recurrences appear.

\subsection{Dynamics of a driven system\label{sec:drivenSystem}}

Within the proposed scheme, one can go beyond single
excitations and investigate non-linear interaction effects. 
One way to study these is by continuously pumping excitations into the system, for example by the incoherent Markovian process described
by the super-operator $\mathcal{L}_{\text{source}}$ introduced in Sec.~\ref{sec:Model}. 
In the following, we concentrate on the properties of the steady state that emerges in the limit $(t-t_{0})\rightarrow\infty$ due to the interplay of the Markovian processes described by $\mathcal{L}_{\text{source}}$, $\mathcal{L}_{\text{diss}}$, and the hopping term $H_{J}$. 
We performed these calculations by numerically searching for the steady state of the Lindblad master equation (\ref{eq:Master equation}) using the QuTiP package \cite{johansson2012qutip}.

\begin{figure}
\begin{overpic}[width=4cm,tics=5]
{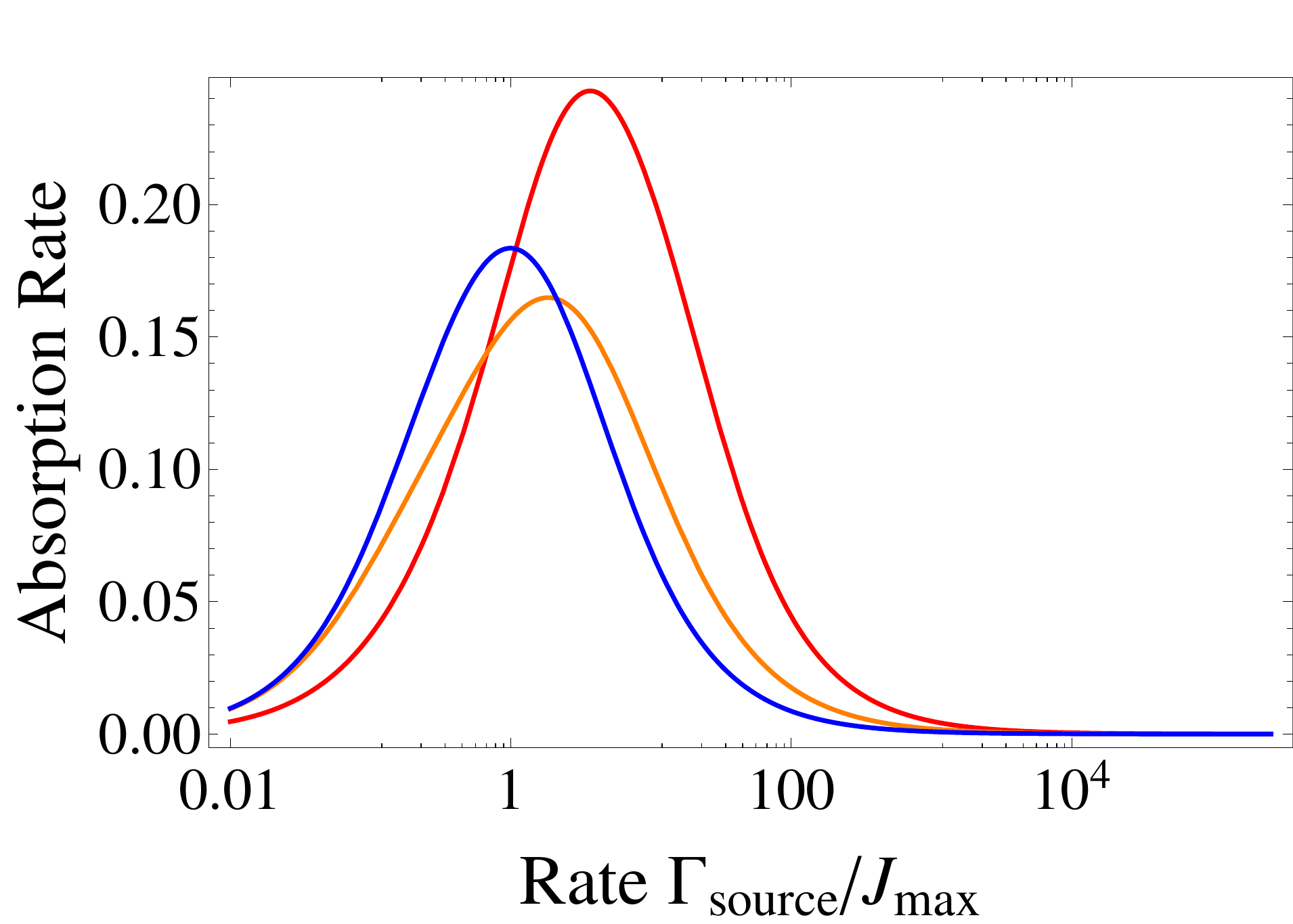}
\put(-10,70){ \large a)} 
\end{overpic}\hspace*{0.5cm}\begin{overpic}[width=4cm,tics=5]
{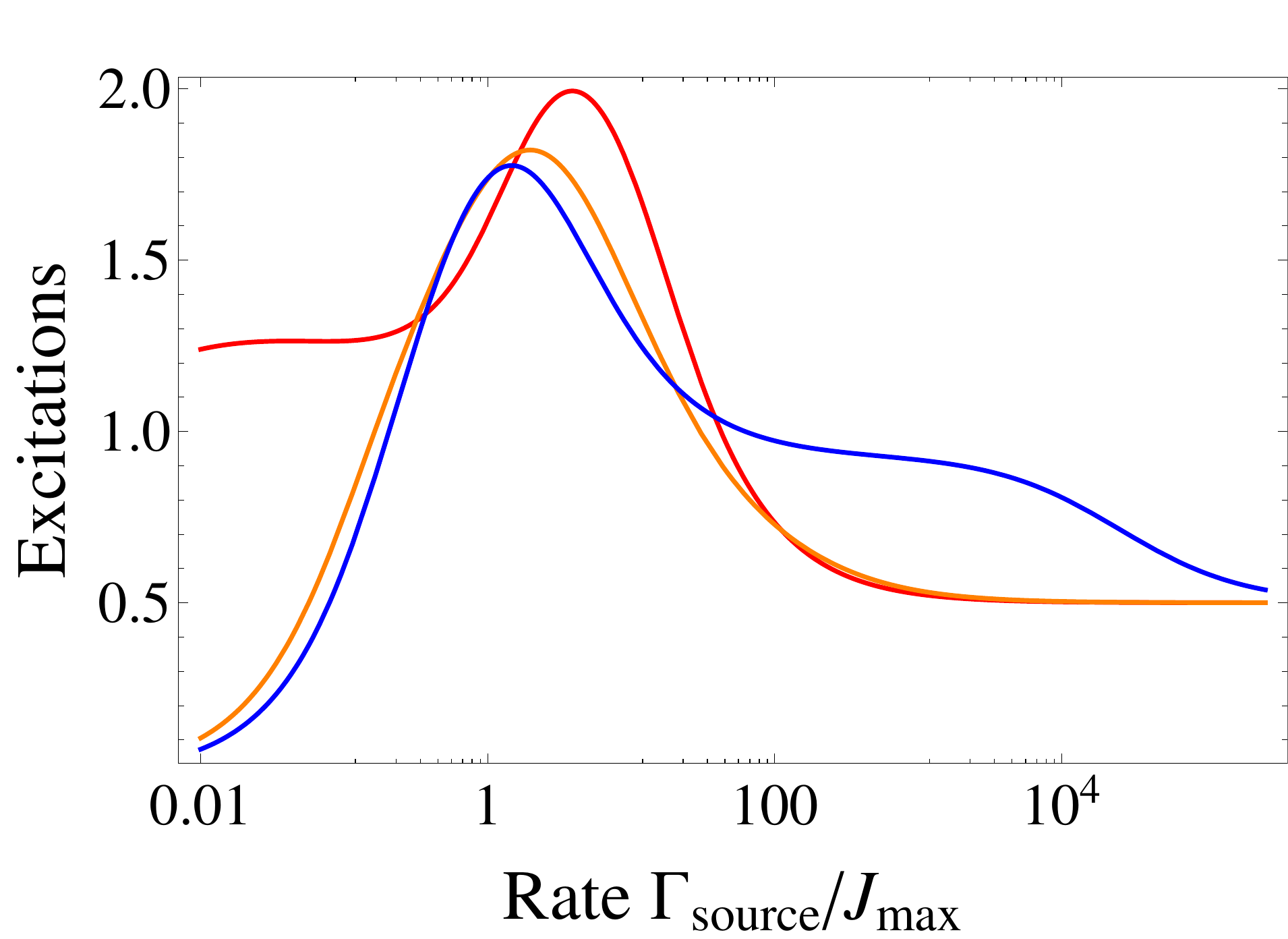}
\put(-10,70){ \large b)} 
\end{overpic}

\caption{
Transfer in a continuously driven system as a function of driving rate, without dephasing and static disorder. Excitations are created at site $i_{\text{source}}=2$ with rate $\Gamma_{\text{source}}$ and the target site is $i_{\text{sink}}=5$, in a system of size $N=6$. 
The couplings are approximate power-laws characterized by $\alpha=0, \, 1.5,\, 3.0$ (red, orange, blue). 
(a) Rate for absorbing excitations at site $i_{\text{sink}}$ 
and (b) total number of excitations in the steady state. 
Due to non-linear effects, the absorption rate is a non-monotonic function of $\Gamma_{\text{source}}$. 
\label{fig:driven_rate_number}}
\end{figure}

\begin{figure}
\begin{overpic}[width=4cm,tics=5]
{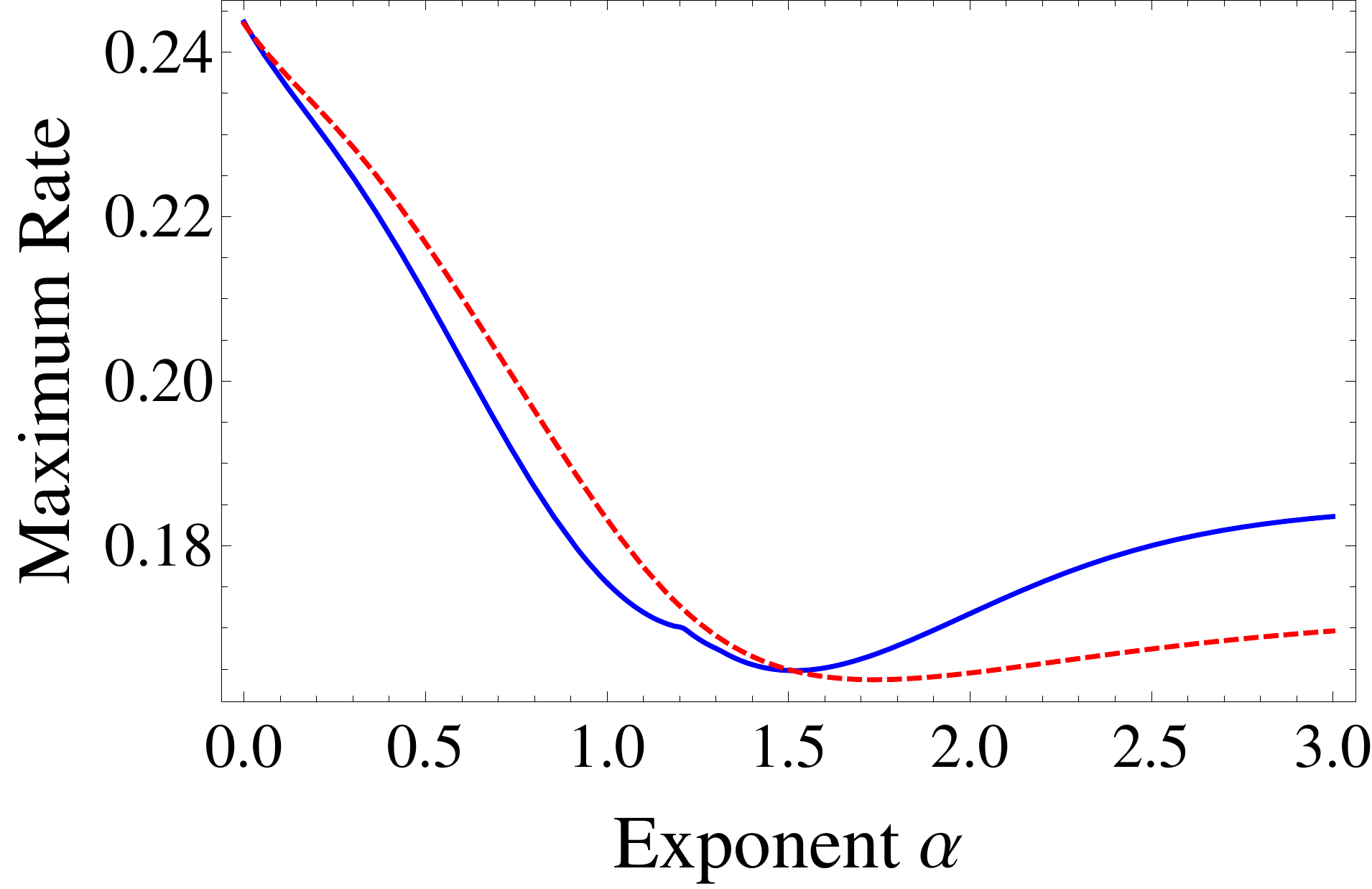}
\put(-10,70){ \large a)} 
\end{overpic}\hspace*{0.5cm}\begin{overpic}[width=4cm,tics=5]
{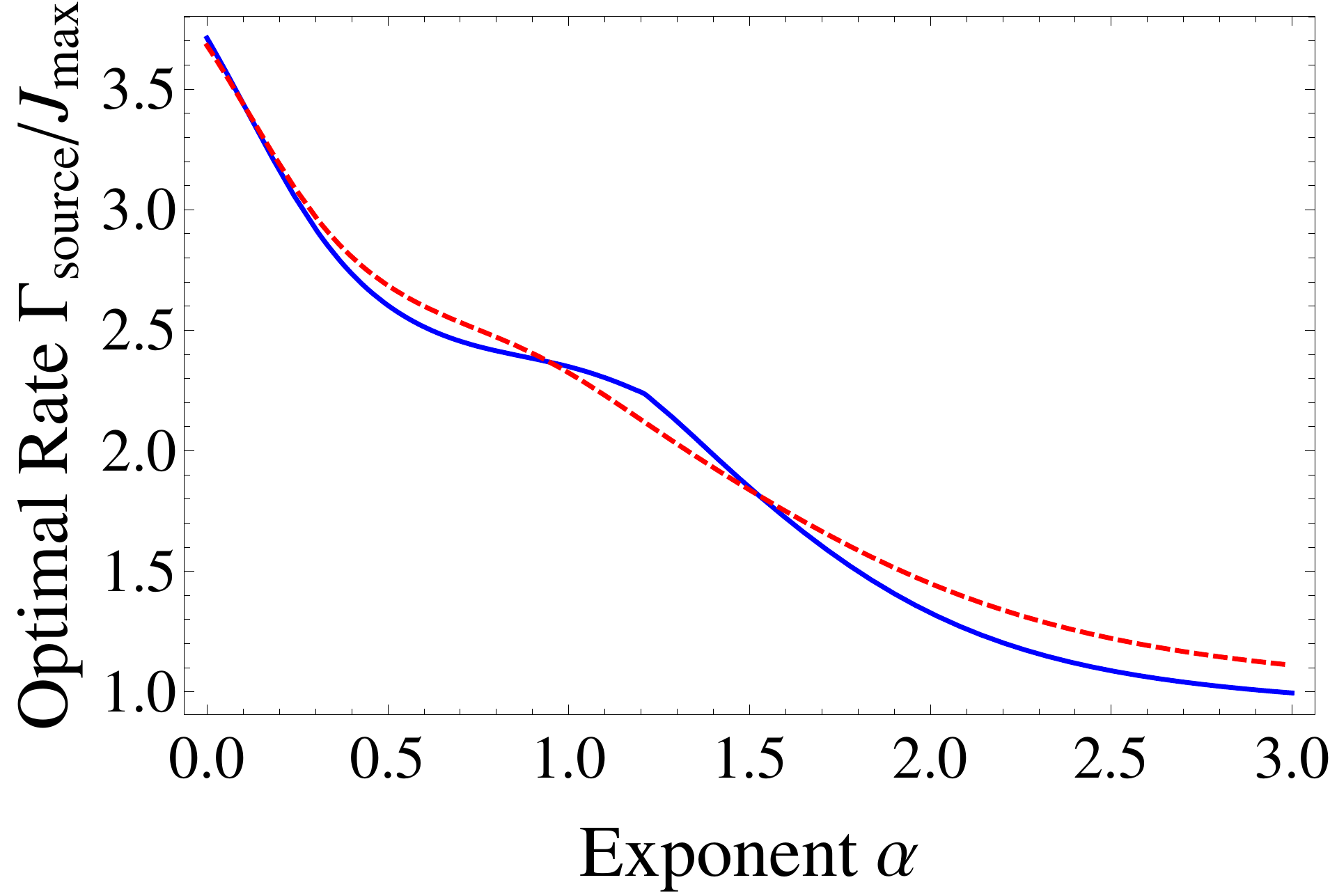}
\put(-12,70){ \large b)} 
\end{overpic}\\
\caption{
Maximum transfer rate in the steady state of a continuously driven system as a function of hopping range, for realistic couplings (solid) and an idealized power-law dependence (dashed), without collective dephasing and static disorder. The source site is $i_{\text{source}}=2$ and the target site $i_{\text{sink}}=5$, with $N=6$. 
(a) The maximum absorption rate of excitations at $i_{\text{sink}}$ has a minimum at intermediate values of $\alpha$. 
(b) The optimal source rate $\Gamma_{\text{source}}^{\text{opt}}$ achieving the maximum transfer rate given in panel (a) decreases monotonically with $\alpha$. 
\label{fig:driven_rmax_lambdamax}}
\end{figure}

In the limit $(t-t_{0})\rightarrow\infty$, the rate at which the excitations are removed from the system at $i_{\text{sink}}$ converges against a constant value, which in a photosynthetic system corresponds to the rate at which excitations are recombined at the target site. 
We numerically evaluate this rate as a function of $\Gamma_{\text{source}}$, which quantifies the coupling strength of site $i_{\text{source}}$ to the thermal reservoir. 
The corresponding results are depicted in Fig.~\ref{fig:driven_rate_number}(a), while the total number of excitations in the system is presented in Fig.~\ref{fig:driven_rate_number}(b). Naively, 
one would expect that the rate for absorbing excitations as well as
the total number of excitations in the system increases monotonically with $\Gamma_{\text{source}}$. However, as Fig.~\ref{fig:driven_rate_number} illustrates, 
there is an optimal injection rate $\Gamma_{\text{source}}^{\text{opt}}$ that yields a maximal rate.
This optimal value depends on $\alpha$ and hence on the connectivity of the network, as depicted in Fig.~\ref{fig:driven_rmax_lambdamax}. 

The existence of an optimal value for $\Gamma_{\text{source}}$ is a manifestation of non-linear effects that come into play as the number of excitation in the spin network grows. 
In order to understand this behaviour, we first focus on $\alpha>0$ in the limiting cases $\Gamma_{\text{source}}\ll J_{\text{max}}$
and $\Gamma_{\text{source}}\gg J_{\text{max}}$. 
In the regime $\Gamma_{\text{source}}\ll J_{\text{max}}$, 
the dynamics of the system is dominated by the hopping term described by $H_{J}$ with its highly delocalized eigenstates, and the driving described by $\mathcal{L}_{\text{source}}$
can be treated as a perturbation. The excitations brought into the system by $\mathcal{L}_{\text{source}}$ delocalize. As the excitations spread over the entire spin network, non-linear effects arising from the fact that each site can only support a single excitation are negligible. 

\begin{figure}
\vspace{0.7cm}

\begin{overpic}[width=4cm,tics=5]
{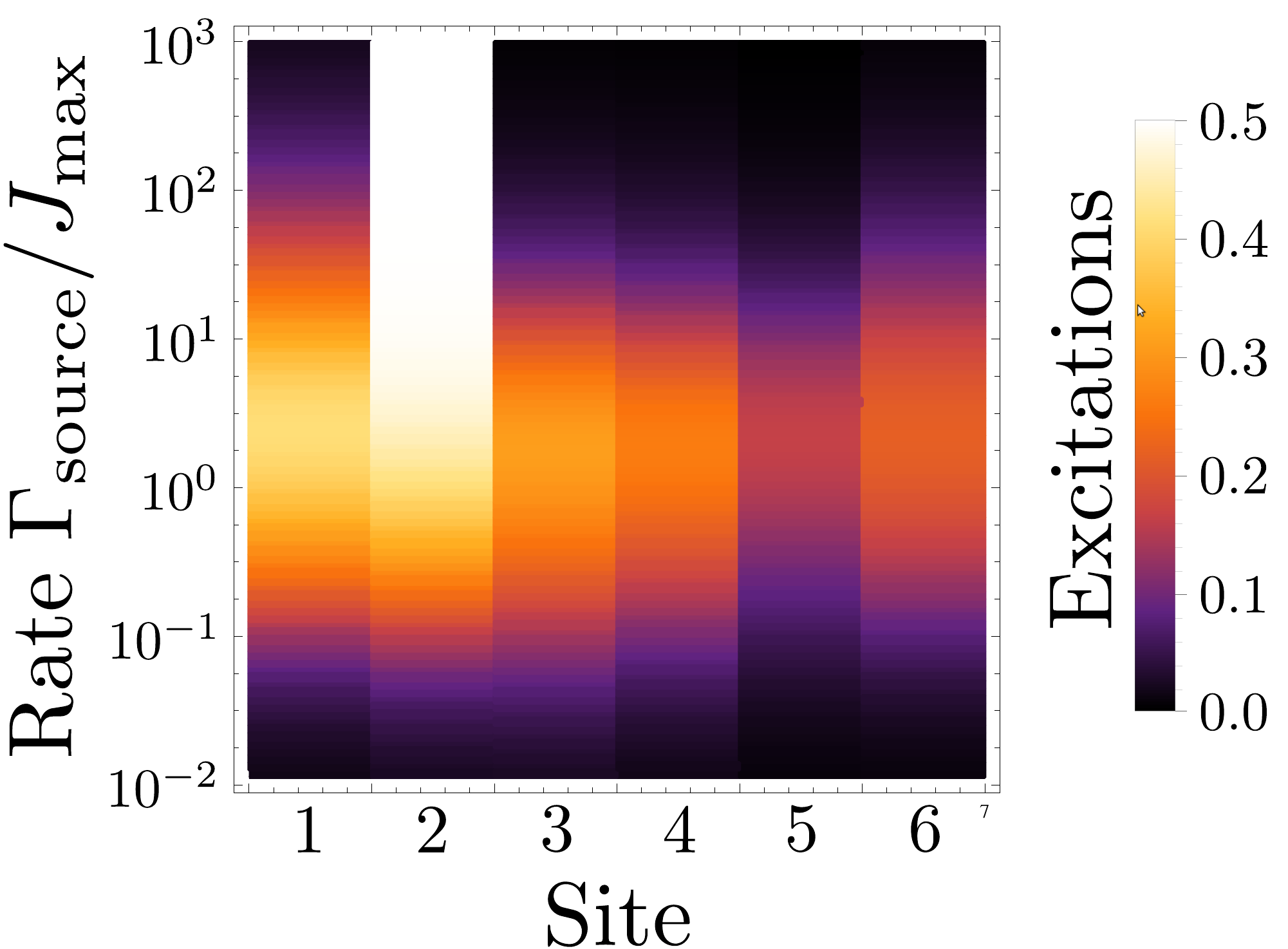}
\put(-0,90){ \large a)}
\put(32,90){ \large {$\alpha=1.5$}} 
\end{overpic}
\hspace*{0.5cm}
\begin{overpic}[width=4cm,tics=5]
{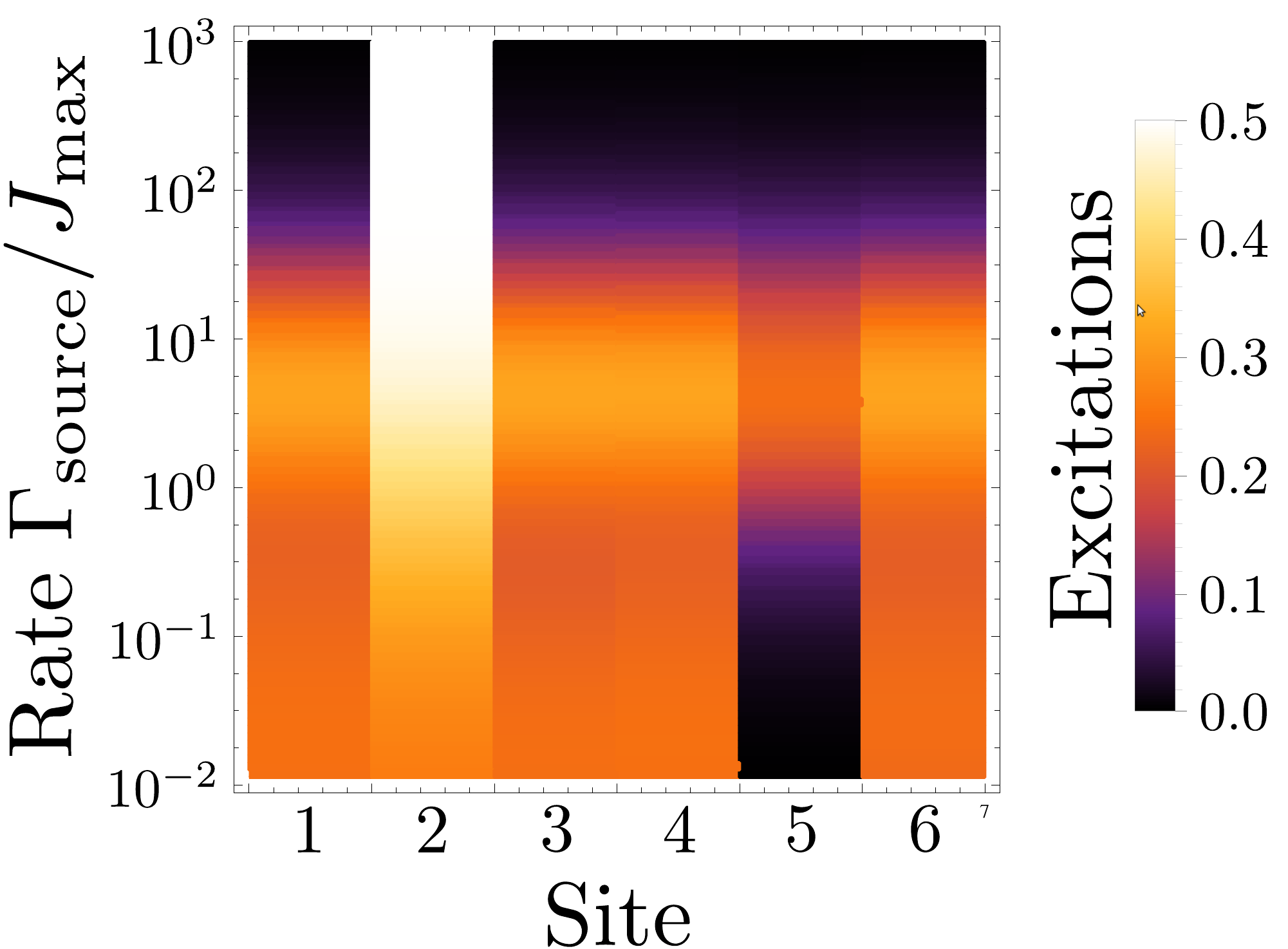}
\put(-0,90){ \large b)}
\put(32,90){ \large $\alpha=0.0$}
\end{overpic}\caption{Average number of excitations per site in the steady state, plotted
against the  rate $\Gamma_{\text{source}}$ for an ion chain with $N=6$
ions. The excitations are created at site $i_{\text{source}}=2$ and
are absorbed at site $i_{\text{sink}}=5$ . The parameters are $\alpha=1.5$ 
(a), $\alpha=0$ (b), and with $\gamma=0$ and no static disorder.\label{fig:driven_distribution}}
\end{figure}

Increasing $\Gamma_{\text{source}}$ brings more excitations into the system and initially improves the absorption rate. As $\Gamma_{\text{source}}$ increases further, however, $\mathcal{L}_{\text{source}}$ can no longer be treated as a perturbation. In the
regime $\Gamma_{\text{source}}\gg J_{\text{max}}$, the structure of the eigenstates of the super-operator $\mathcal{L}_{\text{source}}$ dominates the dynamics, and $H_{J}$ represents a small perturbation.
Since $\mathcal{L}_{\text{source}}$ describes the creation of excitations at $i_{\text{source}}$, its eigenvectors reflect a highly localized dynamics, and in the limit $\Gamma_{\text{source}}\gg J_{\text{max}}$
the excitation remains localized at $i_{\text{source}}$ in a Zeno-like effect. This behaviour is illustrated in Fig.~\ref{fig:driven_distribution}(a) for $\alpha=1.5$. The probability for finding the excitation at $i_{\text{sink}}\neq i_{\text{source}}$ goes to zero as $\Gamma_{\text{source}}\rightarrow\infty$.
In this limit, $i_{\text{sink}}$ decouples from the rest of the system, and the resulting dynamics is that of a single two-level atom coupled to an infinite-temperature heat bath. Consequently, the average number of excitations in the system converges against $0.5$.  

The behavior changes drastically when $\alpha=0$. 
For large values of $\Gamma_{\text{source}}$, the steady state is similar to $\alpha>0$, with 0.5 excitations at $i_{\rm source}$ and vanishing excitation number everywhere else. 
In the limit of small $\Gamma_{\text{source}}$, however, the steady state for $\alpha=0$ deviates significantly from the steady state for non-zero $\alpha$. The reason is again the destructive interference effect, which reduces the number of excitations that reach site $i_{\text{sink}}$ in a given time, and thus decreases the rate for absorbing the excitations when compared to scenarios with $\alpha>0$ (Fig.~\ref{fig:driven_rate_number}(a)). 
As a consequence, excitations can accumulate in the system, and, instead of vanishing numbers of excitations, we find an average excitation number
of $0.25$ at each site, except at $i_{\text{sink}}$ (Fig.~\ref{fig:driven_distribution}(b) and Fig.~\ref{fig:driven_rate_number}(b)). The result is a counterintuitive behavior: while overall more excitations are present in the steady state, fewer of them are absorbed at the target site. 
As these results show, the non-linear behaviour due to interactions between excitations leads to highly non-trivial dynamics. 

\section{Experimental considerations\label{sec:Experimental_Error_Sources}}

In this section, we address possible sources of errors in a realistic experiment and compare their rates to achievable time scales. 
The Mølmer--Sørenson gate at $\alpha\rightarrow0$, employing the axial center-of-mass mode, can be performed with gate times up to
$50\mu s$ ($J_{\text{max}}\approx15.7\cdot10^{3}\,\text{s}^{-1}$) \cite{kirchmair2009deterministic}. 
When addressing the radial modes to obtain tunable long-range interactions, realistic interaction strengths are 
$J_{\text{max}}\approx\{230\,,360\,,125\,,100\}\,\text{s}^{-1}$ 
for
$\alpha=\{0.75,\,1.07,\,1.41,\,1.75\}$ \cite{jurcevic2014quasiparticle}. 
The observation of the phenomena discussed for $\alpha\rightarrow0$ requires times not longer than $T=20/J_{\text{max}}\approx1.25\,\text{ms}$.
The transition between the dynamical regimes $\alpha<1$ and $\alpha>1$ can be observed within times  on the order of $T=\frac{10}{J_{\text{max}}}<0.1\text{\ s}$ for $N=10$ ions and $T=\frac{20}{J_{\text{max}}}<0.2\text{\ s}$ for $N=20$ ions. 
In the following, we compare these time scales to two classes of errors, those which limit the time over which physical qubits can store quantum information, and those which concern the operations performed on the qubit.

\subsection{Qubits as quantum memories}

Two possible error sources are particularly relevant for reducing the coherence time of the qubits in the absence of additional operations: 
Amplitude damping, which is caused by spontaneous decay, and phase damping, which is predominantly caused by relative fluctuations between the frequency of the qubit transition and the laser field used to read out the state of the qubit. 

For an optical qubit in the setup described in \cite{schindler2013quantum}, the lifetime is $\tau_{1}=1.13\,\text{s}$. 
The run times $T$ estimated above are at least an order of magnitude smaller than the lifetime $\tau_{1}$. 
Since the spontaneous decay acts on all ions with the same decay rate of $\tau_{1}^{-1}$, its effect on the absorption probability can be removed in the single excitation sector via the analytical expression 
\begin{equation}
P_{\text{a}}^{\prime}(t)=1-e^{-(t-t_{0})/\tau_{1}}\left[1-P_{\text{a}}(t)\right]\label{eq:res_spontaneous_decay}\,.
\end{equation}
Here, $P_{\text{a}}(t)$ is the probability for having absorbed
the excitation at site $i_{\text{sink}}$ in the absence of spontaneous
decay and $P_{\text{a}}^{\prime}(t)$ is the probability for not
finding the excitation in the quantum network at time $t$ in the presence of spontaneous emission. 
Due to the simple structure of Eq.~(\ref{eq:res_spontaneous_decay}), it is straightforward 
to eliminate the effect of spontaneous decay in the post-processing of the experimental data.

In addition, frequency fluctuations of the qubit transitions and the laser field generate dephasing noise, which, however, has been reported to be almost identical for all the ions \cite{schindler2013quantum}. 
Hence, one can find decoherence-free subspaces (DFSs) in which this type of dephasing is practically absent. 
In our case, subspaces of fixed excitation numbers are decoherence free. 
Decoherence between different subspaces does not affect the proposed quantum simulations, since the coherent Hamiltonian part of the time evolution as well as the (possibly non-Markovian) dephasing noise conserve the number of excitations, 
while the excitation number is only changed by the incoherent Markovian processes $\mathcal{L}_{\text{diss}}$ and $\mathcal{L}_{\text{source}}$.

\subsection{Errors from faulty qubit operations}

A second class of error appears as soon as operations on the qubits are performed, in particular initialization, generation of the dynamics, and readout. 

\emph{Preparation of the initial state:} The first step of the proposed experiment is the preparation of the initial state $\ket{\psi\left(t_{0}\right)}$. 
This can be done via optical pumping, which allows for the preparation of 
the states $\mid\downarrow\rangle_{i}$ and $\mid\uparrow\rangle_{i}$
with a fidelity beyond $F=99.9\%$ \cite{roos2006designer}.
Another important initialization step is the cooling of the vibrational
modes of the ion chain. After a cooling time of $200\text{ \ensuremath{\mu}s}$,
an average steady-state phonon number of $\langle n\rangle=0.5$ per
mode can be achieved, sufficient to implement 
 Mølmer--Sørenson-type interactions with satisfactory fidelity \cite{schindler2013quantum}.

\emph{Time evolution:} 
By implementing the Hamiltonian $\hat{H}$ as well as the dephasing and dissipative processes discussed in Sec.~\ref{sec:Model}, intensity fluctuations of laser fields enter as additional error source. Since the time scale of these fluctuations, on the order of seconds or minutes \cite{schindler2013quantum}, is much longer than a single run of the experiment, this error amounts to a random variation of $J_{\text{max}}$ that remains static within each run. 
The numerical results presented above indicate that the excitation transfer does not crucially depend on the precise value of $J_{\text{max}}$, so this error should not change the results significantly. 
In addition, intensity fluctuations of the addressed beams generating the AC-Stark shifts lead to fluctuations of the local excitation energies $\hbar\omega_{i}$. 
However, these appear only in the form of static disorder and dephasing. 
These will randomize somewhat the amplitude of the disorder strength and the dephasing rate, but since observables depend smoothly on both parameters, the effect should not be significant. 
Thus, our proposed quantum simulation will be robust against the intensity fluctuations of the laser field. 

\begin{figure}
	\includegraphics[scale=0.3]{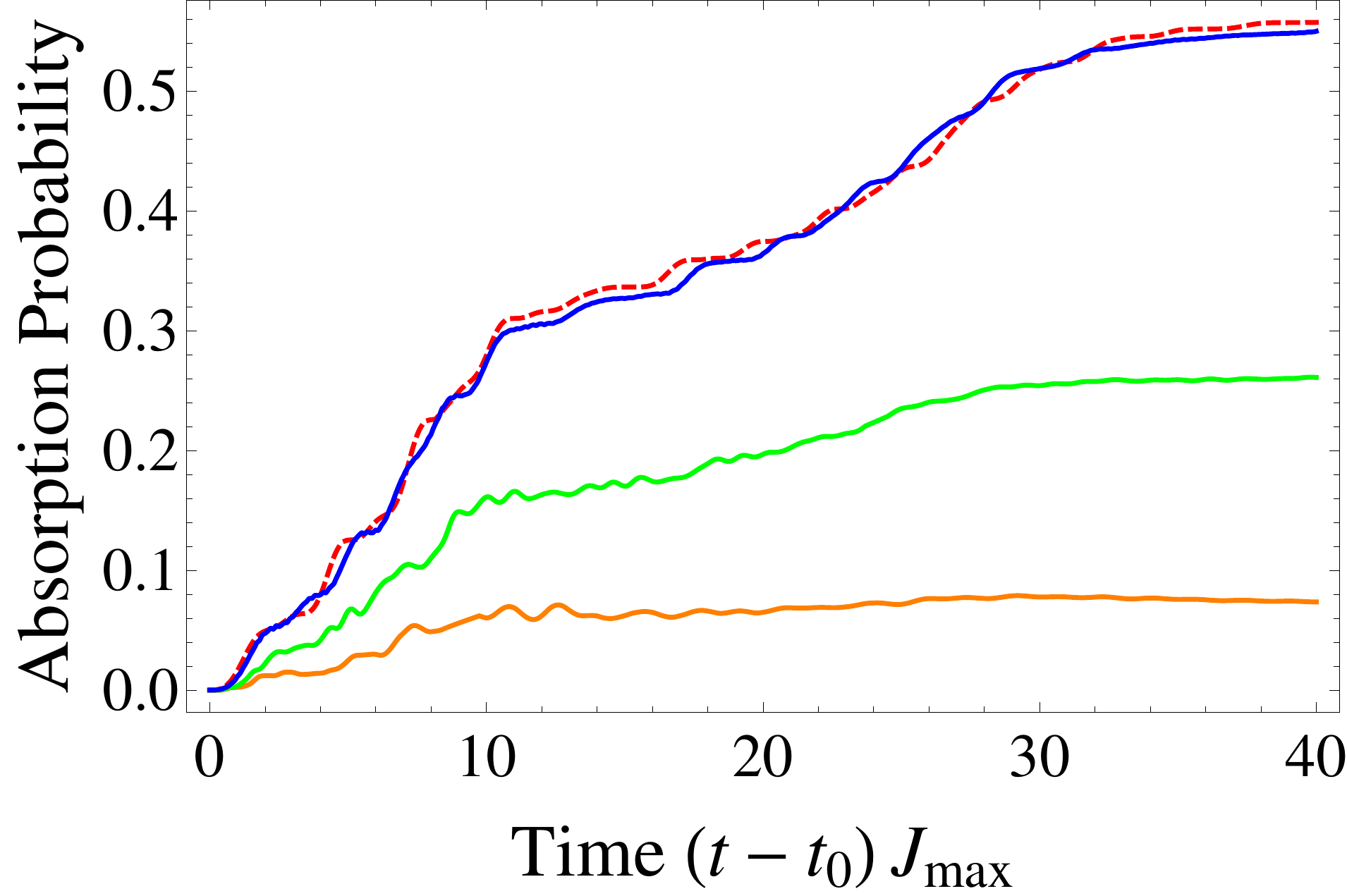}
	
	\caption{
		Comparison of the time evolution with and without off-resonant terms in $\hat{H}_{\text{int}}$, Eq.~\eqref{eq:Hint}.
		Solid curves are from bottom to top for $\omega_{\text{const}}/J_{\text{max}}=1$, 2, 10 (orange, green, blue). 
		Already with $\omega_{\text{const}}/J_{\text{max}}=10$, the result is hardly discernible from the ideal case of $\omega_{\text{const}}/J_{\text{max}}\rightarrow\infty$ (red dashed). 
		The exponent describing the hopping range is $\alpha=1$, system size is $N=10$, and there is no static disorder or dephasing.
		\label{fig:off_resonant_terms}} 
\end{figure}

\emph{Influence of off-resonant terms in the interaction Hamiltonian:}
In Sec.~\ref{sec:Model}, we have proposed to engineer the Hamiltonian $\hat{H}_{J}$ by implementing $\hat{H}_{\text{int}}$ and eliminating off-resonant terms such as $\sigma_i^+ \sigma_j^+$ by working in the regime $\hbar\omega_{\text{const}}\gg J_{ij}$. 
With a finite ratio $\omega_{\text{const}}/J_{ij}$, off-resonant terms may influence the system dynamics by creating and annihilating excitations. In order to quantify the impact of such off-resonant excitations, we performed a simulation taking them into account, with the result depicted in Fig.~\ref{fig:off_resonant_terms}. Already for $\omega_{\text{const}}/J_{\text{max}}=10.0$
the impact of the off-resonant terms in $\hat{H}_{\text{int}}$ is hardly discernible. 

\emph{Measurement:}
In the final step of the experiment, one measures the number of remaining
excitations (number of ions in state $\mid\uparrow\rangle_{i}$). This can be done through electron shelving \cite{nagourney1986shelved} with extremely high accuracy \cite{schindler2013quantum}, such that the errors it introduces can be neglected relative to other error sources discussed above.

\subsection{Accessible parameter regimes}
To estimate the reachable parameter regimes, we compare the relative rates of the ingredients generating Eq.~\eqref{eq:Master equation}. 
As mentioned above, interactions have been experimentally demonstrated with $J_{\text{max}}$ on the order of $100\;\text{s}^{-1}$ up to $15.7\cdot 10^3\;\text{s}^{-1}$. 
Disorder and dephasing amplitudes can be orders of magnitude larger: with detunings of $\text{100 GHz}$ from the
$4^{2}P_{3/2}\leftrightarrow3^{2}D_{5/2}$ transition an AC-Stark shift of $2\pi\cdot1\;\text{MHz}$ is achievable \cite{PhysRevLett.111.180501} and in \cite{lee2016engineering} AC-Stark shifts of up to $10\text{ MHz}$ have been demonstrated. 
Further, dephasing autocorrelation time can be much shorter than the other relevant time scales. For example, by changing the intensity of the corresponding lasers using an acousto-optic modulator one can achieve  switching times on a nanosecond scale, allowing for the realization of dephasing deep in the Markovian regime. 
For $^{171}Yb^+$ ions, with the qubit transition in the microwave range, flat as well as structured dephasing noise has been demonstrated up to a cutoff frequency of $200$Hz \cite{Soare2014}.

\section{Conclusion\label{sec:Conclusion}}

In summary, we have presented a feasible scheme to quantum simulate the transfer of excitations through a quantum network. 
The proposal exploits existing trapped-ion technology and can be robustly implemented in state-of-the-art experiments.  

As we have shown, the main parameters describing ENAQT---disorder strength and dephasing rate---can be tuned over wide ranges, permitting the study of this phenomenon in the interplay between Anderson localization, noise-induced transfer, and freezing due to quantum Zeno dynamics. 
An additional feature of ion chains are tunable long-range interactions. 
In detailed numerical simulations, we have illustrated how the transfer efficiency grows with increasing hopping range at short times, but diminishes at larger times due to a localization phenomenon induced by destructive interference \cite{caruso2009highly}. Counterintuitively, but similar to the case of disorder-induced ENAQT, small amounts of disorder or dephasing destroy the destructive interference and thus enhance the transport efficiencies for large hopping ranges. 

Furthermore, the proposed scheme allows one to study the impact of non-Markovian effects on the excitation transfer, thus complementing recent theoretical investigations into the role of non-Markovian processes for the energy transfer process in biomolecules \cite{Thorwart2009,mohseni2014energy,chen2011excitation,chin2013role,delRey2013,Jesenko2013}. 
In our numerical calculations, we have found that, while the Markovian and a simple non-Markovian dephasing reach similar maximal transfer efficiencies, the non-Markovian noise can hold larger values over a broader parameter range. 
Here, we have been interested in the transfer up to a fixed but finite time, which in realistic experiments will be limited by loss mechanisms. In the limit of infinite waiting times, the efficiency will become independent of non-Markovian effects \cite{Jesenko2013}.

Our scheme also provides a framework for investigating non-linear dynamics, which emerges as larger numbers of excitations are injected into the spin network. 
In this regime, numerical calculations are intractable except for very small systems, thus making it a particularly attractive target for quantum-simulation experiments. 
We have studied this regime for small driven-dissipative systems in which excitations are injected continuously from an infinite-temperature heat bath, resembling the incoherent absorption of photons by photosynthetic systems. 
We have found markedly different behavior for $\alpha=0$ and $\alpha>0$. While for $\alpha=0$ large numbers of excitations may be present, due to the destructive interference effect the rate of absorbing them at the target site may actually be lower than for $\alpha>0$. 
It will be interesting to use interactions between excitations to design so-called `optical ratchet' states. There, the system can continue to absorb excitations while not losing to spontaneous emission those already gained \cite{Higgins2017}.

Experiments may investigate these effects not only in one-dimensional chains, but also in other geometries. Two-dimensional crystals can be realized, e.g., in Penning traps \cite{Britton2012,Bohnet2016} or linear Paul traps with strong axial confinement \cite{Richerme2016}, and segmented Paul traps \cite{Zippilli2014}, nanofabricated surface traps \cite{Schmied2009b}, periodic driving \cite{Nevado2017}, or additional laser frequencies \cite{Korenblit2012} may allow to design arbitrary interaction patterns. 
A further future direction will be to design the dissipation as a true quantum bath instead of classical dephasing. This may be achieved, e.g., by coupling the spin network to the vibrational phonon modes of the ion crystal, as has been proposed for the study of spin-boson models \cite{Porras2008,Juenemann2013}.

To conclude, by exploiting the high level of control of current ion-chain experiments, it will be possible to study many different aspects of excitation transfer in open quantum networks in a highly controllable environment. These studies may permit deeper insights into phenomena such as the energy transfer in photosynthetic systems or the conductance properties of materials.

\emph{Acknowledgements.---}
We acknowledge interesting discussions with Ch.~Roos, Ch.~Maier, P.~Jurcevic, and T.~Brydges, and we thank R.~de J. Le\'on-Montiel and S.~Whitlock for useful comments on the manuscript. 
This work was supported by the DFG as part of the CRC 1119 CROSSING, by the Austrian Science Fund (FWF), through the SFB FoQuS (Project No.\ F4016-N23) and the European Commission via the ERC advanced grant EntangleGen (Project ID 694561).

\bibliographystyle{apsrev4-1}

\begin{thebibliography}{75}%
	\makeatletter
	\providecommand \@ifxundefined [1]{%
		\@ifx{#1\undefined}
	}%
	\providecommand \@ifnum [1]{%
		\ifnum #1\expandafter \@firstoftwo
		\else \expandafter \@secondoftwo
		\fi
	}%
	\providecommand \@ifx [1]{%
		\ifx #1\expandafter \@firstoftwo
		\else \expandafter \@secondoftwo
		\fi
	}%
	\providecommand \natexlab [1]{#1}%
	\providecommand \enquote  [1]{``#1''}%
	\providecommand \bibnamefont  [1]{#1}%
	\providecommand \bibfnamefont [1]{#1}%
	\providecommand \citenamefont [1]{#1}%
	\providecommand \href@noop [0]{\@secondoftwo}%
	\providecommand \href [0]{\begingroup \@sanitize@url \@href}%
	\providecommand \@href[1]{\@@startlink{#1}\@@href}%
	\providecommand \@@href[1]{\endgroup#1\@@endlink}%
	\providecommand \@sanitize@url [0]{\catcode `\\12\catcode `\$12\catcode
		`\&12\catcode `\#12\catcode `\^12\catcode `\_12\catcode `\%12\relax}%
	\providecommand \@@startlink[1]{}%
	\providecommand \@@endlink[0]{}%
	\providecommand \url  [0]{\begingroup\@sanitize@url \@url }%
	\providecommand \@url [1]{\endgroup\@href {#1}{\urlprefix }}%
	\providecommand \urlprefix  [0]{URL }%
	\providecommand \Eprint [0]{\href }%
	\providecommand \doibase [0]{http://dx.doi.org/}%
	\providecommand \selectlanguage [0]{\@gobble}%
	\providecommand \bibinfo  [0]{\@secondoftwo}%
	\providecommand \bibfield  [0]{\@secondoftwo}%
	\providecommand \translation [1]{[#1]}%
	\providecommand \BibitemOpen [0]{}%
	\providecommand \bibitemStop [0]{}%
	\providecommand \bibitemNoStop [0]{.\EOS\space}%
	\providecommand \EOS [0]{\spacefactor3000\relax}%
	\providecommand \BibitemShut  [1]{\csname bibitem#1\endcsname}%
	\let\auto@bib@innerbib\@empty
	\bibitem [{\citenamefont {Fassioli}\ \emph {et~al.}(2013)\citenamefont
		{Fassioli}, \citenamefont {Dinshaw}, \citenamefont {Arpin},\ and\
		\citenamefont {Scholes}}]{Fassioli2013}%
	\BibitemOpen
	\bibfield  {author} {\bibinfo {author} {\bibfnamefont {F.}~\bibnamefont
			{Fassioli}}, \bibinfo {author} {\bibfnamefont {R.}~\bibnamefont {Dinshaw}},
		\bibinfo {author} {\bibfnamefont {P.~C.}\ \bibnamefont {Arpin}}, \ and\
		\bibinfo {author} {\bibfnamefont {G.~D.}\ \bibnamefont {Scholes}},\ }\href
	{\doibase 10.1098/rsif.2013.0901} {\bibfield  {journal} {\bibinfo  {journal}
			{Journal of The Royal Society Interface}\ }\textbf {\bibinfo {volume} {11}}
		(\bibinfo {year} {2013}),\ 10.1098/rsif.2013.0901}\BibitemShut {NoStop}%
	\bibitem [{\citenamefont {Huelga}\ and\ \citenamefont
		{Plenio}(2013)}]{HuelgaPlenio2013}%
	\BibitemOpen
	\bibfield  {author} {\bibinfo {author} {\bibfnamefont {S.}~\bibnamefont
			{Huelga}}\ and\ \bibinfo {author} {\bibfnamefont {M.}~\bibnamefont
			{Plenio}},\ }\href {\doibase 10.1080/00405000.2013.829687} {\bibfield
		{journal} {\bibinfo  {journal} {Contemp. Phys}\ }\textbf {\bibinfo {volume}
			{54}},\ \bibinfo {pages} {181 } (\bibinfo {year} {2013})}\BibitemShut
	{NoStop}%
	\bibitem [{\citenamefont {Lambert}\ \emph {et~al.}(2013)\citenamefont
		{Lambert}, \citenamefont {Chen}, \citenamefont {Cheng}, \citenamefont {Li},
		\citenamefont {Chen},\ and\ \citenamefont {Nori}}]{Lambert2013}%
	\BibitemOpen
	\bibfield  {author} {\bibinfo {author} {\bibfnamefont {N.}~\bibnamefont
			{Lambert}}, \bibinfo {author} {\bibfnamefont {Y.-N.}\ \bibnamefont {Chen}},
		\bibinfo {author} {\bibfnamefont {Y.-C.}\ \bibnamefont {Cheng}}, \bibinfo
		{author} {\bibfnamefont {C.-M.}\ \bibnamefont {Li}}, \bibinfo {author}
		{\bibfnamefont {G.-Y.}\ \bibnamefont {Chen}}, \ and\ \bibinfo {author}
		{\bibfnamefont {F.}~\bibnamefont {Nori}},\ }\href {\doibase
		doi:10.1038/nphys2474} {\bibfield  {journal} {\bibinfo  {journal} {Nature
				Physics}\ }\textbf {\bibinfo {volume} {9}},\ \bibinfo {pages} {10} (\bibinfo
		{year} {2013})}\BibitemShut {NoStop}%
	\bibitem [{\citenamefont {Hu}\ \emph {et~al.}(1997)\citenamefont {Hu},
		\citenamefont {Ritz}, \citenamefont {Damjanovic},\ and\ \citenamefont
		{Schulten}}]{hu1997pigment}%
	\BibitemOpen
	\bibfield  {author} {\bibinfo {author} {\bibfnamefont {X.}~\bibnamefont
			{Hu}}, \bibinfo {author} {\bibfnamefont {T.}~\bibnamefont {Ritz}}, \bibinfo
		{author} {\bibfnamefont {A.}~\bibnamefont {Damjanovic}}, \ and\ \bibinfo
		{author} {\bibfnamefont {K.}~\bibnamefont {Schulten}},\ }\href@noop {}
	{\bibfield  {journal} {\bibinfo  {journal} {The Journal of Physical Chemistry
				B}\ }\textbf {\bibinfo {volume} {101}},\ \bibinfo {pages} {3854} (\bibinfo
		{year} {1997})}\BibitemShut {NoStop}%
	\bibitem [{\citenamefont {Ritz}\ \emph {et~al.}(2001)\citenamefont {Ritz},
		\citenamefont {Park},\ and\ \citenamefont {Schulten}}]{ritz2001kinetics}%
	\BibitemOpen
	\bibfield  {author} {\bibinfo {author} {\bibfnamefont {T.}~\bibnamefont
			{Ritz}}, \bibinfo {author} {\bibfnamefont {S.}~\bibnamefont {Park}}, \ and\
		\bibinfo {author} {\bibfnamefont {K.}~\bibnamefont {Schulten}},\ }\href@noop
	{} {\bibfield  {journal} {\bibinfo  {journal} {The Journal of Physical
				Chemistry B}\ }\textbf {\bibinfo {volume} {105}},\ \bibinfo {pages} {8259}
		(\bibinfo {year} {2001})}\BibitemShut {NoStop}%
	\bibitem [{\citenamefont {Novoderezhkin}\ \emph {et~al.}(2004)\citenamefont
		{Novoderezhkin}, \citenamefont {Palacios}, \citenamefont {Van~Amerongen},\
		and\ \citenamefont {Van~Grondelle}}]{novoderezhkin2004energy}%
	\BibitemOpen
	\bibfield  {author} {\bibinfo {author} {\bibfnamefont {V.~I.}\ \bibnamefont
			{Novoderezhkin}}, \bibinfo {author} {\bibfnamefont {M.~A.}\ \bibnamefont
			{Palacios}}, \bibinfo {author} {\bibfnamefont {H.}~\bibnamefont
			{Van~Amerongen}}, \ and\ \bibinfo {author} {\bibfnamefont {R.}~\bibnamefont
			{Van~Grondelle}},\ }\href@noop {} {\bibfield  {journal} {\bibinfo  {journal}
			{The Journal of Physical Chemistry B}\ }\textbf {\bibinfo {volume} {108}},\
		\bibinfo {pages} {10363} (\bibinfo {year} {2004})}\BibitemShut {NoStop}%
	\bibitem [{\citenamefont {Cho}\ \emph {et~al.}(2005)\citenamefont {Cho},
		\citenamefont {Vaswani}, \citenamefont {Brixner}, \citenamefont {Stenger},\
		and\ \citenamefont {Fleming}}]{cho2005exciton}%
	\BibitemOpen
	\bibfield  {author} {\bibinfo {author} {\bibfnamefont {M.}~\bibnamefont
			{Cho}}, \bibinfo {author} {\bibfnamefont {H.~M.}\ \bibnamefont {Vaswani}},
		\bibinfo {author} {\bibfnamefont {T.}~\bibnamefont {Brixner}}, \bibinfo
		{author} {\bibfnamefont {J.}~\bibnamefont {Stenger}}, \ and\ \bibinfo
		{author} {\bibfnamefont {G.~R.}\ \bibnamefont {Fleming}},\ }\href@noop {}
	{\bibfield  {journal} {\bibinfo  {journal} {The Journal of Physical Chemistry
				B}\ }\textbf {\bibinfo {volume} {109}},\ \bibinfo {pages} {10542} (\bibinfo
		{year} {2005})}\BibitemShut {NoStop}%
	\bibitem [{\citenamefont {Engel}\ \emph {et~al.}(2007)\citenamefont {Engel},
		\citenamefont {Calhoun}, \citenamefont {Read}, \citenamefont {Ahn},
		\citenamefont {Man{\v{c}}al}, \citenamefont {Cheng}, \citenamefont
		{Blankenship},\ and\ \citenamefont {Fleming}}]{engel2007evidence}%
	\BibitemOpen
	\bibfield  {author} {\bibinfo {author} {\bibfnamefont {G.~S.}\ \bibnamefont
			{Engel}}, \bibinfo {author} {\bibfnamefont {T.~R.}\ \bibnamefont {Calhoun}},
		\bibinfo {author} {\bibfnamefont {E.~L.}\ \bibnamefont {Read}}, \bibinfo
		{author} {\bibfnamefont {T.-K.}\ \bibnamefont {Ahn}}, \bibinfo {author}
		{\bibfnamefont {T.}~\bibnamefont {Man{\v{c}}al}}, \bibinfo {author}
		{\bibfnamefont {Y.-C.}\ \bibnamefont {Cheng}}, \bibinfo {author}
		{\bibfnamefont {R.~E.}\ \bibnamefont {Blankenship}}, \ and\ \bibinfo {author}
		{\bibfnamefont {G.~R.}\ \bibnamefont {Fleming}},\ }\href@noop {} {\bibfield
		{journal} {\bibinfo  {journal} {Nature}\ }\textbf {\bibinfo {volume} {446}},\
		\bibinfo {pages} {782} (\bibinfo {year} {2007})}\BibitemShut {NoStop}%
	\bibitem [{\citenamefont {Lee}\ \emph {et~al.}(2007)\citenamefont {Lee},
		\citenamefont {Cheng},\ and\ \citenamefont {Fleming}}]{Lee2007a}%
	\BibitemOpen
	\bibfield  {author} {\bibinfo {author} {\bibfnamefont {H.}~\bibnamefont
			{Lee}}, \bibinfo {author} {\bibfnamefont {Y.-C.}\ \bibnamefont {Cheng}}, \
		and\ \bibinfo {author} {\bibfnamefont {G.~R.}\ \bibnamefont {Fleming}},\
	}\href@noop {} {\bibfield  {journal} {\bibinfo  {journal} {Science}\ }\textbf
		{\bibinfo {volume} {316}},\ \bibinfo {pages} {1462} (\bibinfo {year}
		{2007})}\BibitemShut {NoStop}%
	\bibitem [{\citenamefont {Panitchayangkoon}\ \emph {et~al.}(2010)\citenamefont
		{Panitchayangkoon}, \citenamefont {Hayes}, \citenamefont {Fransted},
		\citenamefont {Caram}, \citenamefont {Harel}, \citenamefont {Wen},
		\citenamefont {Blankenship},\ and\ \citenamefont
		{Engel}}]{Panitchayangkoon2010}%
	\BibitemOpen
	\bibfield  {author} {\bibinfo {author} {\bibfnamefont {G.}~\bibnamefont
			{Panitchayangkoon}}, \bibinfo {author} {\bibfnamefont {D.}~\bibnamefont
			{Hayes}}, \bibinfo {author} {\bibfnamefont {K.~A.}\ \bibnamefont {Fransted}},
		\bibinfo {author} {\bibfnamefont {J.~R.}\ \bibnamefont {Caram}}, \bibinfo
		{author} {\bibfnamefont {E.}~\bibnamefont {Harel}}, \bibinfo {author}
		{\bibfnamefont {J.}~\bibnamefont {Wen}}, \bibinfo {author} {\bibfnamefont
			{R.~E.}\ \bibnamefont {Blankenship}}, \ and\ \bibinfo {author} {\bibfnamefont
			{G.~S.}\ \bibnamefont {Engel}},\ }\href {\doibase
		doi:10.1073/pnas.1005484107} {\bibfield  {journal} {\bibinfo  {journal}
			{PNAS}\ }\textbf {\bibinfo {volume} {107}},\ \bibinfo {pages} {12766}
		(\bibinfo {year} {2010})}\BibitemShut {NoStop}%
	\bibitem [{\citenamefont {Mohseni}\ \emph {et~al.}(2008)\citenamefont
		{Mohseni}, \citenamefont {Rebentrost}, \citenamefont {Lloyd},\ and\
		\citenamefont {Aspuru-Guzik}}]{mohseni2008environment}%
	\BibitemOpen
	\bibfield  {author} {\bibinfo {author} {\bibfnamefont {M.}~\bibnamefont
			{Mohseni}}, \bibinfo {author} {\bibfnamefont {P.}~\bibnamefont {Rebentrost}},
		\bibinfo {author} {\bibfnamefont {S.}~\bibnamefont {Lloyd}}, \ and\ \bibinfo
		{author} {\bibfnamefont {A.}~\bibnamefont {Aspuru-Guzik}},\ }\href@noop {}
	{\bibfield  {journal} {\bibinfo  {journal} {The Journal of chemical physics}\
		}\textbf {\bibinfo {volume} {129}},\ \bibinfo {pages} {174106} (\bibinfo
		{year} {2008})}\BibitemShut {NoStop}%
	\bibitem [{\citenamefont {Plenio}\ and\ \citenamefont
		{Huelga}(2008)}]{plenio2008dephasing}%
	\BibitemOpen
	\bibfield  {author} {\bibinfo {author} {\bibfnamefont {M.~B.}\ \bibnamefont
			{Plenio}}\ and\ \bibinfo {author} {\bibfnamefont {S.~F.}\ \bibnamefont
			{Huelga}},\ }\href@noop {} {\bibfield  {journal} {\bibinfo  {journal} {New
				Journal of Physics}\ }\textbf {\bibinfo {volume} {10}},\ \bibinfo {pages}
		{113019} (\bibinfo {year} {2008})}\BibitemShut {NoStop}%
	\bibitem [{\citenamefont {Caruso}\ \emph {et~al.}(2009)\citenamefont {Caruso},
		\citenamefont {Chin}, \citenamefont {Datta}, \citenamefont {Huelga},\ and\
		\citenamefont {Plenio}}]{caruso2009highly}%
	\BibitemOpen
	\bibfield  {author} {\bibinfo {author} {\bibfnamefont {F.}~\bibnamefont
			{Caruso}}, \bibinfo {author} {\bibfnamefont {A.~W.}\ \bibnamefont {Chin}},
		\bibinfo {author} {\bibfnamefont {A.}~\bibnamefont {Datta}}, \bibinfo
		{author} {\bibfnamefont {S.~F.}\ \bibnamefont {Huelga}}, \ and\ \bibinfo
		{author} {\bibfnamefont {M.~B.}\ \bibnamefont {Plenio}},\ }\href@noop {}
	{\bibfield  {journal} {\bibinfo  {journal} {The Journal of Chemical Physics}\
		}\textbf {\bibinfo {volume} {131}},\ \bibinfo {pages} {105106} (\bibinfo
		{year} {2009})}\BibitemShut {NoStop}%
	\bibitem [{\citenamefont {Thorwart}\ \emph {et~al.}(2009)\citenamefont
		{Thorwart}, \citenamefont {Eckel}, \citenamefont {Reina}, \citenamefont
		{Nalbach},\ and\ \citenamefont {Weiss}}]{Thorwart2009}%
	\BibitemOpen
	\bibfield  {author} {\bibinfo {author} {\bibfnamefont {M.}~\bibnamefont
			{Thorwart}}, \bibinfo {author} {\bibfnamefont {J.}~\bibnamefont {Eckel}},
		\bibinfo {author} {\bibfnamefont {J.~H.}\ \bibnamefont {Reina}}, \bibinfo
		{author} {\bibfnamefont {P.}~\bibnamefont {Nalbach}}, \ and\ \bibinfo
		{author} {\bibfnamefont {S.}~\bibnamefont {Weiss}},\ }\href {\doibase
		10.1016/j.cplett.2009.07.053} {\bibfield  {journal} {\bibinfo  {journal}
			{Chem. Phys. Lett.}\ }\textbf {\bibinfo {volume} {478}},\ \bibinfo {pages}
		{234} (\bibinfo {year} {2009})}\BibitemShut {NoStop}%
	\bibitem [{\citenamefont {Chen}\ and\ \citenamefont
		{Silbey}(2011)}]{chen2011excitation}%
	\BibitemOpen
	\bibfield  {author} {\bibinfo {author} {\bibfnamefont {X.}~\bibnamefont
			{Chen}}\ and\ \bibinfo {author} {\bibfnamefont {R.~J.}\ \bibnamefont
			{Silbey}},\ }\href@noop {} {\bibfield  {journal} {\bibinfo  {journal} {The
				Journal of Physical Chemistry B}\ }\textbf {\bibinfo {volume} {115}},\
		\bibinfo {pages} {5499} (\bibinfo {year} {2011})}\BibitemShut {NoStop}%
	\bibitem [{\citenamefont {Chin}\ \emph {et~al.}(2013)\citenamefont {Chin},
		\citenamefont {Prior}, \citenamefont {Rosenbach}, \citenamefont
		{Caycedo-Soler}, \citenamefont {Huelga},\ and\ \citenamefont
		{Plenio}}]{chin2013role}%
	\BibitemOpen
	\bibfield  {author} {\bibinfo {author} {\bibfnamefont {A.}~\bibnamefont
			{Chin}}, \bibinfo {author} {\bibfnamefont {J.}~\bibnamefont {Prior}},
		\bibinfo {author} {\bibfnamefont {R.}~\bibnamefont {Rosenbach}}, \bibinfo
		{author} {\bibfnamefont {F.}~\bibnamefont {Caycedo-Soler}}, \bibinfo {author}
		{\bibfnamefont {S.}~\bibnamefont {Huelga}}, \ and\ \bibinfo {author}
		{\bibfnamefont {M.}~\bibnamefont {Plenio}},\ }\href@noop {} {\bibfield
		{journal} {\bibinfo  {journal} {Nature Physics}\ }\textbf {\bibinfo {volume}
			{9}},\ \bibinfo {pages} {113} (\bibinfo {year} {2013})}\BibitemShut {NoStop}%
	\bibitem [{\citenamefont {{del Rey}}\ \emph {et~al.}(2013)\citenamefont {{del
				Rey}}, \citenamefont {Chin}, \citenamefont {Huelga},\ and\ \citenamefont
		{Plenio}}]{delRey2013}%
	\BibitemOpen
	\bibfield  {author} {\bibinfo {author} {\bibfnamefont {M.}~\bibnamefont {{del
					Rey}}}, \bibinfo {author} {\bibfnamefont {A.~W.}\ \bibnamefont {Chin}},
		\bibinfo {author} {\bibfnamefont {S.~F.}\ \bibnamefont {Huelga}}, \ and\
		\bibinfo {author} {\bibfnamefont {M.~B.}\ \bibnamefont {Plenio}},\ }\href
	{\doibase DOI: 10.1021/jz400058a} {\bibfield  {journal} {\bibinfo  {journal}
			{J. Phys. Chem. Lett}\ }\textbf {\bibinfo {volume} {4}},\ \bibinfo {pages}
		{903} (\bibinfo {year} {2013})}\BibitemShut {NoStop}%
	\bibitem [{\citenamefont {Mohseni}\ \emph {et~al.}(2014)\citenamefont
		{Mohseni}, \citenamefont {Shabani}, \citenamefont {Lloyd},\ and\
		\citenamefont {Rabitz}}]{mohseni2014energy}%
	\BibitemOpen
	\bibfield  {author} {\bibinfo {author} {\bibfnamefont {M.}~\bibnamefont
			{Mohseni}}, \bibinfo {author} {\bibfnamefont {A.}~\bibnamefont {Shabani}},
		\bibinfo {author} {\bibfnamefont {S.}~\bibnamefont {Lloyd}}, \ and\ \bibinfo
		{author} {\bibfnamefont {H.}~\bibnamefont {Rabitz}},\ }\href@noop {}
	{\bibfield  {journal} {\bibinfo  {journal} {The Journal of chemical physics}\
		}\textbf {\bibinfo {volume} {140}},\ \bibinfo {pages} {035102} (\bibinfo
		{year} {2014})}\BibitemShut {NoStop}%
	\bibitem [{\citenamefont {de~J.~Leon-Montiel}\ \emph
		{et~al.}(2014)\citenamefont {de~J.~Leon-Montiel}, \citenamefont {Kassal},\
		and\ \citenamefont {Torres}}]{Leon2014}%
	\BibitemOpen
	\bibfield  {author} {\bibinfo {author} {\bibfnamefont {R.}~\bibnamefont
			{de~J.~Leon-Montiel}}, \bibinfo {author} {\bibfnamefont {I.}~\bibnamefont
			{Kassal}}, \ and\ \bibinfo {author} {\bibfnamefont {J.~P.}\ \bibnamefont
			{Torres}},\ }\href@noop {} {\bibfield  {journal} {\bibinfo  {journal} {J.
				Phys. Chem. B}\ }\textbf {\bibinfo {volume} {118}},\ \bibinfo {pages} {10588}
		(\bibinfo {year} {2014})}\BibitemShut {NoStop}%
	\bibitem [{\citenamefont {Viciani}\ \emph {et~al.}(2015)\citenamefont
		{Viciani}, \citenamefont {Lima}, \citenamefont {Bellini},\ and\ \citenamefont
		{Caruso}}]{Viciani2015}%
	\BibitemOpen
	\bibfield  {author} {\bibinfo {author} {\bibfnamefont {S.}~\bibnamefont
			{Viciani}}, \bibinfo {author} {\bibfnamefont {M.}~\bibnamefont {Lima}},
		\bibinfo {author} {\bibfnamefont {M.}~\bibnamefont {Bellini}}, \ and\
		\bibinfo {author} {\bibfnamefont {F.}~\bibnamefont {Caruso}},\ }\href@noop {}
	{\bibfield  {journal} {\bibinfo  {journal} {Phys. Rev. Lett.}\ }\textbf
		{\bibinfo {volume} {115}},\ \bibinfo {pages} {083601} (\bibinfo {year}
		{2015})}\BibitemShut {NoStop}%
	\bibitem [{\citenamefont {Biggerstaff}\ \emph {et~al.}(2016)\citenamefont
		{Biggerstaff}, \citenamefont {Heilmann}, \citenamefont {Zecevik},
		\citenamefont {Graefe}, \citenamefont {Broome}, \citenamefont {Fedrizzi},
		\citenamefont {Nolte}, \citenamefont {Szameit}, \citenamefont {White},\ and\
		\citenamefont {Kassal}}]{Biggerstaff2016}%
	\BibitemOpen
	\bibfield  {author} {\bibinfo {author} {\bibfnamefont {D.~N.}\ \bibnamefont
			{Biggerstaff}}, \bibinfo {author} {\bibfnamefont {R.}~\bibnamefont
			{Heilmann}}, \bibinfo {author} {\bibfnamefont {A.~A.}\ \bibnamefont
			{Zecevik}}, \bibinfo {author} {\bibfnamefont {M.}~\bibnamefont {Graefe}},
		\bibinfo {author} {\bibfnamefont {M.~A.}\ \bibnamefont {Broome}}, \bibinfo
		{author} {\bibfnamefont {A.}~\bibnamefont {Fedrizzi}}, \bibinfo {author}
		{\bibfnamefont {S.}~\bibnamefont {Nolte}}, \bibinfo {author} {\bibfnamefont
			{A.}~\bibnamefont {Szameit}}, \bibinfo {author} {\bibfnamefont {A.~G.}\
			\bibnamefont {White}}, \ and\ \bibinfo {author} {\bibfnamefont
			{I.}~\bibnamefont {Kassal}},\ }\href {\doibase 10.1038/ncomms11282}
	{\bibfield  {journal} {\bibinfo  {journal} {Nat. Commun.}\ }\textbf {\bibinfo
			{volume} {7}},\ \bibinfo {pages} {11282} (\bibinfo {year}
		{2016})}\BibitemShut {NoStop}%
	\bibitem [{\citenamefont {{de J. Leon-Montiel}}\ \emph
		{et~al.}(2015)\citenamefont {{de J. Leon-Montiel}}, \citenamefont
		{Quiroz-Juarez}, \citenamefont {Quintero-Torres}, \citenamefont
		{Dominguez-Juarez}, \citenamefont {Moya-Cessa}, \citenamefont {Torres},\ and\
		\citenamefont {Aragon}}]{Leon2015}%
	\BibitemOpen
	\bibfield  {author} {\bibinfo {author} {\bibfnamefont {R.}~\bibnamefont {{de
					J. Leon-Montiel}}}, \bibinfo {author} {\bibfnamefont {M.~A.}\ \bibnamefont
			{Quiroz-Juarez}}, \bibinfo {author} {\bibfnamefont {R.}~\bibnamefont
			{Quintero-Torres}}, \bibinfo {author} {\bibfnamefont {J.~L.}\ \bibnamefont
			{Dominguez-Juarez}}, \bibinfo {author} {\bibfnamefont {H.~M.}\ \bibnamefont
			{Moya-Cessa}}, \bibinfo {author} {\bibfnamefont {J.~P.}\ \bibnamefont
			{Torres}}, \ and\ \bibinfo {author} {\bibfnamefont {J.~L.}\ \bibnamefont
			{Aragon}},\ }\href {\doibase 10.1038/srep17339} {\bibfield  {journal}
		{\bibinfo  {journal} {Sci. Rep.}\ }\textbf {\bibinfo {volume} {5}},\ \bibinfo
		{pages} {17339} (\bibinfo {year} {2015})}\BibitemShut {NoStop}%
	\bibitem [{\citenamefont {Mostame}\ \emph {et~al.}(2012)\citenamefont
		{Mostame}, \citenamefont {Rebentrost}, \citenamefont {Eisfeld}, \citenamefont
		{Kerman}, \citenamefont {Tsomokos},\ and\ \citenamefont
		{Aspuru-Guzik}}]{Mostame2012}%
	\BibitemOpen
	\bibfield  {author} {\bibinfo {author} {\bibfnamefont {S.}~\bibnamefont
			{Mostame}}, \bibinfo {author} {\bibfnamefont {P.}~\bibnamefont {Rebentrost}},
		\bibinfo {author} {\bibfnamefont {A.}~\bibnamefont {Eisfeld}}, \bibinfo
		{author} {\bibfnamefont {A.~J.}\ \bibnamefont {Kerman}}, \bibinfo {author}
		{\bibfnamefont {D.~I.}\ \bibnamefont {Tsomokos}}, \ and\ \bibinfo {author}
		{\bibfnamefont {A.}~\bibnamefont {Aspuru-Guzik}},\ }\href@noop {} {\bibfield
		{journal} {\bibinfo  {journal} {New Journal of Physics}\ }\textbf {\bibinfo
			{volume} {14}},\ \bibinfo {pages} {105013} (\bibinfo {year}
		{2012})}\BibitemShut {NoStop}%
	\bibitem [{\citenamefont {Potocnik}\ \emph {et~al.}(2017)\citenamefont
		{Potocnik}, \citenamefont {Bargerbos}, \citenamefont {Schroeder},
		\citenamefont {Khan}, \citenamefont {Collodo}, \citenamefont {Gasparinetti},
		\citenamefont {Salathe}, \citenamefont {Creatore}, \citenamefont {Eichler},
		\citenamefont {T{\"u}reci}, \citenamefont {Chin},\ and\ \citenamefont
		{Wallraff}}]{Potocnik2017}%
	\BibitemOpen
	\bibfield  {author} {\bibinfo {author} {\bibfnamefont {A.}~\bibnamefont
			{Potocnik}}, \bibinfo {author} {\bibfnamefont {A.}~\bibnamefont {Bargerbos}},
		\bibinfo {author} {\bibfnamefont {F.~A. Y.~N.}\ \bibnamefont {Schroeder}},
		\bibinfo {author} {\bibfnamefont {S.~A.}\ \bibnamefont {Khan}}, \bibinfo
		{author} {\bibfnamefont {M.~C.}\ \bibnamefont {Collodo}}, \bibinfo {author}
		{\bibfnamefont {S.}~\bibnamefont {Gasparinetti}}, \bibinfo {author}
		{\bibfnamefont {Y.}~\bibnamefont {Salathe}}, \bibinfo {author} {\bibfnamefont
			{C.}~\bibnamefont {Creatore}}, \bibinfo {author} {\bibfnamefont
			{C.}~\bibnamefont {Eichler}}, \bibinfo {author} {\bibfnamefont {H.~E.}\
			\bibnamefont {T{\"u}reci}}, \bibinfo {author} {\bibfnamefont {A.~W.}\
			\bibnamefont {Chin}}, \ and\ \bibinfo {author} {\bibfnamefont
			{A.}~\bibnamefont {Wallraff}},\ }\href@noop {} {\bibfield  {journal}
		{\bibinfo  {journal} {arXiv:1710.07466 [quant-ph]}\ } (\bibinfo {year}
		{2017})}\BibitemShut {NoStop}%
	\bibitem [{\citenamefont {Schempp}\ \emph {et~al.}(2015)\citenamefont
		{Schempp}, \citenamefont {Guenter}, \citenamefont {Wuester}, \citenamefont
		{Weidemueller},\ and\ \citenamefont {Whitlock}}]{Schempp2015}%
	\BibitemOpen
	\bibfield  {author} {\bibinfo {author} {\bibfnamefont {H.}~\bibnamefont
			{Schempp}}, \bibinfo {author} {\bibfnamefont {G.}~\bibnamefont {Guenter}},
		\bibinfo {author} {\bibfnamefont {S.}~\bibnamefont {Wuester}}, \bibinfo
		{author} {\bibfnamefont {M.}~\bibnamefont {Weidemueller}}, \ and\ \bibinfo
		{author} {\bibfnamefont {S.}~\bibnamefont {Whitlock}},\ }\href@noop {}
	{\bibfield  {journal} {\bibinfo  {journal} {Phys. Rev. Lett.}\ }\textbf
		{\bibinfo {volume} {115}},\ \bibinfo {pages} {093002} (\bibinfo {year}
		{2015})}\BibitemShut {NoStop}%
	\bibitem [{\citenamefont {Sch\"onleber}\ \emph {et~al.}(2015)\citenamefont
		{Sch\"onleber}, \citenamefont {Eisfeld}, \citenamefont {Genkin},
		\citenamefont {Whitlock},\ and\ \citenamefont {W\"uster}}]{Schoenleber2015}%
	\BibitemOpen
	\bibfield  {author} {\bibinfo {author} {\bibfnamefont {D.}~\bibnamefont
			{Sch\"onleber}}, \bibinfo {author} {\bibfnamefont {A.}~\bibnamefont
			{Eisfeld}}, \bibinfo {author} {\bibfnamefont {M.}~\bibnamefont {Genkin}},
		\bibinfo {author} {\bibfnamefont {S.}~\bibnamefont {Whitlock}}, \ and\
		\bibinfo {author} {\bibfnamefont {S.}~\bibnamefont {W\"uster}},\ }\href
	{\doibase 10.1103/PhysRevLett.114.123005} {\bibfield  {journal} {\bibinfo
			{journal} {Phys. Rev. Lett.}\ }\textbf {\bibinfo {volume} {114}},\ \bibinfo
		{pages} {123005} (\bibinfo {year} {2015})}\BibitemShut {NoStop}%
	\bibitem [{\citenamefont {Genkin}\ \emph {et~al.}(2016)\citenamefont {Genkin},
		\citenamefont {Schoenleber}, \citenamefont {Wuester},\ and\ \citenamefont
		{Eisfeld}}]{Genkin2016}%
	\BibitemOpen
	\bibfield  {author} {\bibinfo {author} {\bibfnamefont {M.}~\bibnamefont
			{Genkin}}, \bibinfo {author} {\bibfnamefont {D.~W.}\ \bibnamefont
			{Schoenleber}}, \bibinfo {author} {\bibfnamefont {S.}~\bibnamefont
			{Wuester}}, \ and\ \bibinfo {author} {\bibfnamefont {A.}~\bibnamefont
			{Eisfeld}},\ }\href@noop {} {\bibfield  {journal} {\bibinfo  {journal} {J.
				Phys. B: At. Mol. Opt. Phys.}\ }\textbf {\bibinfo {volume} {49}},\ \bibinfo
		{pages} {134001} (\bibinfo {year} {2016})}\BibitemShut {NoStop}%
	\bibitem [{\citenamefont {Schoenleber}\ \emph {et~al.}(2016)\citenamefont
		{Schoenleber}, \citenamefont {Bentley},\ and\ \citenamefont
		{Eisfeld}}]{Schoenleber2016}%
	\BibitemOpen
	\bibfield  {author} {\bibinfo {author} {\bibfnamefont {D.~W.}\ \bibnamefont
			{Schoenleber}}, \bibinfo {author} {\bibfnamefont {C.~D.~B.}\ \bibnamefont
			{Bentley}}, \ and\ \bibinfo {author} {\bibfnamefont {A.}~\bibnamefont
			{Eisfeld}},\ }\href@noop {} {\bibfield  {journal} {\bibinfo  {journal}
			{arXiv:1611.02914 [quant-ph]}\ } (\bibinfo {year} {2016})}\BibitemShut
	{NoStop}%
	\bibitem [{\citenamefont {Guenter}\ \emph {et~al.}(2013)\citenamefont
		{Guenter}, \citenamefont {Schempp}, \citenamefont {de~Saint-Vincent},
		\citenamefont {Gavryusev}, \citenamefont {Helmrich}, \citenamefont {Hofmann},
		\citenamefont {Whitlock},\ and\ \citenamefont {Weidemueller}}]{Guenter2013}%
	\BibitemOpen
	\bibfield  {author} {\bibinfo {author} {\bibfnamefont {G.}~\bibnamefont
			{Guenter}}, \bibinfo {author} {\bibfnamefont {H.}~\bibnamefont {Schempp}},
		\bibinfo {author} {\bibfnamefont {M.~R.}\ \bibnamefont {de~Saint-Vincent}},
		\bibinfo {author} {\bibfnamefont {V.}~\bibnamefont {Gavryusev}}, \bibinfo
		{author} {\bibfnamefont {S.}~\bibnamefont {Helmrich}}, \bibinfo {author}
		{\bibfnamefont {C.~S.}\ \bibnamefont {Hofmann}}, \bibinfo {author}
		{\bibfnamefont {S.}~\bibnamefont {Whitlock}}, \ and\ \bibinfo {author}
		{\bibfnamefont {M.}~\bibnamefont {Weidemueller}},\ }\href {\doibase
		10.1126/science.1244843} {\bibfield  {journal} {\bibinfo  {journal}
			{Science}\ }\textbf {\bibinfo {volume} {342}},\ \bibinfo {pages} {954}
		(\bibinfo {year} {2013})}\BibitemShut {NoStop}%
	\bibitem [{\citenamefont {Cheneau}\ \emph {et~al.}(2012)\citenamefont
		{Cheneau}, \citenamefont {Barmettler}, \citenamefont {Poletti}, \citenamefont
		{Endres}, \citenamefont {Schausz}, \citenamefont {Fukuhara}, \citenamefont
		{Gross}, \citenamefont {Bloch}, \citenamefont {Kollath},\ and\ \citenamefont
		{Kuhr}}]{Cheneau2012}%
	\BibitemOpen
	\bibfield  {author} {\bibinfo {author} {\bibfnamefont {M.}~\bibnamefont
			{Cheneau}}, \bibinfo {author} {\bibfnamefont {P.}~\bibnamefont {Barmettler}},
		\bibinfo {author} {\bibfnamefont {D.}~\bibnamefont {Poletti}}, \bibinfo
		{author} {\bibfnamefont {M.}~\bibnamefont {Endres}}, \bibinfo {author}
		{\bibfnamefont {P.}~\bibnamefont {Schausz}}, \bibinfo {author} {\bibfnamefont
			{T.}~\bibnamefont {Fukuhara}}, \bibinfo {author} {\bibfnamefont
			{C.}~\bibnamefont {Gross}}, \bibinfo {author} {\bibfnamefont
			{I.}~\bibnamefont {Bloch}}, \bibinfo {author} {\bibfnamefont
			{C.}~\bibnamefont {Kollath}}, \ and\ \bibinfo {author} {\bibfnamefont
			{S.}~\bibnamefont {Kuhr}},\ }\href@noop {} {\bibfield  {journal} {\bibinfo
			{journal} {Nature}\ }\textbf {\bibinfo {volume} {481}},\ \bibinfo {pages}
		{484} (\bibinfo {year} {2012})}\BibitemShut {NoStop}%
	\bibitem [{\citenamefont {Jurcevic}\ \emph {et~al.}(2014)\citenamefont
		{Jurcevic}, \citenamefont {Lanyon}, \citenamefont {Hauke}, \citenamefont
		{Hempel}, \citenamefont {Zoller}, \citenamefont {Blatt},\ and\ \citenamefont
		{Roos}}]{jurcevic2014quasiparticle}%
	\BibitemOpen
	\bibfield  {author} {\bibinfo {author} {\bibfnamefont {P.}~\bibnamefont
			{Jurcevic}}, \bibinfo {author} {\bibfnamefont {B.}~\bibnamefont {Lanyon}},
		\bibinfo {author} {\bibfnamefont {P.}~\bibnamefont {Hauke}}, \bibinfo
		{author} {\bibfnamefont {C.}~\bibnamefont {Hempel}}, \bibinfo {author}
		{\bibfnamefont {P.}~\bibnamefont {Zoller}}, \bibinfo {author} {\bibfnamefont
			{R.}~\bibnamefont {Blatt}}, \ and\ \bibinfo {author} {\bibfnamefont
			{C.}~\bibnamefont {Roos}},\ }\href {\doibase doi:10.1038/nature13461}
	{\bibfield  {journal} {\bibinfo  {journal} {Nature}\ }\textbf {\bibinfo
			{volume} {511}},\ \bibinfo {pages} {202} (\bibinfo {year}
		{2014})}\BibitemShut {NoStop}%
	\bibitem [{\citenamefont {Richerme}\ \emph {et~al.}(2014)\citenamefont
		{Richerme}, \citenamefont {Gong}, \citenamefont {Lee}, \citenamefont {Senko},
		\citenamefont {Smith}, \citenamefont {Foss-Feig}, \citenamefont {Michalakis},
		\citenamefont {Gorshkov},\ and\ \citenamefont {Monroe}}]{Richerme2014}%
	\BibitemOpen
	\bibfield  {author} {\bibinfo {author} {\bibfnamefont {P.}~\bibnamefont
			{Richerme}}, \bibinfo {author} {\bibfnamefont {Z.-X.}\ \bibnamefont {Gong}},
		\bibinfo {author} {\bibfnamefont {A.}~\bibnamefont {Lee}}, \bibinfo {author}
		{\bibfnamefont {C.}~\bibnamefont {Senko}}, \bibinfo {author} {\bibfnamefont
			{J.}~\bibnamefont {Smith}}, \bibinfo {author} {\bibfnamefont
			{M.}~\bibnamefont {Foss-Feig}}, \bibinfo {author} {\bibfnamefont
			{S.}~\bibnamefont {Michalakis}}, \bibinfo {author} {\bibfnamefont {A.~V.}\
			\bibnamefont {Gorshkov}}, \ and\ \bibinfo {author} {\bibfnamefont
			{C.}~\bibnamefont {Monroe}},\ }\href@noop {} {\bibfield  {journal} {\bibinfo
			{journal} {Nature}\ }\textbf {\bibinfo {volume} {511}},\ \bibinfo {pages}
		{198} (\bibinfo {year} {2014})}\BibitemShut {NoStop}%
	\bibitem [{\citenamefont {Krinner}\ \emph {et~al.}(2015)\citenamefont
		{Krinner}, \citenamefont {Stadler}, \citenamefont {Husmann}, \citenamefont
		{Brantut},\ and\ \citenamefont {Esslinger}}]{Krinner2015}%
	\BibitemOpen
	\bibfield  {author} {\bibinfo {author} {\bibfnamefont {S.}~\bibnamefont
			{Krinner}}, \bibinfo {author} {\bibfnamefont {D.}~\bibnamefont {Stadler}},
		\bibinfo {author} {\bibfnamefont {D.}~\bibnamefont {Husmann}}, \bibinfo
		{author} {\bibfnamefont {J.-P.}\ \bibnamefont {Brantut}}, \ and\ \bibinfo
		{author} {\bibfnamefont {T.}~\bibnamefont {Esslinger}},\ }\href {\doibase
		doi:10.1038/nature14049} {\bibfield  {journal} {\bibinfo  {journal} {Nature}\
		}\textbf {\bibinfo {volume} {517}},\ \bibinfo {pages} {64} (\bibinfo {year}
		{2015})}\BibitemShut {NoStop}%
	\bibitem [{\citenamefont {Roati}\ \emph {et~al.}(2008)\citenamefont {Roati},
		\citenamefont {D'Errico}, \citenamefont {Fallani}, \citenamefont {Fattori},
		\citenamefont {Fort}, \citenamefont {Zaccanti}, \citenamefont {Modugno},
		\citenamefont {Modugno},\ and\ \citenamefont {Inguscio}}]{Roati2008}%
	\BibitemOpen
	\bibfield  {author} {\bibinfo {author} {\bibfnamefont {G.}~\bibnamefont
			{Roati}}, \bibinfo {author} {\bibfnamefont {C.}~\bibnamefont {D'Errico}},
		\bibinfo {author} {\bibfnamefont {L.}~\bibnamefont {Fallani}}, \bibinfo
		{author} {\bibfnamefont {M.}~\bibnamefont {Fattori}}, \bibinfo {author}
		{\bibfnamefont {C.}~\bibnamefont {Fort}}, \bibinfo {author} {\bibfnamefont
			{M.}~\bibnamefont {Zaccanti}}, \bibinfo {author} {\bibfnamefont
			{G.}~\bibnamefont {Modugno}}, \bibinfo {author} {\bibfnamefont
			{M.}~\bibnamefont {Modugno}}, \ and\ \bibinfo {author} {\bibfnamefont
			{M.}~\bibnamefont {Inguscio}},\ }\href@noop {} {\bibfield  {journal}
		{\bibinfo  {journal} {Nature}\ }\textbf {\bibinfo {volume} {453}},\ \bibinfo
		{pages} {895} (\bibinfo {year} {2008})}\BibitemShut {NoStop}%
	\bibitem [{\citenamefont {Billy}\ \emph {et~al.}(2008)\citenamefont {Billy},
		\citenamefont {Josse}, \citenamefont {Zuo}, \citenamefont {Bernard},
		\citenamefont {Hambrecht}, \citenamefont {Lugan}, \citenamefont {Clement},
		\citenamefont {Sanchez-Palencia}, \citenamefont {Bouyer},\ and\ \citenamefont
		{Aspect}}]{Billy2008}%
	\BibitemOpen
	\bibfield  {author} {\bibinfo {author} {\bibfnamefont {J.}~\bibnamefont
			{Billy}}, \bibinfo {author} {\bibfnamefont {V.}~\bibnamefont {Josse}},
		\bibinfo {author} {\bibfnamefont {Z.}~\bibnamefont {Zuo}}, \bibinfo {author}
		{\bibfnamefont {A.}~\bibnamefont {Bernard}}, \bibinfo {author} {\bibfnamefont
			{B.}~\bibnamefont {Hambrecht}}, \bibinfo {author} {\bibfnamefont
			{P.}~\bibnamefont {Lugan}}, \bibinfo {author} {\bibfnamefont
			{D.}~\bibnamefont {Clement}}, \bibinfo {author} {\bibfnamefont
			{L.}~\bibnamefont {Sanchez-Palencia}}, \bibinfo {author} {\bibfnamefont
			{P.}~\bibnamefont {Bouyer}}, \ and\ \bibinfo {author} {\bibfnamefont
			{A.}~\bibnamefont {Aspect}},\ }\href@noop {} {\bibfield  {journal} {\bibinfo
			{journal} {Nature}\ }\textbf {\bibinfo {volume} {453}},\ \bibinfo {pages}
		{891} (\bibinfo {year} {2008})}\BibitemShut {NoStop}%
	\bibitem [{\citenamefont {Kondov}\ \emph {et~al.}(2011)\citenamefont {Kondov},
		\citenamefont {McGehee}, \citenamefont {Zirbel},\ and\ \citenamefont
		{DeMarco}}]{Kondov2011}%
	\BibitemOpen
	\bibfield  {author} {\bibinfo {author} {\bibfnamefont {S.~S.}\ \bibnamefont
			{Kondov}}, \bibinfo {author} {\bibfnamefont {W.~R.}\ \bibnamefont {McGehee}},
		\bibinfo {author} {\bibfnamefont {J.~J.}\ \bibnamefont {Zirbel}}, \ and\
		\bibinfo {author} {\bibfnamefont {B.}~\bibnamefont {DeMarco}},\ }\href@noop
	{} {\bibfield  {journal} {\bibinfo  {journal} {Science}\ }\textbf {\bibinfo
			{volume} {334}},\ \bibinfo {pages} {66} (\bibinfo {year} {2011})}\BibitemShut
	{NoStop}%
	\bibitem [{\citenamefont {Semeghini}\ \emph {et~al.}(2015)\citenamefont
		{Semeghini}, \citenamefont {Landini}, \citenamefont {Castilho}, \citenamefont
		{Roy}, \citenamefont {Spagnolli}, \citenamefont {Trenkwalder}, \citenamefont
		{Fattori}, \citenamefont {Inguscio},\ and\ \citenamefont
		{Modugno}}]{Semeghini2014}%
	\BibitemOpen
	\bibfield  {author} {\bibinfo {author} {\bibfnamefont {G.}~\bibnamefont
			{Semeghini}}, \bibinfo {author} {\bibfnamefont {M.}~\bibnamefont {Landini}},
		\bibinfo {author} {\bibfnamefont {P.}~\bibnamefont {Castilho}}, \bibinfo
		{author} {\bibfnamefont {S.}~\bibnamefont {Roy}}, \bibinfo {author}
		{\bibfnamefont {G.}~\bibnamefont {Spagnolli}}, \bibinfo {author}
		{\bibfnamefont {A.}~\bibnamefont {Trenkwalder}}, \bibinfo {author}
		{\bibfnamefont {M.}~\bibnamefont {Fattori}}, \bibinfo {author} {\bibfnamefont
			{M.}~\bibnamefont {Inguscio}}, \ and\ \bibinfo {author} {\bibfnamefont
			{G.}~\bibnamefont {Modugno}},\ }\href {\doibase doi:10.1038/nphys3339}
	{\bibfield  {journal} {\bibinfo  {journal} {Nature Physics}\ }\textbf
		{\bibinfo {volume} {11}},\ \bibinfo {pages} {554} (\bibinfo {year}
		{2015})}\BibitemShut {NoStop}%
	\bibitem [{\citenamefont {Schreiber}\ \emph {et~al.}(2015)\citenamefont
		{Schreiber}, \citenamefont {Hodgman}, \citenamefont {Bordia}, \citenamefont
		{L\"uschen}, \citenamefont {Fischer}, \citenamefont {Vosk}, \citenamefont
		{Altman}, \citenamefont {Schneider},\ and\ \citenamefont
		{Bloch}}]{Schreiber2015}%
	\BibitemOpen
	\bibfield  {author} {\bibinfo {author} {\bibfnamefont {M.}~\bibnamefont
			{Schreiber}}, \bibinfo {author} {\bibfnamefont {S.~S.}\ \bibnamefont
			{Hodgman}}, \bibinfo {author} {\bibfnamefont {P.}~\bibnamefont {Bordia}},
		\bibinfo {author} {\bibfnamefont {H.~P.}\ \bibnamefont {L\"uschen}}, \bibinfo
		{author} {\bibfnamefont {M.~H.}\ \bibnamefont {Fischer}}, \bibinfo {author}
		{\bibfnamefont {R.}~\bibnamefont {Vosk}}, \bibinfo {author} {\bibfnamefont
			{E.}~\bibnamefont {Altman}}, \bibinfo {author} {\bibfnamefont
			{U.}~\bibnamefont {Schneider}}, \ and\ \bibinfo {author} {\bibfnamefont
			{I.}~\bibnamefont {Bloch}},\ }\href {\doibase doi:10.1126/science.aaa7432}
	{\bibfield  {journal} {\bibinfo  {journal} {Science}\ }\textbf {\bibinfo
			{volume} {349}},\ \bibinfo {pages} {842} (\bibinfo {year}
		{2015})}\BibitemShut {NoStop}%
	\bibitem [{\citenamefont {Smith}\ \emph {et~al.}(2016)\citenamefont {Smith},
		\citenamefont {Lee}, \citenamefont {Richerme}, \citenamefont {Neyenhuis},
		\citenamefont {Hess}, \citenamefont {Hauke}, \citenamefont {Heyl},
		\citenamefont {Huse},\ and\ \citenamefont {Monroe}}]{smith2015many}%
	\BibitemOpen
	\bibfield  {author} {\bibinfo {author} {\bibfnamefont {J.}~\bibnamefont
			{Smith}}, \bibinfo {author} {\bibfnamefont {A.}~\bibnamefont {Lee}}, \bibinfo
		{author} {\bibfnamefont {P.}~\bibnamefont {Richerme}}, \bibinfo {author}
		{\bibfnamefont {B.}~\bibnamefont {Neyenhuis}}, \bibinfo {author}
		{\bibfnamefont {P.}~\bibnamefont {Hess}}, \bibinfo {author} {\bibfnamefont
			{P.}~\bibnamefont {Hauke}}, \bibinfo {author} {\bibfnamefont
			{M.}~\bibnamefont {Heyl}}, \bibinfo {author} {\bibfnamefont {D.}~\bibnamefont
			{Huse}}, \ and\ \bibinfo {author} {\bibfnamefont {C.}~\bibnamefont
			{Monroe}},\ }\href {\doibase doi:10.1038/nphys3783} {\bibfield  {journal}
		{\bibinfo  {journal} {Nature Physics}\ }\textbf {\bibinfo {volume} {6}},\
		\bibinfo {pages} {23} (\bibinfo {year} {2016})}\BibitemShut {NoStop}%
	\bibitem [{\citenamefont {S{\o}rensen}\ and\ \citenamefont
		{M{\o}lmer}(1999)}]{sorensen1999quantum}%
	\BibitemOpen
	\bibfield  {author} {\bibinfo {author} {\bibfnamefont {A.}~\bibnamefont
			{S{\o}rensen}}\ and\ \bibinfo {author} {\bibfnamefont {K.}~\bibnamefont
			{M{\o}lmer}},\ }\href@noop {} {\bibfield  {journal} {\bibinfo  {journal}
			{Physical review letters}\ }\textbf {\bibinfo {volume} {82}},\ \bibinfo
		{pages} {1971} (\bibinfo {year} {1999})}\BibitemShut {NoStop}%
	\bibitem [{\citenamefont {Porras}\ and\ \citenamefont
		{Cirac}(2004)}]{Porras2004a}%
	\BibitemOpen
	\bibfield  {author} {\bibinfo {author} {\bibfnamefont {D.}~\bibnamefont
			{Porras}}\ and\ \bibinfo {author} {\bibfnamefont {J.~I.}\ \bibnamefont
			{Cirac}},\ }\href@noop {} {\bibfield  {journal} {\bibinfo  {journal} {Phys.
				Rev. Lett.}\ }\textbf {\bibinfo {volume} {92}},\ \bibinfo {pages} {207901}
		(\bibinfo {year} {2004})}\BibitemShut {NoStop}%
	\bibitem [{\citenamefont {Friedenauer}\ \emph {et~al.}(2008)\citenamefont
		{Friedenauer}, \citenamefont {Schmitz}, \citenamefont {Glueckert},
		\citenamefont {Porras},\ and\ \citenamefont {Schaetz}}]{Friedenauer2008}%
	\BibitemOpen
	\bibfield  {author} {\bibinfo {author} {\bibfnamefont {A.}~\bibnamefont
			{Friedenauer}}, \bibinfo {author} {\bibfnamefont {H.}~\bibnamefont
			{Schmitz}}, \bibinfo {author} {\bibfnamefont {J.~T.}\ \bibnamefont
			{Glueckert}}, \bibinfo {author} {\bibfnamefont {D.}~\bibnamefont {Porras}}, \
		and\ \bibinfo {author} {\bibfnamefont {T.}~\bibnamefont {Schaetz}},\
	}\href@noop {} {\bibfield  {journal} {\bibinfo  {journal} {Nat. Phys.}\
		}\textbf {\bibinfo {volume} {4}},\ \bibinfo {pages} {757} (\bibinfo {year}
		{2008})}\BibitemShut {NoStop}%
	\bibitem [{\citenamefont {Kim}\ \emph {et~al.}(2009)\citenamefont {Kim},
		\citenamefont {Chang}, \citenamefont {Islam}, \citenamefont {Korenblit},
		\citenamefont {Duan},\ and\ \citenamefont {Monroe}}]{kim2009entanglement}%
	\BibitemOpen
	\bibfield  {author} {\bibinfo {author} {\bibfnamefont {K.}~\bibnamefont
			{Kim}}, \bibinfo {author} {\bibfnamefont {M.-S.}\ \bibnamefont {Chang}},
		\bibinfo {author} {\bibfnamefont {R.}~\bibnamefont {Islam}}, \bibinfo
		{author} {\bibfnamefont {S.}~\bibnamefont {Korenblit}}, \bibinfo {author}
		{\bibfnamefont {L.-M.}\ \bibnamefont {Duan}}, \ and\ \bibinfo {author}
		{\bibfnamefont {C.}~\bibnamefont {Monroe}},\ }\href@noop {} {\bibfield
		{journal} {\bibinfo  {journal} {Physical review letters}\ }\textbf {\bibinfo
			{volume} {103}},\ \bibinfo {pages} {120502} (\bibinfo {year}
		{2009})}\BibitemShut {NoStop}%
	\bibitem [{\citenamefont {Britton}\ \emph {et~al.}(2012)\citenamefont
		{Britton}, \citenamefont {Sawyer}, \citenamefont {Keith}, \citenamefont
		{Wang}, \citenamefont {Freericks}, \citenamefont {Uys}, \citenamefont
		{Biercuk},\ and\ \citenamefont {Bollinger}}]{Britton2012}%
	\BibitemOpen
	\bibfield  {author} {\bibinfo {author} {\bibfnamefont {J.~W.}\ \bibnamefont
			{Britton}}, \bibinfo {author} {\bibfnamefont {B.~C.}\ \bibnamefont {Sawyer}},
		\bibinfo {author} {\bibfnamefont {A.~C.}\ \bibnamefont {Keith}}, \bibinfo
		{author} {\bibfnamefont {C.-C.~J.}\ \bibnamefont {Wang}}, \bibinfo {author}
		{\bibfnamefont {J.~K.}\ \bibnamefont {Freericks}}, \bibinfo {author}
		{\bibfnamefont {H.}~\bibnamefont {Uys}}, \bibinfo {author} {\bibfnamefont
			{M.~J.}\ \bibnamefont {Biercuk}}, \ and\ \bibinfo {author} {\bibfnamefont
			{J.~J.}\ \bibnamefont {Bollinger}},\ }\href {\doibase
		doi:10.1038/nature10981} {\bibfield  {journal} {\bibinfo  {journal} {Nature}\
		}\textbf {\bibinfo {volume} {484}},\ \bibinfo {pages} {489} (\bibinfo {year}
		{2012})}\BibitemShut {NoStop}%
	\bibitem [{\citenamefont {Bermudez}\ \emph {et~al.}(2012)\citenamefont
		{Bermudez}, \citenamefont {Almeida}, \citenamefont {Ott}, \citenamefont
		{Kaufmann}, \citenamefont {Ulm}, \citenamefont {Poschinger}, \citenamefont
		{Schmidt-Kaler}, \citenamefont {Retzker},\ and\ \citenamefont
		{Plenio}}]{Bermudez2012b}%
	\BibitemOpen
	\bibfield  {author} {\bibinfo {author} {\bibfnamefont {A.}~\bibnamefont
			{Bermudez}}, \bibinfo {author} {\bibfnamefont {J.}~\bibnamefont {Almeida}},
		\bibinfo {author} {\bibfnamefont {K.}~\bibnamefont {Ott}}, \bibinfo {author}
		{\bibfnamefont {H.}~\bibnamefont {Kaufmann}}, \bibinfo {author}
		{\bibfnamefont {S.}~\bibnamefont {Ulm}}, \bibinfo {author} {\bibfnamefont
			{U.}~\bibnamefont {Poschinger}}, \bibinfo {author} {\bibfnamefont
			{F.}~\bibnamefont {Schmidt-Kaler}}, \bibinfo {author} {\bibfnamefont
			{A.}~\bibnamefont {Retzker}}, \ and\ \bibinfo {author} {\bibfnamefont
			{M.~B.}\ \bibnamefont {Plenio}},\ }\href@noop {} {\bibfield  {journal}
		{\bibinfo  {journal} {New J. Phys.}\ }\textbf {\bibinfo {volume} {14}},\
		\bibinfo {pages} {093042} (\bibinfo {year} {2012})}\BibitemShut {NoStop}%
	\bibitem [{\citenamefont {Bohnet}\ \emph {et~al.}(2016)\citenamefont {Bohnet},
		\citenamefont {Sawyer}, \citenamefont {Britton}, \citenamefont {Wall},
		\citenamefont {Rey}, \citenamefont {Foss-Feig},\ and\ \citenamefont
		{Bollinger}}]{Bohnet2016}%
	\BibitemOpen
	\bibfield  {author} {\bibinfo {author} {\bibfnamefont {J.~G.}\ \bibnamefont
			{Bohnet}}, \bibinfo {author} {\bibfnamefont {B.~C.}\ \bibnamefont {Sawyer}},
		\bibinfo {author} {\bibfnamefont {J.~W.}\ \bibnamefont {Britton}}, \bibinfo
		{author} {\bibfnamefont {M.~L.}\ \bibnamefont {Wall}}, \bibinfo {author}
		{\bibfnamefont {A.~M.}\ \bibnamefont {Rey}}, \bibinfo {author} {\bibfnamefont
			{M.}~\bibnamefont {Foss-Feig}}, \ and\ \bibinfo {author} {\bibfnamefont
			{J.~J.}\ \bibnamefont {Bollinger}},\ }\href {\doibase
		10.1126/science.aad9958} {\bibfield  {journal} {\bibinfo  {journal}
			{Science}\ }\textbf {\bibinfo {volume} {352}},\ \bibinfo {pages} {1297}
		(\bibinfo {year} {2016})}\BibitemShut {NoStop}%
	\bibitem [{\citenamefont {Richerme}(2016)}]{Richerme2016}%
	\BibitemOpen
	\bibfield  {author} {\bibinfo {author} {\bibfnamefont {P.}~\bibnamefont
			{Richerme}},\ }\href {\doibase 10.1103/PhysRevA.94.032320} {\bibfield
		{journal} {\bibinfo  {journal} {Phys. Rev. A}\ }\textbf {\bibinfo {volume}
			{94}},\ \bibinfo {pages} {032320} (\bibinfo {year} {2016})}\BibitemShut
	{NoStop}%
	\bibitem [{\citenamefont {Zippilli}\ \emph {et~al.}(2014)\citenamefont
		{Zippilli}, \citenamefont {Johanning}, \citenamefont {Giampaolo},
		\citenamefont {Wunderlich},\ and\ \citenamefont {Illuminati}}]{Zippilli2014}%
	\BibitemOpen
	\bibfield  {author} {\bibinfo {author} {\bibfnamefont {S.}~\bibnamefont
			{Zippilli}}, \bibinfo {author} {\bibfnamefont {M.}~\bibnamefont {Johanning}},
		\bibinfo {author} {\bibfnamefont {S.~M.}\ \bibnamefont {Giampaolo}}, \bibinfo
		{author} {\bibfnamefont {C.}~\bibnamefont {Wunderlich}}, \ and\ \bibinfo
		{author} {\bibfnamefont {F.}~\bibnamefont {Illuminati}},\ }\href {\doibase
		10.1103/PhysRevA.89.042308} {\bibfield  {journal} {\bibinfo  {journal} {Phys.
				Rev. A}\ }\textbf {\bibinfo {volume} {89}},\ \bibinfo {pages} {042308}
		(\bibinfo {year} {2014})}\BibitemShut {NoStop}%
	\bibitem [{\citenamefont {Schmied}\ \emph {et~al.}(2009)\citenamefont
		{Schmied}, \citenamefont {Wesenberg},\ and\ \citenamefont
		{Leibfried}}]{Schmied2009b}%
	\BibitemOpen
	\bibfield  {author} {\bibinfo {author} {\bibfnamefont {R.}~\bibnamefont
			{Schmied}}, \bibinfo {author} {\bibfnamefont {J.~H.}\ \bibnamefont
			{Wesenberg}}, \ and\ \bibinfo {author} {\bibfnamefont {D.}~\bibnamefont
			{Leibfried}},\ }\href {\doibase 10.1103/PhysRevLett.102.233002} {\bibfield
		{journal} {\bibinfo  {journal} {Phys. Rev. Lett.}\ }\textbf {\bibinfo
			{volume} {102}},\ \bibinfo {pages} {233002} (\bibinfo {year}
		{2009})}\BibitemShut {NoStop}%
	\bibitem{Nevado2017}%
\BibitemOpen
\bibfield  {author} {
	\bibinfo {author} {\bibfnamefont {P.}~\bibnamefont
		{Nevado}}, 
	\bibinfo {author} {\bibfnamefont {S.}\ \bibnamefont
		{Fernandez-Lorenzo}}, \ and\ 
	\bibinfo {author} {\bibfnamefont {D.}~\bibnamefont
		{Porras}},\ }
{\bibfield
	{journal} {\bibinfo  {journal} {Phys. Rev. Lett.}\ }\textbf {\bibinfo
		{volume} {119}},\ \bibinfo {pages} {210401} (\bibinfo {year}
	{2017})}\BibitemShut {NoStop}%
	\bibitem{Korenblit2012}%
	\BibitemOpen
	\bibfield  {author} {
		\bibinfo {author} {\bibfnamefont {S.}~\bibnamefont
			{Korenblit}}, 
		\bibinfo {author} {\bibfnamefont {D.}\ \bibnamefont
			{Kafri}},
		\bibinfo {author} {\bibfnamefont {W.C.}\ \bibnamefont
			{Campbell}},
		\bibinfo {author} {\bibfnamefont {R.}\ \bibnamefont
			{Islam}}, 
				\bibinfo {author} {\bibfnamefont {E.E.}\ \bibnamefont
			{Edwards}},
		\bibinfo {author} {\bibfnamefont {Z.X.}\ \bibnamefont
			{Gong}},
		\bibinfo {author} {\bibfnamefont {G.D.}\ \bibnamefont
			{Lin}}, 
				\bibinfo {author} {\bibfnamefont {L.M.}\ \bibnamefont
			{Duan}}, 
		\bibinfo {author} {\bibfnamefont {J.}\ \bibnamefont
			{Kim}},
		\bibinfo {author} {\bibfnamefont {K.}\ \bibnamefont
			{Kim}}, \ and\ 
		\bibinfo {author} {\bibfnamefont {C.}~\bibnamefont
			{Monroe}},\ }
		{\bibfield
		{journal} {\bibinfo  {journal} {New J. Phys.}\ }\textbf {\bibinfo
			{volume} {14}},\ \bibinfo {pages} {095024} (\bibinfo {year}
		{2012})}\BibitemShut {NoStop}%
	\bibitem [{\citenamefont {Hauke}\ \emph {et~al.}(2015)\citenamefont {Hauke},
		\citenamefont {Bonnes}, \citenamefont {Heyl},\ and\ \citenamefont
		{Lechner}}]{Hauke2015}%
	\BibitemOpen
	\bibfield  {author} {\bibinfo {author} {\bibfnamefont {P.}~\bibnamefont
			{Hauke}}, \bibinfo {author} {\bibfnamefont {L.}~\bibnamefont {Bonnes}},
		\bibinfo {author} {\bibfnamefont {M.}~\bibnamefont {Heyl}}, \ and\ \bibinfo
		{author} {\bibfnamefont {W.}~\bibnamefont {Lechner}},\ }\href {\doibase
		10.3389/fphy.2015.00021} {\bibfield  {journal} {\bibinfo  {journal} {Front.
				Phys.}\ }\textbf {\bibinfo {volume} {3}},\ \bibinfo {pages} {21} (\bibinfo
		{year} {2015})}\BibitemShut {NoStop}%
	\bibitem [{\citenamefont {Soare}\ \emph {et~al.}(2014)\citenamefont {Soare},
		\citenamefont {Ball}, \citenamefont {Hayes}, \citenamefont {Zhen},
		\citenamefont {Jarratt}, \citenamefont {Sastrawan}, \citenamefont {Uys},\
		and\ \citenamefont {Biercuk}}]{Soare2014}%
	\BibitemOpen
	\bibfield  {author} {\bibinfo {author} {\bibfnamefont {A.}~\bibnamefont
			{Soare}}, \bibinfo {author} {\bibfnamefont {H.}~\bibnamefont {Ball}},
		\bibinfo {author} {\bibfnamefont {D.}~\bibnamefont {Hayes}}, \bibinfo
		{author} {\bibfnamefont {X.}~\bibnamefont {Zhen}}, \bibinfo {author}
		{\bibfnamefont {M.~C.}\ \bibnamefont {Jarratt}}, \bibinfo {author}
		{\bibfnamefont {J.}~\bibnamefont {Sastrawan}}, \bibinfo {author}
		{\bibfnamefont {H.}~\bibnamefont {Uys}}, \ and\ \bibinfo {author}
		{\bibfnamefont {M.~J.}\ \bibnamefont {Biercuk}},\ }\href@noop {} {\bibfield
		{journal} {\bibinfo  {journal} {Phys. Rev. A}\ }\textbf {\bibinfo {volume}
			{89}},\ \bibinfo {pages} {042329} (\bibinfo {year} {2014})}\BibitemShut
	{NoStop}%
	\bibitem [{\citenamefont {Hauke}\ and\ \citenamefont
		{Tagliacozzo}(2013)}]{hauke2013spread}%
	\BibitemOpen
	\bibfield  {author} {\bibinfo {author} {\bibfnamefont {P.}~\bibnamefont
			{Hauke}}\ and\ \bibinfo {author} {\bibfnamefont {L.}~\bibnamefont
			{Tagliacozzo}},\ }\href {\doibase 10.1103/PhysRevLett.111.207202} {\bibfield
		{journal} {\bibinfo  {journal} {Physical Review Letters}\ }\textbf {\bibinfo
			{volume} {111}},\ \bibinfo {pages} {207202} (\bibinfo {year}
		{2013})}\BibitemShut {NoStop}%
	\bibitem [{\citenamefont {Eisert}\ \emph {et~al.}(2013)\citenamefont {Eisert},
		\citenamefont {van~den Worm}, \citenamefont {Manmana},\ and\ \citenamefont
		{Kastner}}]{Eisert2013}%
	\BibitemOpen
	\bibfield  {author} {\bibinfo {author} {\bibfnamefont {J.}~\bibnamefont
			{Eisert}}, \bibinfo {author} {\bibfnamefont {M.}~\bibnamefont {van~den
				Worm}}, \bibinfo {author} {\bibfnamefont {S.~R.}\ \bibnamefont {Manmana}}, \
		and\ \bibinfo {author} {\bibfnamefont {M.}~\bibnamefont {Kastner}},\
	}\href@noop {\bibfield
	{journal} {\bibinfo  {journal} {Physical Review Letters}\ }\textbf {\bibinfo
		{volume} {111}},\ \bibinfo {pages} {260401} (\bibinfo {year}
	{2013})}\BibitemShut {NoStop}%
%
	\bibitem [{\citenamefont {Foss-Feig}\ \emph {et~al.}(2015)\citenamefont
		{Foss-Feig}, \citenamefont {Gong}, \citenamefont {Clark},\ and\ \citenamefont
		{Gorshkov}}]{Foss-Feig2015}%
	\BibitemOpen
	\bibfield  {author} {\bibinfo {author} {\bibfnamefont {M.}~\bibnamefont
			{Foss-Feig}}, \bibinfo {author} {\bibfnamefont {Z.-X.}\ \bibnamefont {Gong}},
		\bibinfo {author} {\bibfnamefont {C.~W.}\ \bibnamefont {Clark}}, \ and\
		\bibinfo {author} {\bibfnamefont {A.~V.}\ \bibnamefont {Gorshkov}},\ }\href
	{\doibase 10.1103/PhysRevLett.114.157201} {\bibfield  {journal} {\bibinfo
			{journal} {Phys. Rev. Lett.}\ }\textbf {\bibinfo {volume} {114}},\ \bibinfo
		{pages} {157201} (\bibinfo {year} {2015})}\BibitemShut {NoStop}%
	\bibitem [{\citenamefont {Cevolani}\ \emph {et~al.}(2015)\citenamefont
		{Cevolani}, \citenamefont {Carleo},\ and\ \citenamefont
		{Sanchez-Palencia}}]{Cevolani2015}%
	\BibitemOpen
	\bibfield  {author} {\bibinfo {author} {\bibfnamefont {L.}~\bibnamefont
			{Cevolani}}, \bibinfo {author} {\bibfnamefont {G.}~\bibnamefont {Carleo}}, \
		and\ \bibinfo {author} {\bibfnamefont {L.}~\bibnamefont {Sanchez-Palencia}},\
	}\href {\doibase 10.1103/PhysRevA.92.041603} {\bibfield  {journal} {\bibinfo
			{journal} {Physical Review A}\ }\textbf {\bibinfo {volume} {92}},\ \bibinfo
		{pages} {041603(R)} (\bibinfo {year} {2015})}\BibitemShut {NoStop}%
	\bibitem [{\citenamefont {Lieb}\ and\ \citenamefont
		{Robinson}(1972)}]{Lieb1972}%
	\BibitemOpen
	\bibfield  {author} {\bibinfo {author} {\bibfnamefont {E.}~\bibnamefont
			{Lieb}}\ and\ \bibinfo {author} {\bibfnamefont {D.}~\bibnamefont
			{Robinson}},\ }\href@noop {} {\bibfield  {journal} {\bibinfo  {journal}
			{Commun. Math. Phys.}\ }\textbf {\bibinfo {volume} {28}},\ \bibinfo {pages}
		{251} (\bibinfo {year} {1972})}\BibitemShut {NoStop}%
	\bibitem [{\citenamefont {Anderson}(1958)}]{anderson1958absence}%
	\BibitemOpen
	\bibfield  {author} {\bibinfo {author} {\bibfnamefont {P.~W.}\ \bibnamefont
			{Anderson}},\ }\href@noop {} {\bibfield  {journal} {\bibinfo  {journal}
			{Physical review}\ }\textbf {\bibinfo {volume} {109}},\ \bibinfo {pages}
		{1492} (\bibinfo {year} {1958})}\BibitemShut {NoStop}%
	\bibitem [{\citenamefont {Neyenhuis}\ \emph {et~al.}(2016)\citenamefont
		{Neyenhuis}, \citenamefont {Smith}, \citenamefont {Lee}, \citenamefont
		{Zhang}, \citenamefont {Richerme}, \citenamefont {Hess}, \citenamefont
		{Gong}, \citenamefont {Gorshkov},\ and\ \citenamefont
		{Monroe}}]{neyenhuis2016observation}%
	\BibitemOpen
	\bibfield  {author} {\bibinfo {author} {\bibfnamefont {B.}~\bibnamefont
			{Neyenhuis}}, \bibinfo {author} {\bibfnamefont {J.}~\bibnamefont {Smith}},
		\bibinfo {author} {\bibfnamefont {A.}~\bibnamefont {Lee}}, \bibinfo {author}
		{\bibfnamefont {J.}~\bibnamefont {Zhang}}, \bibinfo {author} {\bibfnamefont
			{P.}~\bibnamefont {Richerme}}, \bibinfo {author} {\bibfnamefont
			{P.}~\bibnamefont {Hess}}, \bibinfo {author} {\bibfnamefont {Z.-X.}\
			\bibnamefont {Gong}}, \bibinfo {author} {\bibfnamefont {A.}~\bibnamefont
			{Gorshkov}}, \ and\ \bibinfo {author} {\bibfnamefont {C.}~\bibnamefont
			{Monroe}},\ }\href@noop 
		{} {\bibfield  {journal} {\bibinfo  {journal} {arXiv
				preprint arXiv:1608.00681}\ } (\bibinfo {year} {2016})}
			\BibitemShut {NoStop}%
	\bibitem [{\citenamefont {Misra}\ and\ \citenamefont
		{Sudarshan}(1977)}]{misra1977zeno}%
	\BibitemOpen
	\bibfield  {author} {\bibinfo {author} {\bibfnamefont {B.}~\bibnamefont
			{Misra}}\ and\ \bibinfo {author} {\bibfnamefont {E.~C.~G.}\ \bibnamefont
			{Sudarshan}},\ }\href@noop {} {\bibfield  {journal} {\bibinfo  {journal}
			{Journal of Mathematical Physics}\ }\textbf {\bibinfo {volume} {18}},\
		\bibinfo {pages} {756} (\bibinfo {year} {1977})}\BibitemShut {NoStop}%
	\bibitem [{\citenamefont {Jesenko}\ and\ \citenamefont
		{Znidaric}(2013)}]{Jesenko2013}%
	\BibitemOpen
	\bibfield  {author} {\bibinfo {author} {\bibfnamefont {S.}~\bibnamefont
			{Jesenko}}\ and\ \bibinfo {author} {\bibfnamefont {M.}~\bibnamefont
			{Znidaric}},\ }\href {\doibase 10.1063/1.4802816} {\bibfield  {journal}
		{\bibinfo  {journal} {The Journal of Chemical Physics}\ }\textbf {\bibinfo
			{volume} {138}},\ \bibinfo {pages} {174103} (\bibinfo {year}
		{2013})}\BibitemShut {NoStop}%
	\bibitem [{\citenamefont {Goldstein}(1951)}]{goldstein1951diffusion}%
	\BibitemOpen
	\bibfield  {author} {\bibinfo {author} {\bibfnamefont {S.}~\bibnamefont
			{Goldstein}},\ }\href@noop {} {\bibfield  {journal} {\bibinfo  {journal} {The
				Quarterly Journal of Mechanics and Applied Mathematics}\ }\textbf {\bibinfo
			{volume} {4}},\ \bibinfo {pages} {129} (\bibinfo {year} {1951})}\BibitemShut
	{NoStop}%
	\bibitem [{\citenamefont {Masoliver}\ \emph {et~al.}(1989)\citenamefont
		{Masoliver}, \citenamefont {Lindenberg},\ and\ \citenamefont
		{Weiss}}]{masoliver1989continuous}%
	\BibitemOpen
	\bibfield  {author} {\bibinfo {author} {\bibfnamefont {J.}~\bibnamefont
			{Masoliver}}, \bibinfo {author} {\bibfnamefont {K.}~\bibnamefont
			{Lindenberg}}, and\ \bibinfo {author} {\bibfnamefont {G.~H.}\ \bibnamefont
			{Weiss}},\ }\href@noop {} {\bibfield  {journal} {\bibinfo  {journal} {Physica
				A: Statistical Mechanics and its Applications}\ }\textbf {\bibinfo {volume}
			{157}},\ \bibinfo {pages} {891} (\bibinfo {year} {1989})}\BibitemShut
	{NoStop}%
	\bibitem [{\citenamefont {Schindler}\ \emph {et~al.}(2013)\citenamefont
		{Schindler}, \citenamefont {Nigg}, \citenamefont {Monz}, \citenamefont
		{Barreiro}, \citenamefont {Martinez}, \citenamefont {Wang}, \citenamefont
		{Quint}, \citenamefont {Brandl}, \citenamefont {Nebendahl}, \citenamefont
		{Roos} \emph {et~al.}}]{schindler2013quantum}%
	\BibitemOpen
	\bibfield  {author} {\bibinfo {author} {\bibfnamefont {P.}~\bibnamefont
			{Schindler}}, \bibinfo {author} {\bibfnamefont {D.}~\bibnamefont {Nigg}},
		\bibinfo {author} {\bibfnamefont {T.}~\bibnamefont {Monz}}, \bibinfo {author}
		{\bibfnamefont {J.~T.}\ \bibnamefont {Barreiro}}, \bibinfo {author}
		{\bibfnamefont {E.}~\bibnamefont {Martinez}}, \bibinfo {author}
		{\bibfnamefont {S.~X.}\ \bibnamefont {Wang}}, \bibinfo {author}
		{\bibfnamefont {S.}~\bibnamefont {Quint}}, \bibinfo {author} {\bibfnamefont
			{M.~F.}\ \bibnamefont {Brandl}}, \bibinfo {author} {\bibfnamefont
			{V.}~\bibnamefont {Nebendahl}}, \bibinfo {author} {\bibfnamefont {C.~F.}\
			\bibnamefont {Roos}},  \emph {et~al.},\ }\href@noop {} {\bibfield  {journal}
		{\bibinfo  {journal} {New Journal of Physics}\ }\textbf {\bibinfo {volume}
			{15}},\ \bibinfo {pages} {123012} (\bibinfo {year} {2013})}\BibitemShut
	{NoStop}%
	\bibitem [{\citenamefont {James}(1998)}]{james1998quantum}%
	\BibitemOpen
	\bibfield  {author} {\bibinfo {author} {\bibfnamefont {D.~F.}\ \bibnamefont
			{James}},\ }\href@noop {} {\bibfield  {journal} {\bibinfo  {journal} {Applied
				Physics B: Lasers and Optics}\ }\textbf {\bibinfo {volume} {66}},\ \bibinfo
		{pages} {181} (\bibinfo {year} {1998})}\BibitemShut {NoStop}%
	\bibitem [{\citenamefont {Jurcevic}\ \emph {et~al.}(2015)\citenamefont
		{Jurcevic}, \citenamefont {Hauke}, \citenamefont {Maier}, \citenamefont
		{Hempel}, \citenamefont {Lanyon}, \citenamefont {Blatt},\ and\ \citenamefont
		{Roos}}]{Jurcevic2015}%
	\BibitemOpen
	\bibfield  {author} {\bibinfo {author} {\bibfnamefont {P.}~\bibnamefont
			{Jurcevic}}, \bibinfo {author} {\bibfnamefont {P.}~\bibnamefont {Hauke}},
		\bibinfo {author} {\bibfnamefont {C.}~\bibnamefont {Maier}}, \bibinfo
		{author} {\bibfnamefont {C.}~\bibnamefont {Hempel}}, \bibinfo {author}
		{\bibfnamefont {B.}~\bibnamefont {Lanyon}}, \bibinfo {author} {\bibfnamefont
			{R.}~\bibnamefont {Blatt}}, \ and\ \bibinfo {author} {\bibfnamefont
			{C.}~\bibnamefont {Roos}},\ }\href {\doibase 10.1103/PhysRevLett.115.100501}
	{\bibfield  {journal} {\bibinfo  {journal} {Phys. Rev. Lett.}\ }\textbf
		{\bibinfo {volume} {115}},\ \bibinfo {pages} {100501} (\bibinfo {year}
		{2015})}\BibitemShut {NoStop}%
	\bibitem [{\citenamefont {Nevado}\ and\ \citenamefont
		{Porras}(2016)}]{Nevado2014}%
	\BibitemOpen
	\bibfield  {author} {\bibinfo {author} {\bibfnamefont {P.}~\bibnamefont
			{Nevado}}\ and\ \bibinfo {author} {\bibfnamefont {D.}~\bibnamefont
			{Porras}},\ }\href {\doibase 10.1103/PhysRevA.93.013625} {\bibfield
		{journal} {\bibinfo  {journal} {Phys. Rev. A}\ }\textbf {\bibinfo {volume}
			{93}},\ \bibinfo {pages} {013625} (\bibinfo {year} {2016})}\BibitemShut
	{NoStop}%
	\bibitem {eberly1984noise}%
	\BibitemOpen
	\bibfield  {author} {\bibinfo {author} {\bibfnamefont {J.}~\bibnamefont
			{Eberly}}, \bibinfo {author} {\bibfnamefont {K.}~\bibnamefont
			{Wodkiewicz}}, and\ \bibinfo {author} {\bibfnamefont {B.}~\bibnamefont
			{Shore}},\ }\href@noop {} {\bibfield  {journal} {\bibinfo  {journal} {Physical Review A}\ }\textbf {\bibinfo {volume}
			{30}},\ \bibinfo {pages} {2381} (\bibinfo {year} {1984})}\BibitemShut
	{NoStop}%
	\bibitem {wodkiewicz1985random}%
	\BibitemOpen
	\bibfield  {author} {\bibinfo {author} {\bibfnamefont {K.}~\bibnamefont
			{Wodkiewicz}} and\ \bibinfo {author} {\bibfnamefont {J.}~\bibnamefont
			{Eberly}},\ }\href@noop {} {\bibfield  {journal} {\bibinfo  {journal} {Physical Review A}\ }\textbf {\bibinfo {volume}
			{32}},\ \bibinfo {pages} {992} (\bibinfo {year} {1985})}\BibitemShut
	{NoStop}%
		\bibitem {Rivas2014}%
	\BibitemOpen
	\bibfield  {author} {\bibinfo {author} {\bibfnamefont {A.}~\bibnamefont
			{Rivas}}, \bibinfo {author} {\bibfnamefont {S. F.}~\bibnamefont {Huelga}}, \
		and\ \bibinfo {author} {\bibfnamefont {M. B.}~\bibnamefont {Plenio}},\ }\href@noop
	{} {\bibfield  {journal} {\bibinfo  {journal} {Rep. Prog. Phys.}\ }\textbf {\bibinfo {volume} {77}},\ \bibinfo {pages} {094001}
		(\bibinfo {year} {2014})}\BibitemShut {NoStop}%
	\bibitem {Breuer2016}%
	\BibitemOpen
	\bibfield  {author} {\bibinfo {author} {\bibfnamefont {H.-P.}~\bibnamefont
			{Breuer}}, \bibinfo {author} {\bibfnamefont {E.-M.}~\bibnamefont {Laine}}, 
		\bibinfo {author} {\bibfnamefont {J.}~\bibnamefont {Piilo}},
		\
		and\ \bibinfo {author} {\bibfnamefont {B.}~\bibnamefont {Vacchini}},\ }\href@noop
	{} {\bibfield  {journal} {\bibinfo  {journal} {Rev. Mod. Phys.}\ }\textbf {\bibinfo {volume} {88}},\ \bibinfo {pages} {021002}
		(\bibinfo {year} {2016})}\BibitemShut {NoStop}%
	\bibitem [{\citenamefont {Johansson}\ \emph {et~al.}(2012)\citenamefont
		{Johansson}, \citenamefont {Nation},\ and\ \citenamefont
		{Nori}}]{johansson2012qutip}%
	\BibitemOpen
	\bibfield  {author} {\bibinfo {author} {\bibfnamefont {J.}~\bibnamefont
			{Johansson}}, \bibinfo {author} {\bibfnamefont {P.}~\bibnamefont {Nation}}, \
		and\ \bibinfo {author} {\bibfnamefont {F.}~\bibnamefont {Nori}},\ }\href@noop
	{} {\bibfield  {journal} {\bibinfo  {journal} {Computer Physics
				Communications}\ }\textbf {\bibinfo {volume} {183}},\ \bibinfo {pages} {1760}
		(\bibinfo {year} {2012})}\BibitemShut {NoStop}%
	\bibitem [{\citenamefont {Lee}\ \emph {et~al.}(2016)\citenamefont {Lee},
		\citenamefont {Smith}, \citenamefont {Richerme}, \citenamefont {Neyenhuis},
		\citenamefont {Hess}, \citenamefont {Zhang},\ and\ \citenamefont
		{Monroe}}]{lee2016engineering}%
	\BibitemOpen
	\bibfield  {author} {\bibinfo {author} {\bibfnamefont {A.~C.}\ \bibnamefont
			{Lee}}, \bibinfo {author} {\bibfnamefont {J.}~\bibnamefont {Smith}}, \bibinfo
		{author} {\bibfnamefont {P.}~\bibnamefont {Richerme}}, \bibinfo {author}
		{\bibfnamefont {B.}~\bibnamefont {Neyenhuis}}, \bibinfo {author}
		{\bibfnamefont {P.~W.}\ \bibnamefont {Hess}}, \bibinfo {author}
		{\bibfnamefont {J.}~\bibnamefont {Zhang}}, \ and\ \bibinfo {author}
		{\bibfnamefont {C.}~\bibnamefont {Monroe}},\ }\href {\doibase
		10.1103/PhysRevA.94.042308} {\bibfield  {journal} {\bibinfo  {journal} {Phys.
				Rev. A}\ }\textbf {\bibinfo {volume} {94}},\ \bibinfo {pages} {042308}
		(\bibinfo {year} {2016})}\BibitemShut {NoStop}%
	\bibitem {Johri2012}%
	\BibitemOpen
	\bibfield  {author} {\bibinfo
		{author} {\bibfnamefont {S.}~\bibnamefont {Johri}}, \ and\ \bibinfo
		{author} {\bibfnamefont {R.N.}~\bibnamefont {Bhatt}},\ }\href@noop {}
	{\bibfield  {journal} {\bibinfo  {journal} {Phys. Rev. Lett.}\ }\textbf
		{\bibinfo {volume} {109}},\ \bibinfo {pages} {076402} (\bibinfo {year}
		{2012})}\BibitemShut {NoStop}%
	\bibitem [{\citenamefont {Kirchmair}\ \emph {et~al.}(2009)\citenamefont
		{Kirchmair}, \citenamefont {Benhelm}, \citenamefont {Z{\"a}hringer},
		\citenamefont {Gerritsma}, \citenamefont {Roos},\ and\ \citenamefont
		{Blatt}}]{kirchmair2009deterministic}%
	\BibitemOpen
	\bibfield  {author} {\bibinfo {author} {\bibfnamefont {G.}~\bibnamefont
			{Kirchmair}}, \bibinfo {author} {\bibfnamefont {J.}~\bibnamefont {Benhelm}},
		\bibinfo {author} {\bibfnamefont {F.}~\bibnamefont {Z{\"a}hringer}}, \bibinfo
		{author} {\bibfnamefont {R.}~\bibnamefont {Gerritsma}}, \bibinfo {author}
		{\bibfnamefont {C.}~\bibnamefont {Roos}}, \ and\ \bibinfo {author}
		{\bibfnamefont {R.}~\bibnamefont {Blatt}},\ }\href@noop {} {\bibfield
		{journal} {\bibinfo  {journal} {New Journal of Physics}\ }\textbf {\bibinfo
			{volume} {11}},\ \bibinfo {pages} {023002} (\bibinfo {year}
		{2009})}\BibitemShut {NoStop}%
	\bibitem [{\citenamefont {Roos}\ \emph {et~al.}(2006)\citenamefont {Roos},
		\citenamefont {Chwalla}, \citenamefont {Kim}, \citenamefont {Riebe},\ and\
		\citenamefont {Blatt}}]{roos2006designer}%
	\BibitemOpen
	\bibfield  {author} {\bibinfo {author} {\bibfnamefont {C.}~\bibnamefont
			{Roos}}, \bibinfo {author} {\bibfnamefont {M.}~\bibnamefont {Chwalla}},
		\bibinfo {author} {\bibfnamefont {K.}~\bibnamefont {Kim}}, \bibinfo {author}
		{\bibfnamefont {M.}~\bibnamefont {Riebe}}, \ and\ \bibinfo {author}
		{\bibfnamefont {R.}~\bibnamefont {Blatt}},\ }\href@noop {} {\bibfield
		{journal} {\bibinfo  {journal} {Nature}\ }\textbf {\bibinfo {volume} {443}},\
		\bibinfo {pages} {316} (\bibinfo {year} {2006})}\BibitemShut {NoStop}%
	\bibitem [{\citenamefont {Nagourney}\ \emph {et~al.}(1986)\citenamefont
		{Nagourney}, \citenamefont {Sandberg},\ and\ \citenamefont
		{Dehmelt}}]{nagourney1986shelved}%
	\BibitemOpen
	\bibfield  {author} {\bibinfo {author} {\bibfnamefont {W.}~\bibnamefont
			{Nagourney}}, \bibinfo {author} {\bibfnamefont {J.}~\bibnamefont {Sandberg}},
		\ and\ \bibinfo {author} {\bibfnamefont {H.}~\bibnamefont {Dehmelt}},\
	}\href@noop {} {\bibfield  {journal} {\bibinfo  {journal} {Physical Review
				Letters}\ }\textbf {\bibinfo {volume} {56}},\ \bibinfo {pages} {2797}
		(\bibinfo {year} {1986})}\BibitemShut {NoStop}%
	\bibitem [{\citenamefont {Sherman}\ \emph {et~al.}(2013)\citenamefont
		{Sherman}, \citenamefont {Curtis}, \citenamefont {Szwer}, \citenamefont
		{Allcock}, \citenamefont {Imreh}, \citenamefont {Lucas},\ and\ \citenamefont
		{Steane}}]{PhysRevLett.111.180501}%
	\BibitemOpen
	\bibfield  {author} {\bibinfo {author} {\bibfnamefont {J.~A.}\ \bibnamefont
			{Sherman}}, \bibinfo {author} {\bibfnamefont {M.~J.}\ \bibnamefont {Curtis}},
		\bibinfo {author} {\bibfnamefont {D.~J.}\ \bibnamefont {Szwer}}, \bibinfo
		{author} {\bibfnamefont {D.~T.~C.}\ \bibnamefont {Allcock}}, \bibinfo
		{author} {\bibfnamefont {G.}~\bibnamefont {Imreh}}, \bibinfo {author}
		{\bibfnamefont {D.~M.}\ \bibnamefont {Lucas}}, \ and\ \bibinfo {author}
		{\bibfnamefont {A.~M.}\ \bibnamefont {Steane}},\ }\href@noop {} {\bibfield
		{journal} {\bibinfo  {journal} {Phys. Rev. Lett.}\ }\textbf {\bibinfo
			{volume} {111}},\ \bibinfo {pages} {180501} (\bibinfo {year}
		{2013})}\BibitemShut {NoStop}%
	\bibitem [{\citenamefont {Higgins}\ \emph {et~al.}(2017)\citenamefont
		{Higgins}, \citenamefont {Lovett},\ and\ \citenamefont
		{Gauger}}]{Higgins2017}%
	\BibitemOpen
	\bibfield  {author} {\bibinfo {author} {\bibfnamefont {K.~D.~B.}\
			\bibnamefont {Higgins}}, \bibinfo {author} {\bibfnamefont {B.~W.}\
			\bibnamefont {Lovett}}, \ and\ \bibinfo {author} {\bibfnamefont {E.~M.}\
			\bibnamefont {Gauger}},\ }\href@noop {} {\bibfield  {journal} {\bibinfo
			{journal} {J. Phys. Chem. C}\ }\textbf {\bibinfo {volume} {121}},\ \bibinfo
		{pages} {20714} (\bibinfo {year} {2017})}\BibitemShut {NoStop}%
	\bibitem [{\citenamefont {Porras}\ \emph {et~al.}(2008)\citenamefont {Porras},
		\citenamefont {Marquardt}, \citenamefont {von Delft},\ and\ \citenamefont
		{Cirac}}]{Porras2008}%
	\BibitemOpen
	\bibfield  {author} {\bibinfo {author} {\bibfnamefont {D.}~\bibnamefont
			{Porras}}, \bibinfo {author} {\bibfnamefont {F.}~\bibnamefont {Marquardt}},
		\bibinfo {author} {\bibfnamefont {J.}~\bibnamefont {von Delft}}, \ and\
		\bibinfo {author} {\bibfnamefont {J.~I.}\ \bibnamefont {Cirac}},\ }\href@noop
	{} {\bibfield  {journal} {\bibinfo  {journal} {Phys. Rev. A}\ }\textbf
		{\bibinfo {volume} {78}},\ \bibinfo {pages} {010101(R)} (\bibinfo {year}
		{2008})}\BibitemShut {NoStop}%
	\bibitem [{\citenamefont {Juenemann}\ \emph {et~al.}(2013)\citenamefont
		{Juenemann}, \citenamefont {Cadarso}, \citenamefont {Perez-Garcia},
		\citenamefont {Bermudez},\ and\ \citenamefont
		{Garcia-Ripoll}}]{Juenemann2013}%
	\BibitemOpen
	\bibfield  {author} {\bibinfo {author} {\bibfnamefont {J.}~\bibnamefont
			{Juenemann}}, \bibinfo {author} {\bibfnamefont {A.}~\bibnamefont {Cadarso}},
		\bibinfo {author} {\bibfnamefont {D.}~\bibnamefont {Perez-Garcia}}, \bibinfo
		{author} {\bibfnamefont {A.}~\bibnamefont {Bermudez}}, \ and\ \bibinfo
		{author} {\bibfnamefont {J.}~\bibnamefont {Garcia-Ripoll}},\ }\href@noop {}
	{\bibfield  {journal} {\bibinfo  {journal} {Phys. Rev. Lett.}\ }\textbf
		{\bibinfo {volume} {111}},\ \bibinfo {pages} {230404} (\bibinfo {year}
		{2013})}\BibitemShut {NoStop}%
\end{thebibliography}

\end{document}